\documentclass{aa} %

\usepackage{bm, txfonts}
\usepackage[T1]{fontenc}
\usepackage[colorlinks=true,citecolor=blue,linkcolor=blue,urlcolor=blue,breaklinks]{hyperref} %
\usepackage{natbib, graphicx}
\usepackage[usenames,dvipsnames]{xcolor}
\usepackage[nolist,printonlyused,nohyperlinks]{acronym}
\usepackage{comment, subcaption, stfloats, multirow, multicol, tabularx, booktabs, lipsum, upgreek} %

\iftrue  %
  \def\jvl#1{\textcolor{blue}{\bf[#1 -- JvL]}\xspace}
  \def\lco#1{\textcolor{orange}{\bf[#1 -- LCO]}\xspace}
  
  \definecolor{armygreen}{rgb}{0.29, 0.33, 0.13}
  \def\stv#1{\textcolor{armygreen}{\bf[#1 -- STV]}\xspace}
  \def\sms#1{{\bf\color{magenta}[#1 -- SMS]}\xspace}
  \def\ym#1{\textcolor{brown}{\bf[#1 -- YM]}\xspace}
  \def\ab#1{\textcolor{purple}{\bf[#1 -- AB]}\xspace}
  \def\ipm#1{\textcolor{OliveGreen}{\bf[#1 -- IPM]}\xspace}
  
  \def\ap#1{\textcolor{armygreen}{#1}\xspace}
\else
  \def\ipm#1{}
  \def\jvl#1{}
  \def\lco#1{}
  \def\sms#1{}
  \def\ym#1{}
  \def\ab#1{}
  \def\stv#1{}
  \def\lc#1{}
  \def\dv#1{}
  \def\ap#1{}
  \renewcommand{\textbf}[1]{{#1}}
\fi

\newcommand{\ppcc}{\ensuremath{{\rm pc\,cm^{-3}}}\xspace}
\newcommand{\pccm}{\ensuremath{{\rm pc\,cm^{-3}}}\xspace}
\newcommand{\mhzpms}{\ensuremath{{\rm MHz\,ms}^{-1}}\xspace}
\newcommand{\unisq}{Uniboard$^{2}$}

\newcommand*{\Msun}{\ensuremath{\mathrm{M}_\odot}}

\newcommand{\accom}[2]{\aclu{#1} (\acs{#1}; #2)}
\newcommand{\acpcom}[2]{\aclp{#1} \acused{#1} (\acsp{#1}; #2)}

\newcommand*{\psrdada}{{\sc psrdada}\xspace}

\newcommand*{\dm}{{\sc DARCMaster}\xspace}
\newcommand*{\al}{{\sc AMBERListener}\xspace}
\newcommand*{\amc}{{\sc AMBERClustering}\xspace}
\newcommand*{\dt}{{\sc DADATrigger}\xspace}
\newcommand*{\lt}{{\sc LOFARTrigger}\xspace}
\newcommand*{\vog}{{\sc VOEventGenerator}\xspace}
\newcommand*{\op}{{\sc OfflineProcessing}\xspace}
\newcommand*{\sbg}{{\sc SBGenerator}\xspace}
\newcommand*{\sw}{{\sc StatusWebsite}\xspace}

\graphicspath{{Figures/}}

\makeatletter
\renewcommand*\aa@pageof{, page \thepage{} of \pageref*{LastPage}}
\makeatother

\acrodef{ADC}{Analogue-to-Digital Converter}
\acrodef{ADU}{Analogue to Digital Unit}
\acrodef{AGN}{Active Galactic Nucleus}
\acrodefplural{AGN}[AGNs]{Active Galactic Nuclei}
\acrodef{ALERT}{Apertif-LOFAR Exploration of the Radio Transient sky}
\acrodef{ALTA}{Apertif Long-Term Archive}
\acrodef{AMBER}[{\sc AMBER}]{Apertif Monitor for Bursts Encountered in Real-Time}
\acrodef{Apertif}{APERture Tile In Focus}
\acrodef{Apertif-X}{Apertif correlator}
\acrodef{FEBF}{Apertif Front-End Beamformer}
\acrodef{Artamis}[ARTAMIS]{the All-Round Telescope Array Monitoring and Information System}
\acrodef{ARTS}{the Apertif Radio Transient System}
\acrodef{ARTS0}{ARTS Pulsar timing machine}
\acrodef{ATDB}{Apertif Task DataBase}
\acrodef{beam}{Group of beamlets or beamlet-channels that point in the same direction}
\acrodef{beamlet}{Beam formed subband by the Apertif FE BF, a small beam spanning one subband}
\acrodef{BER}{Bit Error Rate}
\acrodef{BF}{Beamformer}
\acrodef{BG}{Block Generator}
\acrodef{BN}{Back Node}
\acrodef{bps}{bits per second}
\acrodef{Bps}{bytes per second}
\acrodef{BRAM}{Block RAM} 
\acrodef{BSN}{Block Sequence Number}
\acrodef{BW}{BandWidth}
\acrodef{CB}{Compound Beam}
\acrodef{CCU}{Central Control Unit}
\acrodef{CDR}{Critical Design Review}
\acrodef{CHAN}{Channel, a frequency channel within a beamlet}
\acrodef{CGM}{Circumgalactic Medium}
\acrodef{channel}{Unit frequency band within a beamlet}
\acrodef{cint}{Complex integer (type cast)}
\acrodef{Corner-turn}{Transposes data in time (in series)}
\acrodef{COTS}{Commercial Of The Shelf}
\acrodef{CPU}{Central Processing Unit}
\acrodef{CR}{Change Request}
\acrodef{CRC}{Cyclic Redundancy Check}
\acrodef{Cross-connect}{Transposes data in space (in parallel)}
\acrodef{CW}{Carrier Wave (single frequency signal)}
\acrodef{DARC}[{\sc DARC}]{Data Analysis of Real-time Candidates}
\acrodef{DB}{Data Buffer}
\acrodef{DM}{dispersion measure}
\acrodef{DNS}{double neutron star}
\acrodef{DP}{Data Path (streaming interface)}
\acrodef{DSP}{Digital Signal Processing}
\acrodef{DT}{Delay Tracking}
\acrodef{eop}{End of Packet (or frame, or block)}
\acrodef{ESD}{Electro Static Discharge}
\acrodef{FB}{FilterBank}
\acrodef{Fchan_a}{Channel filterbank in Arts BF for SC3 and SC4}
\acrodef{Fchan_x}{Channel filterbank in Apertif X}
\acrodef{FF}{Flip flop (1 bit storage in FPGA)}
\acrodef{FFT}{Fast Fourier Transform}
\acrodef{FIFO}{First In First Out (buffer in FPGA)}
\acrodef{FIR}{Finite Impulse Response}
\acrodef{FN}{Front Node}
\acrodef{FoV}{Field of View}
\acrodef{FPA}{Focal Plane Array (= PAF)}
\acrodef{FPGA}{Field-Programmable Gate Array}
\acrodef{FR}{Functional Requirement}
\acrodef{FRB}{Fast Radio Burst} 
\acrodef{FS}{Fringe Stopping}
\acrodef{Fsub}{Subband filterbank in Apertif FE BF}
\acrodef{FW}{Firmware}
\acrodef{GbE}{Gigabit Ethernet}
\acrodef{GPU}{Graphics processing unit}
\acrodef{HDL}{Hardware Description Language}
\acrodef{HF}{High Frequency}
\acrodef{HEM}{HMC Extension Board (provides UniBoard2 with HMC and extra optical IO)}
\acrodef{HMC}{Hybrid Memory Cube}
\acrodef{HW}{Hardware}
\acrodef{IAB}{Incoherent-Array Beam}
\acrodef{iablet}{A incoherently array beamformed beamlet using both polarizations}
\acrodef{IBF}{Incoherent BeamFormer}
\acrodef{iblet}{Fixed index of a data slot that carries a selected subband or a selected beamlet}
\acrodef{IGM}{InterGalactic Medium}
\acrodef{Im}{Imaginary}
\acrodef{int}{Signed integer}
\acrodef{I/O}{Input/Output}
\acrodef{IO}[I/O]{Input/Output}
\acrodef{IP}{Internet Protocol}
\acrodef{ISM}{InterStellar Medium}
\acrodef{LCU}{Local Control Unit}
\acrodef{LNA}{Low-Noise Amplifier}
\acrodef{LOFAR}{LOw Frequency ARray}
\acrodef{LSbit}{Least Significant bit}
\acrodef{LUT}{Look Up Table (boolean logic in FPGA)}
\acrodef{M20K}{Block RAM in Arria10 FPGA, 1024 x 20b so about 2 kByte}
\acrodef{M9K}{Block RAM in Stratix IV FPGA, 1024 x 9b so about 1 kByte}
\acrodef{ML}{Machine Learning}
\acrodef{MAC}{Monitoring And Control}
\acrodef{MFFE}{Multi Frequency Front End}
\acrodef{MISO}{Master In Slave Out (for memory mapped interface) [14]}
\acrodef{MM}{Memory Mapped}
\acrodef{MOSI}{Master Out Slave In (for memory mapped interface) [14]}
\acrodef{MSbit}{Most Significant bit}
\acrodef{node}{Processing node (PN), typically one FPGA chip}
\acrodef{Nof}{Number of}
\acrodef{OEB}{Optical-Electrical Board}
\acrodef{PAF}{Phased Array Feed}
\acrodef{PBS}{Product Breakdown Structure}
\acrodef{PDR}{Preliminary Design Review}
\acrodef{PFB}{Polyphase Filter Bank}
\acrodef{PL}{Pipeline processing}
\acrodef{PN}{Processing Node}
\acrodef{PN2}{Processing Node on UniBoard2)}
\acrodef{power beam}{Full Stokes power values: I, Q, U, V}
\acrodef{PPS}{Pulse Per Second}
\acrodef{PSR}{Pulsar}
\acrodef{PT}{Phase Tracking}
\acrodef{Pulsar}{Pulsating radio star}
\acrodef{Rband}{Reorder TAB output for Arts BF - PL interface}
\acrodef{Rbeam}{Reorder and select beamlets for KCB = 40 compound beam directions Aperitif BF}
\acrodef{Re}{Real}
\acrodef{RF}{Radio Frequency}
\acrodef{RFI}{Radio Frequency Interference}
\acrodef{RO}{Radio Observatory}
\acrodef{RoHS}{Restriction of Hazardous Substances}
\acrodef{Rsub}{Reorder and select subbands for CBBW = 300 MHz in Aperitif BF}
\acrodef{RT}{Radio Telescope}
\acrodef{Rx}{Receive}
\acrodef{SB}{Synthesized Beam}
\acrodef{SC}{Science Case}
\acrodef{SEFD}{System-Equivalent Flux Density}
\acrodef{SFR}{star formation rate}
\acrodef{SISO}{Source In Sink Out (for streaming interface) [14]}
\acrodef{SKA}{Square Kilometre Array}
\acrodef{SST}{Subband statistics}
\acrodef{S/N}{Signal-to-Noise ratio}
\acrodef{SNR}{Signal to Noise Ratio}
\acrodef{SODIMM}{Small Outline Dual In-line Memory Module}
\acrodef{sop}{Start of Packet (or frame, or block)}
\acrodef{SOSI}{Source Out Sink In (for streaming interface) [14]}
\acrodef{SP}{Signal Path, 1 CB consists of Npol = 2 SP}
\acrodef{sps}{Samples per second}
\acrodef{SR}{Science Requirement}
\acrodef{stream}{One stream of data describe by a SOSI signal}
\acrodef{subband}{Coarse channel frequency band, unit output of the Apertif FE BF filterbank}
\acrodef{SW}{Software}
\acrodef{TAB}{Tied-Array Beam}
\acrodef{TABF}{ARTS Tied-Array Beamformer}
\acrodef{tablet}{A coherently array beamformed beamlet, a grating or pencil beam within the CB }
\acrodef{Tant}{Transpose: split over subbands and group S = 64 (>= Nant) antenna elements in the PAF }
\acrodef{Tarray}{Transpose: split over beamlets and group Ntp = Npol * Ndish paths in the array (= Tdish+ Tpol)}
\acrodef{TB}{Tracking Beam}
\acrodef{Tband}{Transpose: split over tablets and group Nband = 16 bands for the full CBBW}
\acrodef{TBB}{Transient Buffer Board}
\acrodef{TCP}{Transmission Control Protocol}
\acrodef{Tdish}{Transpose: split over beamlets and group Ndish = 12 dishes}
\acrodef{Telescope}{The whole WSRT array}
\acrodef{Tint}{Transpose: split over beamlets and group Nint_x = 800000 beamlet time samples}
\acrodef{Tint_x}{Transpose: split over beamlets and group Nint_x = 800000 beamlet time samples}
\acrodef{Tinv}{Inverse transpose of Tint_x at Arts BF output }
\acrodef{TP}{Telescope Path, 1 TP contains all NCB = 37 SP for one single polarization from one dish}
\acrodef{Tpol}{Transpose: split over beamlets and group Npol = 2 polarizations}
\acrodef{Transpose}{Swaps indices in an signal array [15]}
\acrodef{Tx}{Transmit}
\acrodef{UCP}{UniBoard Control Protocol}
\acrodef{UDP}{User Datagram Protocol}
\acrodef{uint}{Unsigned integer}
\acrodef{UNB}{Uniboard}
\acrodef{UNB2}{Uniboard$^{2}$}
\acrodef{UniBoard}{Digital processing board with 8 Altera Stratix IV FPGAs}
\acrodef{UniBoard2}{Digital processing board with 4 Altera Arria 10 FPGAs}
\acrodef{UPE}{UniBoard Python Environment}
\acrodef{UTC}{Coordinated Universal Time}
\acrodef{VLBI}{Very Long Baseline Interferometry}
\acrodef{VO}{Virtual Observatory}
\acrodef{voltage beam}{Dual polarization sample values with phase information:  Xre, Xim, Yre, Yim}
\acrodef{WSRT}{Westerbork Synthesis Radio Telescope}
\acrodef{X}{Correlator}
 
\begin{document} 

\title{The Apertif Radio Transient System (ARTS):\\
Design, Commissioning, Data Release, and Detection of the first~5~Fast Radio Bursts}

\authorrunning{van Leeuwen et al.}
\titlerunning{ARTS -- System Overview and Detection of the first 5 FRBs}
\author{%
     Joeri~van~Leeuwen    \inst{\ref{astron} \and \ref{uva}}
\and Eric~Kooistra        \inst{\ref{astron}}
\and Leon~Oostrum         \inst{\ref{astron} \and \ref{uva} \and \ref{escience}}
\and Liam~Connor          \inst{\ref{astron} \and \ref{caltech}}
\and Jonathan~E.~Hargreaves     \inst{\ref{astron}}
\and Yogesh~Maan          \inst{\ref{ncra} \and \ref{astron}}
\and In{\'e}s~Pastor-Marazuela \inst{\ref{uva} \and \ref{astron}}
\and Emily~Petroff        \inst{\ref{uva} \and \ref{veni} \and \ref{mu}}
\and Daniel~van~der~Schuur \inst{\ref{astron}}
\and Alessio~Sclocco      \inst{\ref{escience}}
\and Samayra~M.~Straal    \inst{\ref{nyuad} \and \ref{nyuadcfa}}
\and Dany~Vohl            \inst{\ref{uva} \and \ref{astron}}
\and Stefan~J.~Wijnholds  \inst{\ref{astron}}
\and Elizabeth~A.~K.~Adams \inst{\ref{astron} \and \ref{kapteyn}}
\and Bj{\"o}rn~Adebahr    \inst{\ref{airub}}
\and Jisk~Attema          \inst{\ref{escience}}
\and Cees~Bassa        \inst{\ref{astron}}
\and Jeanette~E.~Bast     \inst{\ref{astron}}
\and Anna~Bilous          \inst{\ref{astron}}
\and W.~J.~G.~de~Blok     \inst{\ref{astron} \and \ref{cpt} \and \ref{kapteyn}}
\and Oliver~M.~Boersma \inst{\ref{uva}}
\and Wim~A.~van~Cappellen \inst{\ref{astron}}
\and Arthur~H.~W.~M.~Coolen \inst{\ref{astron}}
\and Sieds~Damstra        \inst{\ref{astron}}
\and Helga~D{\'e}nes      \inst{\ref{astron}}
\and Ger~N.~J.~van~Diepen \inst{\ref{astron}}
\and David~W.~Gardenier   \inst{\ref{astron} \and \ref{uva}}
\and Yan~G.~Grange        \inst{\ref{astron}}
\and Andr{\'e}~W.~Gunst   \inst{\ref{astron}}
\and Kelley~M.~Hess       \inst{\ref{iaa} \and \ref{astron} \and \ref{kapteyn}}
\and Hanno~Holties     \inst{\ref{astron}}
\and Thijs~van~der~Hulst  \inst{\ref{kapteyn}}
\and Boudewijn~Hut        \inst{\ref{astron}}
\and Alexander~Kutkin     \inst{\ref{astron}}
\and G.~Marcel~Loose      \inst{\ref{astron}}
\and Danielle~M.~Lucero   \inst{\ref{virginiatech}}
\and {\'A}gnes~Mika       \inst{\ref{astron}}
\and Klim~Mikhailov       \inst{\ref{uva} \and \ref{astron}}
\and Raffaella~Morganti   \inst{\ref{astron} \and \ref{kapteyn}}
\and Vanessa~A.~Moss      \inst{\ref{csiro} \and \ref{sydney} \and \ref{astron}}
\and Henk~Mulder          \inst{\ref{astron}}
\and Menno~J.~Norden     \inst{\ref{astron}}
\and Tom~A.~Oosterloo     \inst{\ref{astron} \and \ref{kapteyn}}
\and Emaneula~Orr{\'u}    \inst{\ref{astron}}
\and Zsolt~Paragi         \inst{\ref{jive}}
\and Jan-Pieter~R.~de~Reijer \inst{\ref{astron}}
\and Arno~P.~Schoenmakers \inst{\ref{astron}}
\and Klaas~J.~C.~Stuurwold \inst{\ref{astron}}
\and Sander~ter~Veen      \inst{\ref{astron}}
\and Yu-Yang~Wang         \inst{\ref{uva}}
\and Alwin~W.~Zanting     \inst{\ref{astron}}
\and Jacob~Ziemke         \inst{\ref{astron} \and \ref{oslocit}}
}

\institute{ASTRON, the Netherlands Institute for Radio Astronomy, Oude Hoogeveensedijk 4,7991 PD Dwingeloo, The Netherlands\label{astron}
  \and
Anton Pannekoek Institute, University of Amsterdam, Postbus 94249, 1090 GE Amsterdam, The Netherlands\label{uva}
  \and
Netherlands eScience Center, Science Park 402, 1098 XH Amsterdam, The Netherlands\label{escience}
  \and
Cahill Center for Astronomy, California Institute of Technology, Pasadena, CA, USA\label{caltech}
  \and
National Centre for Radio Astrophysics, Tata Institute of Fundamental Research, Pune 411007, Maharashtra, India\label{ncra}
  \and
Veni Fellow\label{veni}
  \and
Department of Physics, McGill University, 3600 rue University, Montr{\'e}al, QC H3A 2T8, Canada\label{mu}
  \and
NYU Abu Dhabi, PO Box 129188, Abu Dhabi, United Arab Emirates\label{nyuad}
  \and
Center for Astro, Particle, and Planetary Physics (CAP$^3$), NYU Abu Dhabi, PO Box 129188, Abu Dhabi, United Arab Emirates\label{nyuadcfa}
  \and
Kapteyn Astronomical Institute, University of Groningen, PO Box 800, 9700 AV Groningen, The Netherlands\label{kapteyn}
  \and
  Astronomisches Institut der Ruhr-Universit{\"a}t Bochum (AIRUB), Universit{\"a}tsstrasse 150, 44780 Bochum, Germany\label{airub}
  \and
Dept.\ of Astronomy, Univ.\ of Cape Town, Private Bag X3, Rondebosch 7701, South Africa\label{cpt}
  \and
Instituto de Astrof{\'i}sica de Andaluc{\'i}a (CSIC), Glorieta de la Astronom{\'i}a s/n, 18008 Granada,  Spain\label{iaa}
  \and
Department of Physics, Virginia Polytechnic Institute and State University, 50 West Campus Drive, Blacksburg, VA 24061, USA\label{virginiatech}
  \and
CSIRO Astronomy and Space Science, Australia Telescope National Facility, PO Box 76, Epping NSW 1710, Australia\label{csiro}
  \and
Sydney Institute for Astronomy, School of Physics, University of Sydney, Sydney, New South Wales 2006, Australia\label{sydney}
  \and
Joint Institute for VLBI ERIC (JIVE), Oude Hoogeveensedijk 4, 7991 PD Dwingeloo, The Netherlands\label{jive}
  \and
University of Oslo Center for Information Technology, P.O. Box 1059, 0316 Oslo, Norway\label{oslocit}}

\abstract{
Fast Radio Bursts must be powered by uniquely energetic emission mechanisms.
This requirement has eliminated a number of possible source types,
but several remain.
Identifying the physical nature of Fast Radio Burst (FRB) emitters arguably requires
good localisation of more detections,
and broadband studies enabled by real-time alerting. 
We here present \acf{ARTS},
a supercomputing radio-telescope instrument that performs real-time FRB detection and localisation
on the \acf{WSRT} interferometer. %
It reaches coherent-addition sensitivity over the entire field of the view of the primary-dish beam.
After commissioning results verified the system performed as planned,
we initiated the Apertif FRB survey (ALERT).
Over the first 5 weeks we observed at design sensitivity in 2019, we detected 5 new FRBs,
and interferometrically localised each of these to 0.4$-$10\,sq.\,arcmin.
All detections are broad band and very narrow, of order 1\,ms duration, and unscattered. 
Dispersion measures are generally high. %
Only through the very high time and frequency resolution of ARTS
are these hard-to-find FRBs detected, producing an unbiased view of the intrinsic population properties.
Most localisation regions are small enough to rule out the presence of associated persistent radio sources.
Three FRBs cut through the halos of M31 and M33. 
We demonstrate that Apertif can localise one-off FRBs with an accuracy that maps 
 magneto-ionic material along well-defined lines of sight.
The rate of 1 every $\sim$7\,days next ensures a considerable number of new sources are detected for such study. 
The combination of detection rate and localisation accuracy exemplified by the 5 first ARTS FRBs
thus marks a new phase in which a growing number of bursts can be used to probe our Universe. 
}

\keywords{FRBs -- pulsars: general -- instrumentation}
\maketitle

\section{Introduction}
\label{sec:1}

\label{sec:1b}

\begin{figure}
  \centering
  \includegraphics[width=0.74\columnwidth]{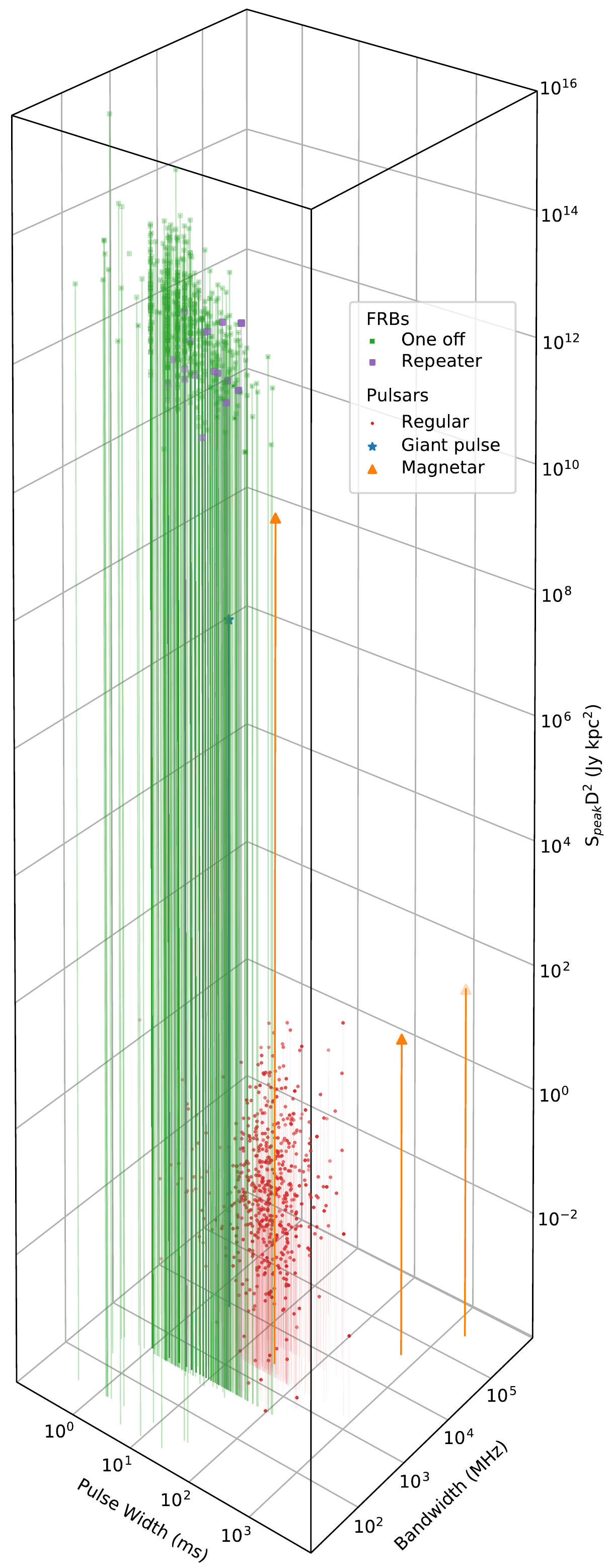} %
  \caption{Phase space diagram of emission duration and emission bandwidth
    versus pseudo luminosity
    ($S_{pk}D^2$), for different fast radio transients at peak flux $S_{pk}$ and distance $D$.
    FRBs are similar to other fast transients in their short and relatively broadband emission;
    but in pseudo luminosity they strongly stand out from the pulsars, magnetars, and giant pulses also plotted.
    Catalog data, per 2022 Feb 1, taken from
    FRBCAT \citep{2016PASA...33...45P},
    the Transient Name Server (TNS; \url{www.wis-tns.org}),
    ATNF \citep{mhth05}, and
    EPN 
    were complemented by individual data from
    \citet{2006Natur.442..892C}, \citet{2019ApJ...882L...9M}, \citet{2019ApJ...874L..14D}, \citet{2018ApJ...866..160P},
    \citet{2016ApJ...832..212M}, \citet{2019MNRAS.490L..12B}, and \citet{bochenek_atel_2020}.
  } %
  \label{fig:phase}
\end{figure}

Of all the time variable sources observable with modern radio telescopes, none have produced such a flurry of excitement
in recent years as \acp{FRB}. These bursts were discovered in \citeyear{lbm+07} by \citeauthor{lbm+07} and have since
emerged as a unique source class, characterized by bright, short radio pulses (Fig.~\ref{fig:phase}) with fluences
$\mathcal{F} \sim 1$ Jy\,ms. Thanks to large-scale, high time resolution radio surveys the population of known FRBs has
grown rapidly, with more than 600 FRB sources now published \citep[see][for recent reviews]{phl19,Petroff_2021_arXiv}.

While the population of FRBs has grown rapidly, and even though a $\sim$4\% subset were discovered to  repeat
\citep{spitler-2016,2019ApJ...885L..24C,2020arXiv200103595F}, their progenitors  remain a mystery.
The detection in 2020 of a bright FRB-like burst from a magnetar in our own Galaxy
\citep{CHIME/FRBCollaboration_2020_Natur,Bochenek_2020_Natur} favors theories that produce FRBs in
highly magnetised neutron stars, perhaps in extreme environments, in distant galaxies.
However, a wide range of other theories have also been proposed, from neutron stars to black
holes to stellar explosions \citep[see][for a theory review]{pww+19}.

\label{sec:1b.1}

Much observational progress has occurred since FRBs were first discovered; however,
many fundamental questions about their
nature remain. The total fraction of FRBs that repeat is unknown,
with many
attempts made to model the underlying population \citep{2021A&A...647A..30G,James_2020_MNRAS}.
Only sustained FRB follow-up %
can  detect repeats; whether all FRBs eventually reappear drives investigation in ongoing and upcoming surveys.

New surveys also target other FRB properties. The polarization behavior
may provide important insights into the FRB emission mechanism,
but only for a subset of FRBs could  polarization
information be recorded.
Of those, many are highly linearly polarized
\citep{2018Natur.553..182M,bannister19,clo+20};
others, however, show distinct circular polarization \citep{pbb+15,Kumar_2021_arXiv}. Some linearly
polarized FRBs %
reside in highly magnetized \citep{2018Natur.553..182M} and highly variable \citep{Hilmarsson_2021_ApJL} environments. 

Recently, periodic behavior on several timescales has also been observed for a number of FRBs.
Two sources, FRB 20121102A
and FRB 20180916B, have distinct windows of active emission, periodic on timescales of $\sim$160 and 16.3 days,
respectively \citep{Rajwade_2020_MNRAS,Cruces_2021_MNRAS,CHIME/FRBCollaboration_2020_Natur}. The activity window of FRB
20180916B is frequency dependent \citep{2020arXiv201208348P, Pleunis_2021_ApJL}, perhaps indicative of precession, or occultation by a
binary wind.
Sub-burst periodicity (periodic spacing of sub-components within a single burst),
has also been observed at high significance for one FRB and at lower significance for three others \citep{CHIME/FRBCollaboration_2021_arXiv,Pastor-Marazuela_2022_arXiv}.

Despite the high all-sky FRB rate, estimated to be a few thousand FRBs sky$^{-1}$ day$^{-1}$ \citep{Chawla_2017_ApJ},
current survey instruments face many challenges. Blind searches require
covering a large range of pulse duration and  \ac{DM} trials.
The enormous search parameter space requires dedicated hardware and software,
typically housed on massive on-site compute clusters. %

Effective FRB searches further require a large instantaneous \ac{FoV} survey instrument,
to maximise the probability
of observing an FRB where it goes off.
In the past, this has come at the expense of localization accuracy, and many FRBs are only localized to
$10' \times 10'$.
The study of FRB host galaxies is of increasing interest, but so far only two dozen have been identified to
sufficiently precise localization, %
provided by a
radio interferometer, or even \acl{VLBI}  \citep[VLBI; e.g.,][]{2020Natur.577..190M}.
\acused{VLBI}

Next-generation
telescopes such as the Square Kilometre Array (SKA) can no longer preserve raw survey data for
offline searches \citep{2015aska.confE..55M}. New sources need to be identified in real time.
This requires development of automated FRB search techniques and pipelines taking advantage of classification and machine learning tools  \citep{connor-2018b}.

\newlength{\introspacefixer}
\setlength{\introspacefixer}{0ex}

\vspace{\introspacefixer}
\label{sec:1d}

To address all these challenges, new FRB search efforts are increasingly employing interferometers to survey the sky
\citep{2017MNRAS.468.3746C,2017ApJ...841L..12B,2017arXiv170906104M,2018ApJS..236....8L,2018ApJ...863...48C}. Interferometers,
coherently or incoherently combining signals from many smaller elements or dishes, have the advantage of a large
instantaneous field of view. Recent technological advances have further led to
\acl{PAF}s (PAFs; Sect.~\ref{sec:3}), which place  additional elements at the focus of each dish in the array.
\acused{PAF}

\vspace{\introspacefixer}
\label{sec:1d.1}

One of the largest challenges of interferometric radio astronomy is  the  computation. Beamforming the \ac{FoV} 
 requires powerful signal processing to combine all elements in phase. This is exacerbated when combining  multi-element
 \ac{PAF} systems.
 Next searching the resulting time streams for impulsive radio signals such as FRBs is an additional challenge. 

Recently, faster and more agile processing has made it possible to form more beams and search
them in real time. Many searches now employ large, dedicated compute clusters to deal with the massive
amounts of data streaming from the telescope, and distribute it over numerous processing nodes (Sect.~\ref{sec:4}).

\vspace{\introspacefixer}
\label{sec:1d.2}

The raw localisation ability of an interferometer depends on its baseline length,
and ability to form all resolution elements.
Even more precise
localisation is possible for brighter signals that appear in several beams (cf.~Sect.~\ref{sec:6b.4}),
or 
when refined  correlation of voltage streams is possible offline.

Only interferometers can provide the arcsecond or better localisation needed to identify an FRB host galaxy unambiguously
\citep{2018ApJ...860...73E}.
More than 20 FRBs were traced back to their hosts using arrays such as ASKAP, DSA-10, the EVN, and the VLA \citep[e.g.,][]{Chatterjee2017,bannister19,2019Natur.572..352R,2020Natur.577..190M}. 

\vspace{\introspacefixer}
\label{sec:1e}

Since repeat characteristics are unknown, FRBs generally need to be localised from the discovery pulse.
Identifying it in real time in the incoming  stream can aid such localisation efforts,
by  preserving data at higher resolution and/or with the  full Stokes
parameters \citep[e.g.,][]{pbb+15,2020arXiv200103595F}.
  
A real-time FRB detection  also enables  triggering of other telescopes to conduct multi-wavelength or multi-messenger
follow-up.
Several theories \citep{pww+19} predict emission in e.g., optical, X-ray, or gravitational waves.
As yet, no such emission was observed for extragalactic FRBs. X-ray emission from the Galactic magnetar SGR 1935+2154
coincident with its bright radio burst \citep{Mereghetti_2020_ApJL} suggests counterparts may be detectable if follow-up
observations are conducted promptly. Triggers on FRBs are now possible with Virtual Observatory Events \citep[VOEvents;][]{Petroff_2017_arXiv}.

The frequency range over which FRBs are detectable remains unknown.
Early %
searches
with the Low Frequency Array (LOFAR) and the Murchison Widefield Array (MWA) at frequencies of 100$-$200 MHz were
unsuccessful \citep{clh+14,2018ApJ...867L..12S,2020arXiv200402862C}. However, studies of repeating FRBs at low
frequencies, including commensal observing between WSRT and LOFAR, have proven more fruitful. LOFAR detected several bursts from
repeating FRB 20180916B, leading to the 
discovery of its frequency-dependent activity window 
\citep{2020arXiv201208348P,Pleunis_2021_ApJL}.
Finding coincident emission at high and low radio frequencies generally requires real-time classification.

\label{sec:1f}

To investigate some of these
outstanding core questions on FRBs, 
we designed the Apertif Radio Transient System (ARTS;   \citealt{leeu14})\acused{ARTS}
to harness
Apertif, the APERture Tile In Focus \acused{Apertif}\citep{cho+20} on 
the \acf{WSRT}
for finding and studying FRBs.

The system operates on an interferometer, the \ac{WSRT},
to provide a spatial resolution
of order $10\arcmin \times 10\arcsec$. 
 Apertif, the
new \ac{PAF} front-end system at Westerbork,
delivers a large  instantaneous \ac{FoV}, of 8.2\,sq.\,deg.
Where the previous  system produced a single primary beam, 
the new front ends generate 40. Together these increase the  \ac{FoV}  by a factor 30.
Although the antenna elements and amplifiers are not cryogenically cooled,
Apertif operates at a nominal system temperature of 70\,K \citep{2009wska.confE..70O}.
The bandwidth is 300\,MHz around 1.4\,GHz, in two linear polarizations.
Of the 12 dishes outfitted with PAFs, the 8 equidistant ones that are most efficiently combined
 deliver a gain of $~\sim$1\,K/Jy. 
That sensitivity is comparable to the successful Parkes FRB surveys, but at much larger \ac{FoV}.
A real-time search system identifies FRBs at high time and frequency resolution (82\,${\mu}$s / 195\,kHz).
Apertif thus has the potential to deeply survey the entire fast-transient sky in radio.

In this paper we describe \ac{ARTS} and its first discoveries.
We %
outline its science potential (Sect.~\ref{sec:3}),
describe the design and implementation (Sects.~\ref{sec:4}$-$\ref{sec:6}),
and present the commissioning results,
performance, and 
survey planning (Sects.~\ref{sec:8}$-$\ref{sec:10}).
In Sect.~\ref{sec:11} and \ref{sec:disc}
we present and discuss our first science results, which include the discovery of
5 new \acp{FRB}.
We close with data releases (Sect.~\ref{sec:12}), future plans (Sect.~\ref{sec:13}), and
conclusions (Sect.~\ref{sec:14}).

The above sections outline the complete system,
providing the reader with the  overall rationale, setup and results.
More specialist readers are suggested to start with the relevant sections
and optionally follow back the references there.
Astronomers mostly interested in the newest results could read
Sects.~\ref{sec:3}, \ref{sec:8}, \ref{sec:11}, \ref{sec:disc} and \ref{sec:14}.
Time-domain specialists could add Sects.~\ref{sec:4}$-$\ref{sec:6} and \ref{sec:9}.
Astronomers interested in FRB populations and surveys should add Sects.~\ref{sec:9}, \ref{sec:10} and \ref{sec:13}.
For reproducing or improving the system and results,
 use Sect.~\ref{sec:12} and Appendix \ref{sec:app:design:i}$-$\ref{sec:app:beams}.

\section{Apertif for Time Domain: \ac{ARTS}}
\label{sec:3}

The Apertif time-domain potential is realised through \ac{ARTS}, an instrument that can
search wide fields for fast transients.  

Downstream from the \acp{PAF}, high-throughput and high-performance processing hardware
enables analysis of this full
field and bandwidth. Processing revolves around \acp{FPGA}
for channelisers, correlators, and a hierarchy of beamformers (\acsp{BF}),
and around CPUs and \acp{GPU} for  pipelines. 
These are directed by modern firmware and software,
powering a real-time detection system that combines the new,
large \ac{FoV} with the high angular resolution of the array. 
ARTS is directly connected to LOFAR \citep{2013A&A...556A...2V}, also operated by ASTRON.

\subsection{Science Motivation}
\label{sec:3b}

The system was designed around %
our goals of 
discovering and characterizing \acp{FRB}, primarily;
and our aims of finding and studying neutron stars,
and the time-domain counterparts of slow, image-domain transients, secondarily.

\subsubsection{\aclp{FRB}}
Using FRBs as cosmological tools and understanding the underlying source population(s) requires a large number of detections --
thus, a survey with good sensitivity, \ac{FoV}, and time
on sky.
These three characteristics are provided by \ac{ARTS}.
As Apertif is a full-time survey machine,
\ac{ARTS} detects an FRB roughly every week of
observing \citep[see 
  Sect.~\ref{sec:13} and][]{2017arXiv170906104M}.
The instrument combines high time and frequency resolution with the ability to capture full-Stokes polarization data.
It features a new GPU FRB search pipeline, and a machine learning classifier
to better identify and trigger on FRB candidates.

Potentially most important in determining the formation of \acp{FRB}
is localising the bursts.
If \acp{FRB} are formed by young neutron stars, the  galaxy in which they reside will need to have recently been forming massive
stars.
To identify the host galaxy  with high confidence,  the FRB position error box must be small enough to hold only a
single candidate host. 
For nearby bursts,
sub-arcsec localisation, using \ac{VLBI},
can connect the (repeating) FRB emitters with features \emph{within} the host galaxy.
\ac{ARTS} contains the hardware to connect the \ac{WSRT} for \ac{VLBI}
(Sect.~\ref{sec:5b.singleTAB}); but more importantly,
the addition of Apertif onto an interferometer with a baseline of over 1\,km
provides good \emph{instantaneous} \ac{FRB} localisation
(Sect.~\ref{sec:9.loc}).

\subsubsection{Neutron Stars}

The surface gravity of these extremely compact stars, about 10$^9$ times
the gravity on Earth, is the largest of any object visible in the
Universe. The internal densities of ten times nuclear 
density have not existed elsewhere since the Universe was $\sim$1\,ms old. 

That combination %
turns 
pulsars, radio-emitting neutron stars, into near-perfect cosmic time keepers \citep[e.g.][]{ht75b}.
Performing high precision pulsar timing on %
\ac{DNS} systems,
informs us of the underlying binary evolution \citep[e.g.][]{2015ApJ...798..118V}, and
enables tests of general relativity \citep{ksm+06,2019Sci...365.1013D}.

Three groups of neutron stars are only very sporadically active in radio:
rotating radio transients (RRATs), intermittent pulsars,
and radio-transient magnetars. Given the odds against their detection, the number of such transient
neutron stars must be comparable to that of radio pulsars \citep{kkl+11}.

\ac{ARTS} performs full-\ac{FoV} searches for single pulses of such pulsars, in a survey mode.
It can also  provide a high-time-resolution data stream,
with real time coherent-dedispersion, and online
folding, for a timing mode.  Together these allow for both searches for, and studies of, radio-emitting neutron stars.

\subsubsection{Prompt and slow emission from neutron-star mergers}

The orbits of \acp{DNS} decay,  due to the emission of gravitational waves, until eventually
the neutron stars merge.
This coalescence produces radio emission, possibly on different timescales
\citep[see, e.g.,][]{2016MNRAS.459..121C}. 
Prompt emission could possibly form at the merger, or %
at the collapse of an intermediary supramassive neutron star. %
Incoherent radio emission will %
produce an image-domain, slow radio transient \citep{2017Sci...358.1579H}.
Apertif allows for the study of both kinds of emission. Afterglow searches are carried out in image domain
\citep{2021arXiv210404280B} while  prompt emission could be detected in time domain.

\subsection{Key Apertif features  for time domain studies}
\label{sec:3c}

\subsubsection{Regular interferometer} %
\label{sec:3c.regular}

Wide-field transient surveys at both full (coherent-addition) sensitivity
{\it and}
high time resolution are almost always limited by the
large compute demands.
In the LOFAR surveys, the
\acp{TAB}
cover about 1/9th of the potential \ac{FoV} (the station beam; \citealt{clh+14,2019A&A...626A.104S}).
In the \ac{SKA}, the planned 1500 SKA-Mid beams cover
1/3rd of the primary beam \citep{sks+09}.
 No sparse two-dimensional or irregularly laid-out interferometer
 has the capability to find sub-ms transients at full sensitivity,
 over its entire \ac{FoV}.
 One unique feature of the \ac{WSRT} is its linear and regularly-spaced, E-W layout.
This produces full-sensitivity \acp{TAB} that are not single, small circular beams,
but have a large area in the N-S direction, and multiple sensitive sidelobes \citep{jsb+09}.
This means coherent beams do not need to computed over the entire 2-D \ac{FoV},
but only over the 1-D distance between the main lobe and the first sidelobe (Sect.~\ref{sec:3d.beam}).
Exploiting this fact allows us to overcome the compute limitations of full-\ac{FoV} surveys with %
e.g., LOFAR and SKA.

A down side of the elongated TABs is the 
modest N-S localisation precision for short-lived transients.
The instantaneous E-W localisation is very good ($\sim$10\arcsec). %
Additionally, every time a source repeats
the localisation ellipse has rotated on the sky,
thus producing an accurate final localisation
of about
$10\arcsec \times 10\arcsec$
\citep[see, e.g.,
][and Sect.\,\ref{sec:9.loc}]{2017arXiv170906104M}.

\subsubsection{Steerable dishes} %
\label{sec:3c.1}
For FRB and slow transient science, the ability 
of WSRT to point its dishes holds  
significant advantage over other wide-field 
surveys in the northern hemisphere. 
A fixed instrument such as CHIME is able 
to search its full \ac{FoV} in part because budget for  compute was prioritized over steerable elements.
This comes with the limitation that sources can only be observed for a relatively short time, at transit.
Because the WSRT can point, and track much longer than
fixed apertures, Apertif is able to dedicate significant 
follow-up time to repeating FRBs that are known 
to be active \citep[cf.][]{2020arXiv201208348P}.
\subsubsection{Digital backend: commensal systems}
\label{sec:3c.commensal}

The pulsar surveys in which the first FRBs were found
\citep[e.g.,][]{lbm+07,tsb+13,2014ApJ...790..101S} operated only during
a relatively small fraction of the total available telescope time.
And yet, for rare bursts such as these, the total time on sky is exceedingly 
important.
Systems that search for FRBs in a secondary mode -- 
commensal to other, primary projects that determine, e.g., the pointing --
are thus very valuable.
Commensal systems such as CRAFT \citep{2010PASA...27..272M,2017ApJ...841L..12B}
and CHIME/FRB \citep{2019ApJ...885L..24C} have found FRBs with this approach.
Exploiting the fact that the incoming Apertif \ac{PAF} data and
the modular hardware can be easily duplicated, 
ARTS was designed and built
to run in parallel\footnote{At time of writing this mode, however, is not commissioned.}
to all other imaging Apertif surveys
\defcitealias{Hess2020}{Hess \mbox{et~al.} 2022}
\citep{2019NatAs...3..188A,Hess2020}.
Coherently combining the 8 WSRT dishes that are equidistant in the imaging ``Maxi-Short'' configuration,
 would produce a commensal transient search with the same wide field, and similar sensitivity, as the dedicated survey.

\subsubsection{Digital backed: high resolution}
\label{sec:3c.hires}

The high time and frequency resolution (82\,${\mu}$s, 195\,kHz) of 
the ARTS  backend offers an 
advantage for both the study of known pulsars and FRBs \citep[see, e.g.,][respectively]{2022A&A...658A.143B,oml+20}
as well as the detection of new sources. 
Many FRBs appear to be sub-millisecond in duration, 
which means a large number of bursts are currently 
missed due to the deleterious effects of 
instrumental smearing \citep{connor-2019}. ARTS
searches filterbank intensity data with more 
favorable smearing properties than  
comparable instruments, boosting its detection 
rate and enabling the discovery of higher-DM, narrower
FRBs.

\begin{figure*}[b]
\centering
\includegraphics[width=0.92\textwidth]{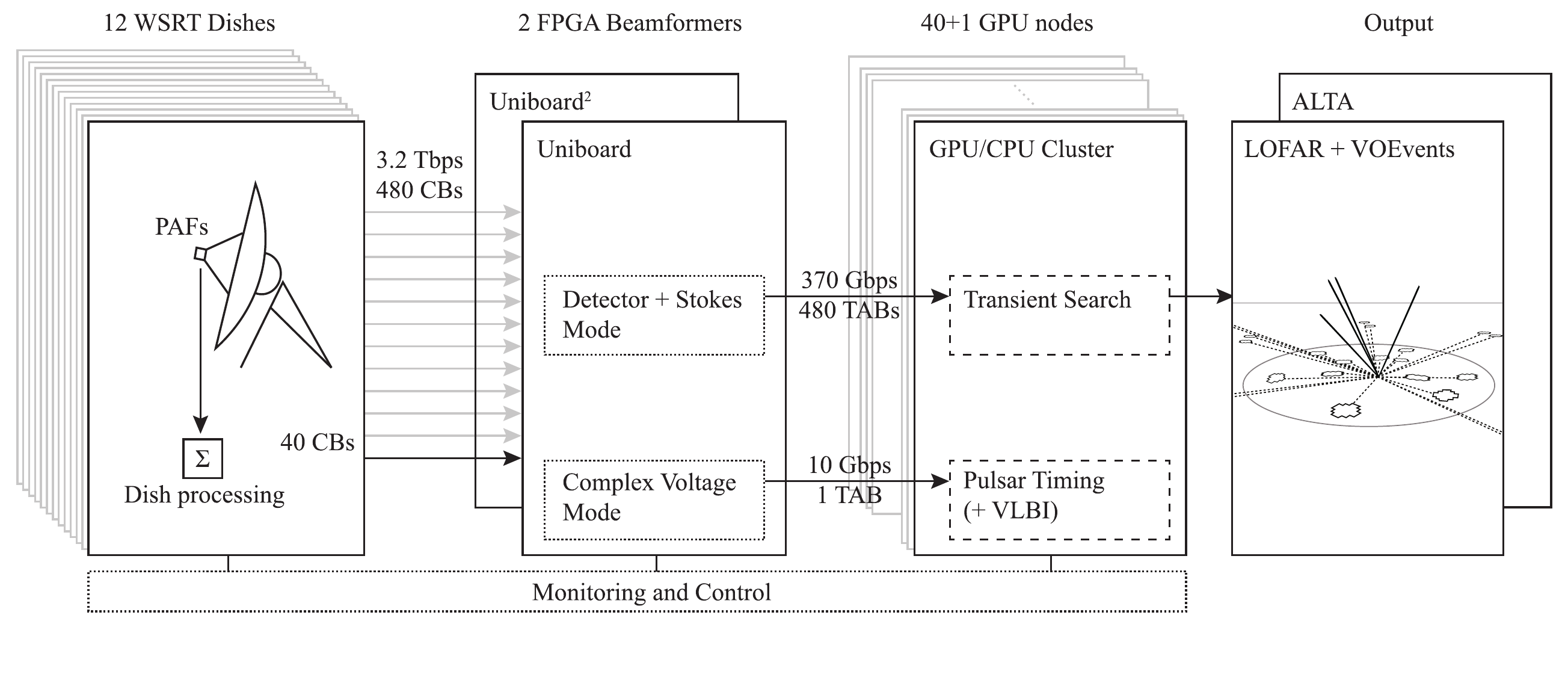}
\vspace{-5mm}%
\caption{
  Top-level diagram for ARTS.
  Different line styles indicate different types of subsystems, each described
  in their own Section. 
  Full lines are hardware elements (Sect.~\ref{sec:4}).
  Dotted boxes depict firmware and software (Sect.~\ref{sec:5}).
  Dashed items are science pipelines (Sect.~\ref{sec:6}).
  The LOFAR and ALTA output are described in Section %
  \ref{sec:12}.
}
\label{fig:toplevel}
\end{figure*}

\subsubsection{Connection to LOFAR}
\label{sec:3c.3}

WSRT can localise an FRB very well in the E-W direction but only to modest N-S
precision (cf.~Sect.~\ref{sec:3d.beam}).
But if it emits over a broad band, this same FRB will arrive later at low frequencies.  Triggering the LOFAR
all-sky \aclp{TBB} \citep[TBBs;][]{sha+11, veen15, 2019A&A...621A..57T} on the FRB, and detecting it at 150\,MHz, would
allow for localisation using the two-dimensional array of LOFAR stations over The Netherlands and Europe,
providing
arcsec precision.

Previous blind LOFAR searches  \citep{clh+14, 2015MNRAS.452.1254K,2019A&A...626A.104S, 2019A&A...621A..57T,2020A&A...634A...3V}
have not found FRBs.
This could partly be caused by the residual dispersion smearing that remains after trial dedispersion at such low
frequencies. In the triggered case, the \ac{DM} is known, allowing for more accurate dedisperion plus a significant
reduction of the number of DM trials, improving detection confidence levels.

Still, also when the DM was known, low-frequency FRB observations were unsuccessful before 2021
\citep[see, e.g.,][]{2019A&A...623A..42H}.
Only with the detection of FRB~20180916B were FRBs seen for the first time
below 200\,MHz; and while we found that
the all-sky low-frequency FRB rate is appreciable   \citep{2020arXiv201208348P},
no single FRB burst was ever seen to emit at both 150\,MHz and 1.4\,GHz. 
The simultaneous ARTS-LOFAR observing that enabled these detections is detailed in \citet{2020arXiv201208348P}.

\subsection{High level system overview}
\label{sec:3d}

Time-domain observations with Apertif
fall under four main science cases (SCs; detailed in \citealt{straal18}):
(1) Pulsar Timing, (2) VLBI, (3) Fast-transient
searching commensal to imaging, and (4) Dedicated transient searching.
These were numbered to reflect increasing complexity; their priority is the inverse,
the dedicated search ranking highest (Table~\ref{tab:modes}).
\begin{table}
  \caption{The observing modes, the science cases they enable,
    and the observing time total for 2019 (this paper).}
    \label{tab:modes}
    \centering
    \begin{tabularx}{\columnwidth}{llll} 
    Mode               & Science Case & Parallel to & Time (hr)  \\
      \toprule 
    Dedicated Search & SC4 & -       & 800  \\
    Commensal Search & SC3 & Imaging & 0    \\
    Timing           & SC1 & -       & 50   \\
     \bottomrule
    \end{tabularx}
\end{table}
Their scientific and resulting technical requirements
led to designs that were reviewed and implemented.
After commissioning (see, \citealt{mikhailov18,straal18,oostrum20}, and
Sect.~\ref{sec:8}) the system was brought into
operation.
Together, a number of subsystems (Sect.~\ref{sec:3d.ss})
form and process
an hierarchical series of beams (Sect.~\ref{sec:3d.beam}).
Transient detections in
\ac{ARTS} produce triggers to allow for follow-up in close to real time, and data products stored in the \acf{ALTA}, where they are publicly
available.

\subsubsection{Three sets of subsystems}
\label{sec:3d.ss}

\ac{ARTS} %
comprises the following major sets of 
subsystems. %
These work
together as illustrated in Fig.~\ref{fig:toplevel}.

The first subsystem set is the hardware platform (Sect.~\ref{sec:4}).
It consists, first, of the
\ac{WSRT} dishes  and the Apertif
\acp{PAF}.
Front-end beamformers next provide dish processing.
Tied-array beamformers built using \acp{FPGA}
on high-performance processing boards
(UniBoard and UniBoard$^2$),
are connected through fast networking to a \ac{GPU} cluster.

The second set is the firmware and software sub systems (Sect.~\ref{sec:5})
that control
and produce one or multiple \acp{TAB}, or `pencil beams'.
These can be in Nyquist sampled,
complex-voltage format for pulsar timing (and \ac{VLBI}).
The data in these \acp{TAB} can also be `detected', i.e., converted to the four
Stokes parameters, allowing subsequent   partial   integration
to reduce  data rates.
 That way, many hundreds of beams can be streamed out.

The third set of subsystems comprises the ARTS pipelines (Sect.~\ref{sec:6}).
These perform
transient searching and pulsar timing.
For timing, the central TAB 
is coherently dedispersed and folded in real time, on a single multi-GPU node.
For the transient search,
all Stokes-I \acp{TAB} are
cleaned of \ac{RFI},
dedispersed over a number of trial \acp{DM},
corrected for chromatic effects,
and searched for transient events.
Good candidates immediately trigger data dumps
from a ring buffer of full Stokes-IQUV data.
A deep learning implementation further classifies all candidates.
Results and data are public %
to the outside world through VOEvents 
 and the archive  (Sect.~\ref{sec:12}).
\subsubsection{Hierarchical beam forming}
\label{sec:3d.beam}

One of the innovative aspects of \ac{ARTS} is its use of hierarchical
beamforming to allow for searches throughout the entire
primary-beam \ac{FoV},
at coherent-addition sensitivity.
That is a challenge in many wide-field fast-transient instruments
such as \ac{SKA} \citep{bac99} and LOFAR \citep{ls10}.
We give a short
conceptual overview of these beam forming steps here, because they are important in all subsystem sets.
Appendix \ref{sec:app:beams} contains a quantitative description. 

\begin{figure*}
\centering
\includegraphics[width=\textwidth]{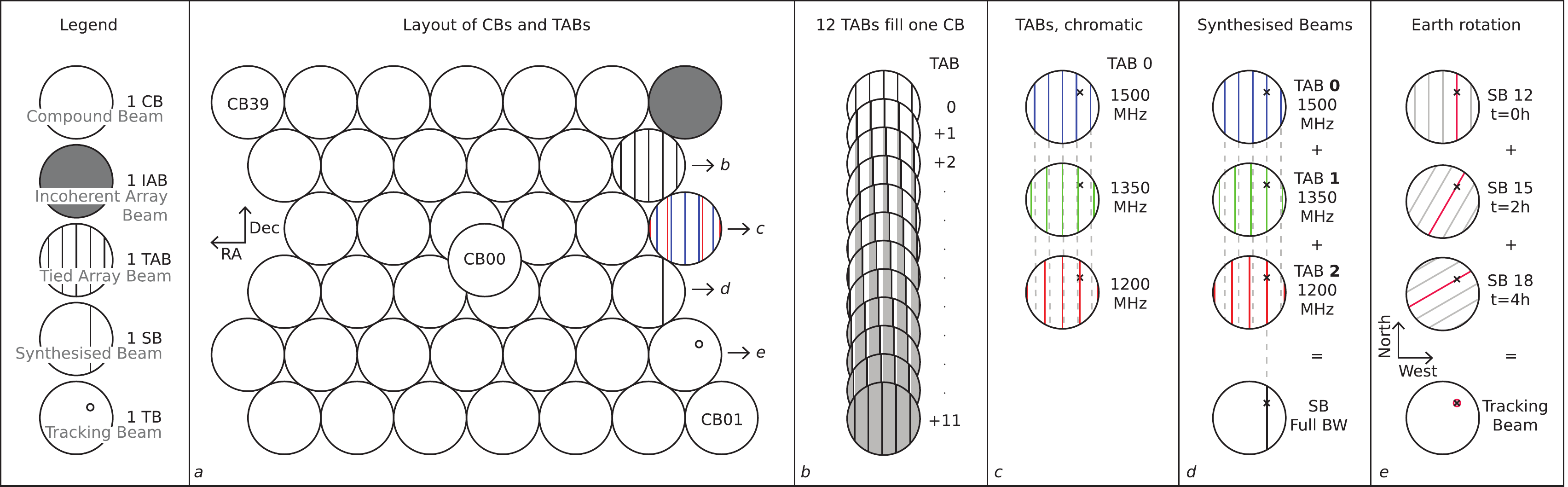}
\caption{
  An overview of the beam hierarchy in ARTS. Panel a) shows the 40 \acp{CB} formed by each PAF. The output signals of
  these CBs can be combined into \acp{IAB} or coherent \acp{TAB},
  whose grating responses fill the entire CB as shown in panel b). The frequency-dependence of these grating responses
  shown in panel c) can be exploited to disambiguate them by combining signals from multiple TABs to form a \ac{SB} as
  illustrated in panel d). Finally, a \ac{TB} can be formed towards a specific locus  by combining multiple SBs over time as shown in panel e).
}
\label{fig:highlevelbeams}
\end{figure*}

Figure~\ref{fig:highlevelbeams} provides an overview of all constituents of the beamforming hierarchy. The base is
formed by the 40 \acfp{CB} formed by the PAF in each Apertif dish. Each CB has the same \ac{FoV} as the primary dish
beam, with a diameter of $\sim$$0.5^\circ$, and produces a full-bandwidth, Nyquist sampled output data stream. Together, the CBs cover the  \ac{FoV} of the Apertif system as shown in Fig.~\ref{fig:highlevelbeams}a.

The data streams of corresponding CBs of multiple dishes are combined using either an incoherent beamformer or a
coherent beamformer. Using the incoherent beamformer, 40 \acfp{IAB}, one per \ac{CB},
are formed that together cover the compound
\ac{FoV} of Apertif with a sensitivity %
scaling with the square root of the number of dishes used.
When using \emph{coherent} beamforming, the peak sensitivity is higher, scaling linearly with the number of dishes,
but more data streams need to be analysed as the \ac{FoV} per TAB is smaller.
For a regularly spaced 1-D array like WSRT,
each TAB is not a single field: all sidelobes reach the same sensitivity as the central peak, and add to the TAB \ac{FoV}.
Together, 12 \acp{TAB} now completely cover the \ac{FoV} of their CB.
This high TAB filling factor  is illustrated in Fig.~\ref{fig:highlevelbeams}b.
It was first used by \citet{jsb+09}, to find new pulsars.
Such full \ac{FoV} beam forming requires that we only use those dishes that are spaced
equidistantly. These are the 8 dishes RT2$-$RT9, plus the movable dishes RTA and B
if they are located on the common baseline grid.

While we exploit the TAB grating to increase the TAB \ac{FoV},
it also presents a degeneracy in the location of a detected event. %
We may not know through which grating sidelobe the signal is coming in.
Fortunately the grating responses %
are  frequency dependent
-- see Fig.~\ref{fig:highlevelbeams}c.
A source slightly further from the phase center
than the specific grating response of a certain \ac{TAB} at e.g. 1500\,MHz, %
may fall exactly in the
grating response of that TAB at 1200\,MHz.
Thus, %
most grating responses will %
detect a source over limited frequency range.
To make a detection over the full bandwidth %
the correct frequency-dependent TABs are 
combined to form a so-called \acf{SB},
as illustrated in panel~\ref{fig:highlevelbeams}d. In ARTS, 71 SBs are formed in each CB.
As it is integrated into the subband dedispersion (Sect.~\ref{sec:6b.2}),
this frequency re-organisation itself comes with no additional computational cost;
but the increased number of beams (12 TABs to 71 SBs) do require more computing down stream. 
The SBs provide instantaneous localisation with a resolution
determined by the array size in one direction, and the size of the CB in the other. 

The orientation of the grating responses on the sky changes with time due to Earth rotation.
A specific locus thus %
moves through different SBs %
(Fig.~\ref{fig:highlevelbeams}e).
By combining the best positioned SBs in time, %
we form the last constituent of our hierarchy of beams: the \acf{TB}.
Over a 12-hour observation, $\sim$3000 unique loci form %
within each CB.

\section{Description of Time Domain System: I. Hardware and Function}
\label{sec:4}

A general description of Apertif is available in \citet{cho+20}. 
In the next three sections we describe the time-domain capabilities.
A number of subsystems in Apertif are shared between imaging, time-domain and VLBI modes.
Where required for context, we first provide high-level descriptions of these general subsystems.
This  section covers function and hardware, with additional detail available in Appendix \ref{sec:app:design:i}
and \ref{sec:app:design:hw}, respectively.

\subsection{Relevant general Apertif hardware and function}

\subsubsection{Dishes in an E-W Interferometer}
\label{sec:4c}

The \ac{WSRT} consists of fourteen 25-m dishes in a linear East-West array.
It started operations in 1970
(for a recent overview see \citealt{strom201850}, marking its 50th anniversary).
The accuracy of the steel structure, the
parabolic surface of the dish, and the size of the mesh,
mean \ac{WSRT} can be used up to $\sim$8\,GHz.
It was, however, designed for 21-cm wavelength observing.
At the installation of \ac{Apertif} the dishes too were refurbished and they operate as new. 
The mounts are equatorial. 
The 10 most western dishes, \ac{RT} 0$-$9, are fixed at redundant, 144\,m intervals.
RTA and B are immediately east of RT9, and can be moved on a rail track.
RTC and D are on a track that is 1.3\,km further east.
Because of this dominant common baseline, the instantaneous interferometric beam has highly sensitive sidelobes.
\subsubsection{\acf{PAF}}
\label{sec:4b}

Twelve of the \ac{WSRT} dishes, RT2$-$RTD, are outfitted with the \ac{Apertif} \acp{PAF}.
Each feed contains 121 receiver elements,
as opposed to the previous generation of receivers, 
the \aclp{MFFE} (MFFEs; cf.~\citealt{1991ASPC...19...42T}, \citealt{2018wtfa.confE..11B}), that were single pixel.
The Vivaldi antenna elements in the \acp{PAF} are optimized for 1500\,MHz.
The elements also host the \ac{LNA}, and are uncooled.
The \acp{LNA} includes a filter that suppresses the strong \ac{RFI} present below
the nominal bottom of the Apertif band, at 1130\,MHz.
This does, however, contribute  $\sim$20\,K to the total system
temperature of $\sim$70\,K \citep{cho+20}.
For the 12 dishes combined, and for the central, most sensitive \ac{CB},
this produces an \ac{SEFD} of $\sim$45\,Jy at 0.75 aperture efficiency.
For the 8 dishes generally used in \ac{ARTS},
we find a  median \ac{SEFD} over all \acp{CB} of  85\,Jy (Sect.~\ref{sec:9b}).

\subsubsection{Front-end digital processing}
\label{sec:4b.1}

At the 12 dishes, the \acpcom{FEBF}{Sect.~\ref{sec:app:design:febf}}
sample the \ac{RF} data from the \ac{PAF} at 800 MHz.
This yields a timing accuracy of 1.25\,ns, at a total input data rate of 9.8 Tb/s. 
The sampled bandwidth is a contiguous 400\,MHz band, tuneable between 1130$-$1720\,MHz.
The digitized data of all antenna elements is separated into 512 subbands, each 0.781250\,MHz wide.
At each dish,
a rack of eight \aclp{UNB} (\acsp{UNB}; \citealt{2010evn..confE..98S})
beamforms the subband signals over the PAF elements, into so-called \emph{beamlets}.\acused{UNB}

The \ac{CB} pattern formed by these beamlets
is laid out in a configuration that maximises even coverage over multiple, adjacent
pointings \citep{Hess2020}.
Initially Apertif was designed for 37 \acp{CB} \citep{2009wska.confE..70O}.
Because the Apertif and ARTS hardware produce \acp{CB} in multiples of 8, the surveys now use 40.

Next, 384 out of the 512 subbands are selected for output, to achieve the final 300\,MHz bandwidth.
When expressed as 6-bit numbers, these %
 fill the data transport capacity over the 384 links of 10\,Gpbs each,
 to the central building for correlation and/or tied-array beamforming.
In total, the twelve Apertif-equipped dishes produce 12 $\times$ 40\,CBs.
This data stream (called CB480) contains $3.5\,$Tb/s of compound-beam data
(Sect.~\ref{sec:app:design:febf}, \ref{sec:app:design:hw:febf}; and Table~\ref{tab:app:datarates}).

\subsection{The \acf{TABF} and Apertif \acf{X} subsystems}

\begin{figure}
  \centering
  \includegraphics[width=\columnwidth]{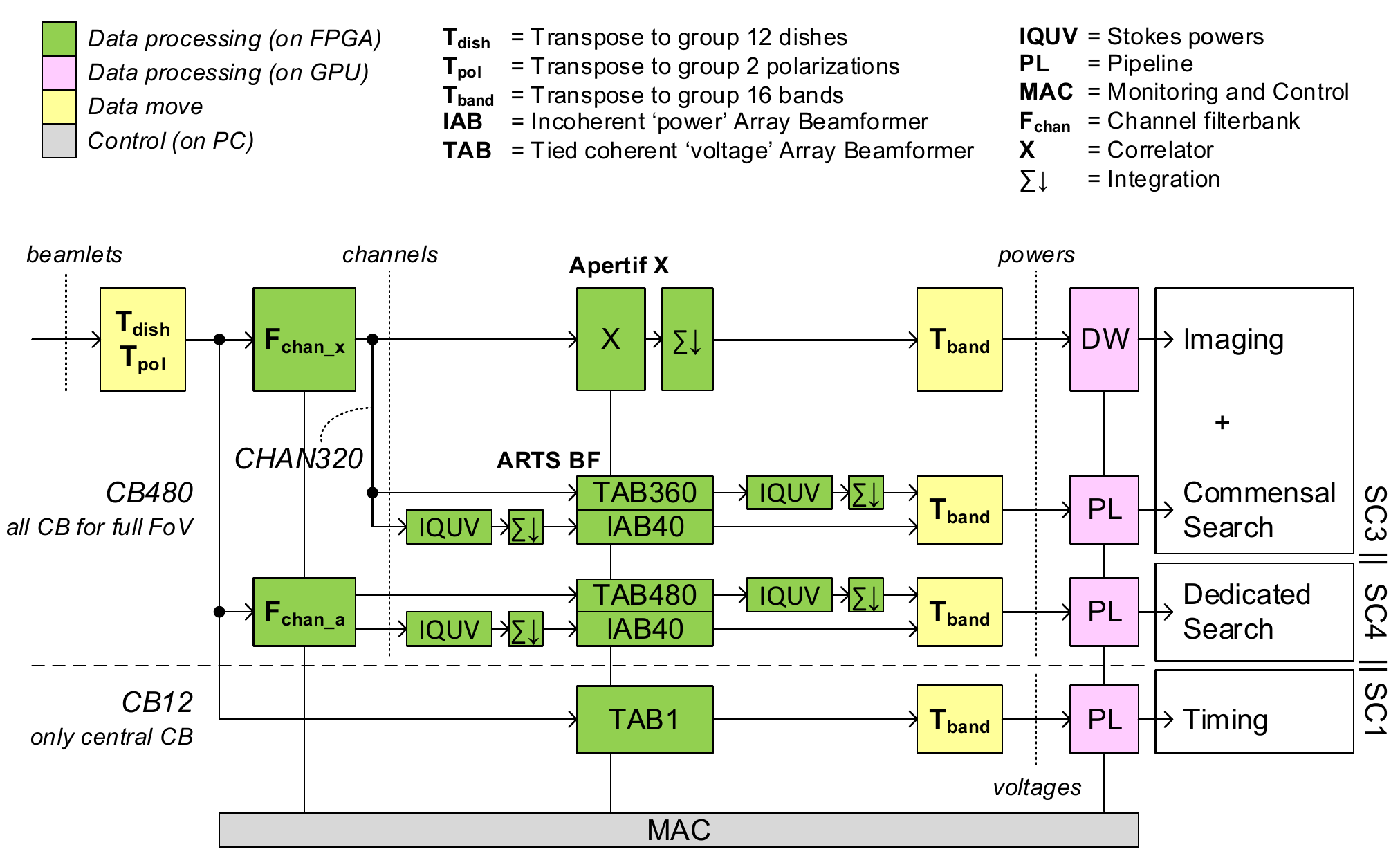}
  \caption{The ARTS \ac{TABF} subsystem and the Apertif \ac{X} subsystem. The modes for the 3 SCs (Table~\ref{tab:modes}) are mutually exclusive.
  \label{fig:app:XBF}
  }
\end{figure}

The Apertif \acl{X} (``\acs{X}'') and ARTS \ac{TABF} both use the same Apertif \ac{FEBF} CB480 data.
In Fig.~\ref{fig:app:XBF} we show how these subsystems relate.
The transpose
T$_\mathrm{array}$ = T$_\mathrm{dish}$ + T$_\mathrm{pol}$ groups the data from all N$_\mathrm{tp}$ =
N$_\mathrm{pol}$ $\times$ N$_\mathrm{dish}$ = 24 single
polarization telescope paths (TP) in the \ac{WSRT} array.
The Apertif \ac{X} creates visibilities by cross correlating the compound beams between all dish pairs.
The ARTS \ac{TABF} creates array beams by summing compound
beams over the dishes. %
These array beams can be coherent
\acp{TAB}
or \acp{IAB}, for either
voltage, power or full Stokes power data (Sect.~\ref{sec:app:design:dataformats}, \ref{sec:app:design:bfalgos}).

\subsection{The \acf{TABF}}
\label{sec:4b.1.tabf}

The function %
of the \ac{TABF} is to create the \ac{IAB} and \ac{TAB} data, and 
to  transpose the data  for the Arts \accom{PL}{Sect.~\ref{sec:6}}.
For the search data, 
the \ac{TABF} also functions to reduce data rates, 
by integrating in time.

For searching, the ARTS \ac{TABF} first separates the beamlets into 64 subchannels, using a similar filterbank
(F$_\mathrm{chan\_x}$ in Fig.~\ref{fig:app:XBF}) as Apertif \ac{X}.  The search modes use all 40 \acp{CB}.
The \ac{TABF} can combine all 12 dishes (480 \acp{CB}) but in practice only
includes dishes that are equidistant in the array
configuration at the time.
The output of the IQUV data is integrated over 16 subchannels to achieve the required
81.92\,$\mu$s time and 195\,kHz frequency resolution (Sect.~\ref{sec:app:design:rf}).  The standard \ac{TABF} output are
integrated full Stokes IQUV array power beams, 480 for dedicated search and 360 for commensal search. Optionally, IABs
can be produced. In that case the order of beamforming and detection are switched.

For timing-mode (and VLBI) output, the
 \ac{FEBF} transposing is bypassed (see~Sect.~\ref{sec:app:design:febf} and Fig.~\ref{td-febf-map-v01})
and the beamlets are used directly.
The central \ac{CB} is voltage beam-formed into a single TAB (TAB1 in Fig.~\ref{fig:app:XBF}).

The central beam-forming hardware
needs to handle the large amount of input links, and the high data
rate of $\sim$3.5\,Tb/s, produced by the Apertif dish processing (Sect.~\ref{sec:4b.1}).
The relatively low number of operations per bit  led to the implementation of the \acp{TABF}
on \acp{FPGA}.
In dedicated ARTS mode, the time-domain survey determines the baselines and the pointing of the \ac{WSRT} dishes.
As this mode runs \emph{instead} of Apertif \ac{X}, %
the \ac{TABF}
can
 use the same 16 UniBoards used by the correlator, as detailed in the next subsection.

In
commensal mode (Sect.~\ref{sec:3c.commensal}), the Apertif imaging science case is in control, and the correlator runs.
Firmware for both does not fit together on  16 UniBoards, but
needs to run on a separate set of FPGA boards. 
Four \acp{UNB2} can provide the required \ac{IO} and compute in a setup described in Sect.~\ref{sec:4e}.

\begin{figure}[b]
  \centering
  \includegraphics[width=\columnwidth]{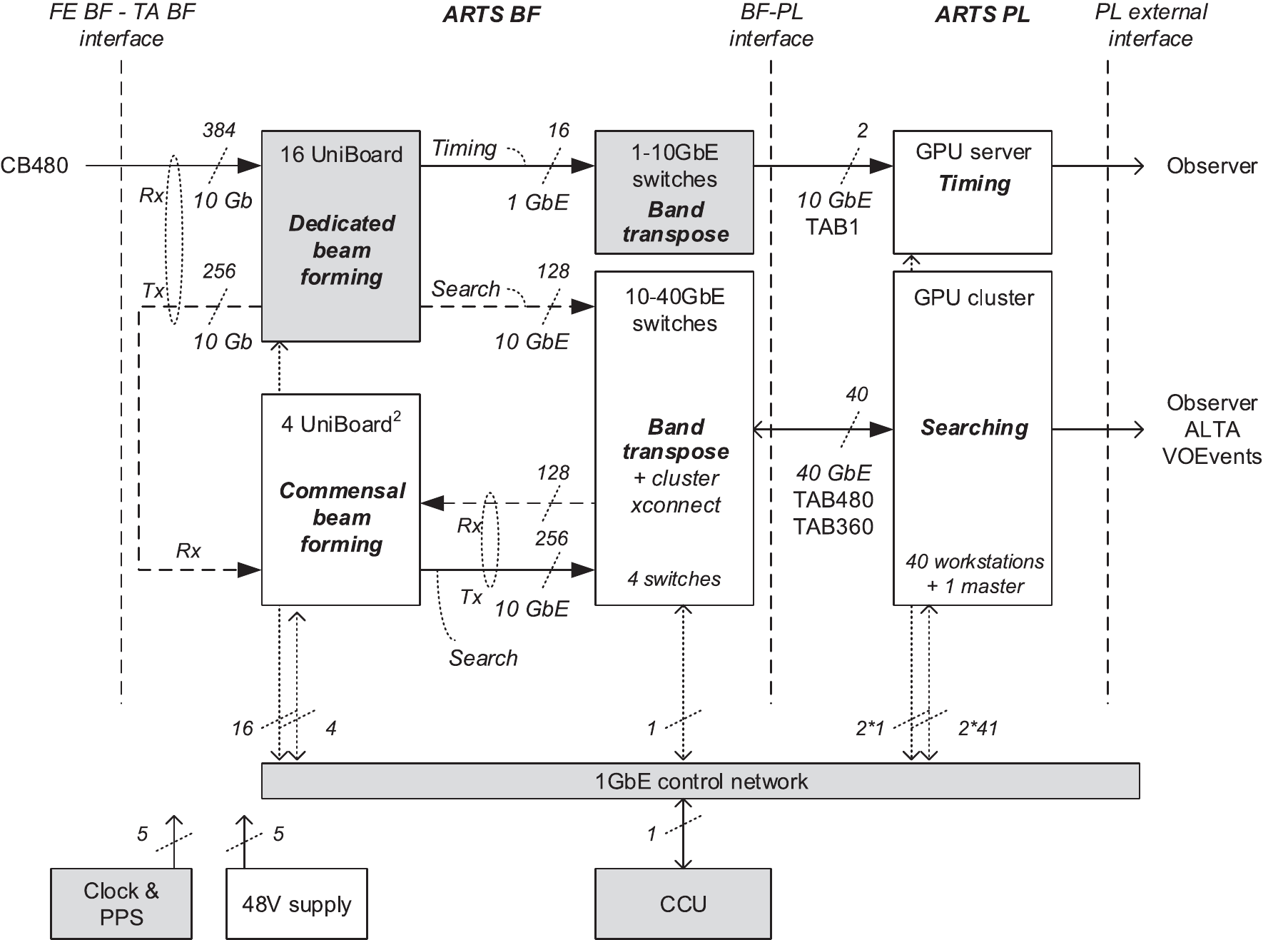}
  \caption{The mapping of the ARTS system on the \acp{UNB}, \acp{UNB2}, switch, and CPU/GPU cluster hardware.
Only one of either the commensal or the dedicated beam forming platforms will be running at a given time. 
    The dashed Tx links are unidirectional links. The dashed Rx links are full duplex links that are combined with
  the \ac{UNB2} Tx links. Parts in gray are hardware used by or shared with Apertif X.
  \label{td-bf-map-v01}
    }
\end{figure}

Fig.~\ref{td-bf-map-v01}  shows how the ARTS FPGA beamformer maps on Uniboard and \ac{UNB2},
and how the boards are interconnected. 

\subsection{Dedicated ARTS \ac{TABF} on Central Uniboards}
\label{sec:4d}

The central ARTS beamformer consists
of 16 \acp{UNB}
with a  total of 128 \acp{FPGA} (for details, see Appendix~\ref{sec:app:design:hw:u}).
Each \ac{UNB} processes 1/16$^\mathrm{th}$ of the bandwidth (i.e., 18.75\,MHz) of all dishes.

The \acp{UNB} are equipped with  \acpcom{OEB}{Sect.~\ref{sec:app:design:hw:u}}
to handle the demanding \ac{IO}.
All 384 10-GE ports are required for data that comes in.
For \ac{X}, outgoing data can fit on the 1 Gbps ports
because the time integration strongly reduce the data rate.
But for the \ac{TABF} the complete usage of the links by the input is a challenge.
The ARTS dedicated search  makes 12 TABs/CB that 
also require 128 %
10-GE links, for transport and transpose of the data to the \ac{PL}.
For all 384 ports, the Rx and Tx are thus physically split.
The \acp{UNB} receive Apertif \ac{FEBF} data over Rx and send beam formed data over Tx to the \ac{PL}.

During dedicated FRB survey observations, the Uniboards generate 480 \acp{TAB} with a time resolution of $81.92\,\upmu$s and 1536 frequency channels over a bandwidth of 300\,MHz. For each TAB, both a Stokes I and a Stokes-IQUV data stream are created, for a total output data rate of ${\sim}360\,$Gb/s.

\subsection{Commensal ARTS \ac{TABF} on Central {\unisq}s}
\label{sec:4e}

In the central building,
a set of four \acp{UNB2} \citep{2019JAI.....850003S} is also installed,
to be able to run imaging and time-domain at the same
time.
As that commensal search uses as input
only the 8 dishes  that are equidistant in the imaging Maxi Short configuration,
it  requires making only 9 TABs/CB.
Four \ac{UNB2}s each with four Arria10 FPGAs (see Sect.~\ref{sec:app:design:fpgares})
can create these beams, but then a channel filterbank does not also fit. 
Therefore the \ac{TABF} for commensal search uses the output from the channel filterbank in the Apertif \ac{X},
that will then be running on the \acp{UNB}.
The filterbank in the Apertif \ac{X} correlator already makes 64 channels, the same as the dedicated search.
The {\unisq}s receive the CHAN320 data (8 dishes $\times$ 40\,\acp{CB}) from the 16 \acp{UNB}, as visualized in
Fig.~\ref{fig:app:XBF} and detailed in  Appendix~\ref{sec:app:design:hw:u2}.

\subsection{Networking and interconnect}
\label{sec:4.networking}

The central \acp{UNB} receive the CB480 input from the Apertif \ac{FEBF} (Fig.~\ref{td-bf-map-v01})
over the 384 10G Rx links (Sect.~\ref{sec:4d}).
The 10G Tx part of these same links is used to offload the output TAB data, or to pass on data  to the 4 \acp{UNB2}.
For the  dedicated search  360\,Gbps (Table~\ref{tab:app:datarates}) of
TAB power data is carried via 128 10G links.
Each \acp{UNB} or  \ac{UNB2} beam forms  all telescopes and \acp{CB} but only for part of the bandwidth of these TAB
power data.
For searching, these data need to be transposed such that all bands
for a single set of \acp{TAB} converge in a single \ac{PL} GPU node (Fig.~\ref{td-interconnect-map}). 
In the data network, this band transpose function T$_\mathrm{band}$ is implemented using
\ac{COTS} 10/40 GbE switches (called the transposer in Fig.~\ref{td-bf-map-v01} and in Sect.~\ref{sec:app:design:transpose}).

In commensal-search mode, the \acp{UNB} operate as correlator, and only pass on the 
 CHAN320 data to the \acp{UNB2}. For timing, the modest data rate of 9.6\,Gbps (2\,pols $\times$ 300\,MHz  $\times$ 2 complex  $\times$  8\,bit; Table~\ref{tab:app:datarates})
is carried via 16 1GbE links.

\subsection{ARTS GPU cluster}
\label{sec:4f}

Data for pulsar timing is coherently dedispersed and folded by a single GPU-accelerated server (``ARTS-0'').
This machine contains two 6-core CPUs, two Titan-X GPUs, two 10GE Network Interface Cards,
and %
16 active disks in a JBOD configuration.%

Data for searching is much higher volume, and is processed by a GPU cluster
consisting of one login head node
and %
40 worker nodes, all with identical hardware
 (Table~\ref{tab:cluster_hardware}). Together the latter provide a total of 160 GPUs, 1600 CPU cores, 5\,TB RAM, and ${\sim}$1.3\,PB storage, for a theoretical peak performance of ${\sim}$2 PFLOP/s.

Each %
worker node processes the data from one compound beam.
That way, cross detections can immediately be compared between TABs in a CB; and tracking beams made.
The incoming data rates per CB are 1.8\,Gbps of
Stokes I data, and 7.2\,Gbps of Stokes IQUV data.
Half of the storage is available for the incoming Stokes I data, which when writing continuously fills up in
${\sim}20\,$hrs. Hence any processing that requires access to these data should take at most a few hours to run.

Transient data buffers were initially designed to reside on both the \acp{UNB} and \acp{UNB2}. 
The GPU cluster, however, could hold more DDR3 memory more economically, in a single location,
with simpler  transient-dump control functions.
Thus, the Arts BF now outputs full Stokes IQUV (for dumping), instead of only the Stokes I data (for searching).
This Stokes IQUV data is only stored upon a trigger from the transient pipeline, so even though the Stokes IQUV data rate is four times that of Stokes I, its storage will fill up very slowly.

Both Stokes I and IQUV data are buffered in RAM, for 10 and 15\,s, respectively, and made available to the pipelines. The system buffers about 1\,TB of data for triggering overall.

\subsection{Housing}

\ac{ARTS} is located in an \ac{RF}-shielded \acl{HF}
cabin in the central building of \ac{WSRT}.
The cabin contains coolers with air-water heat exchangers.
This water is pumped through an outside chiller,
 that operates  passively when possible, and only uses compressors when required.
In the cabin, three cabinets house the \acp{UNB} and switch equipment.
These are air cooled through the cabin ambient cooling.
The ARTS GPU servers and \acp{UNB2} are distributed over 6 racks in a single row.
Three large dedicated reverse coolers create a cold and hot isle.
Cold air flows through the servers to the hot isle. The CPUs and GPUs are cooled against this air.
The \acp{FPGA} on \ac{UNB2} are
liquid cooled, and exchange heat with the air flow outside the \ac{UNB2} box.
When operational, the total ARTS system uses 50\,kW of power.

\begin{table}
    \caption{Hardware overview of a GPU cluster node.}
    \label{tab:cluster_hardware}
    \centering
    \begin{tabularx}{\columnwidth}{lX} 
    \toprule 
     CPU & 2 $\times$ Intel Xeon E5-2640 v4 (40 cores total) \\
     GPU & 4 $\times$ Nvidia 1080 Ti \\
     RAM & 128\,GB \\
     Network & 40\,Gbps, full-duplex \\
     Storage & 32\,TB in 2 independent hardware RAID0\\
     \bottomrule
    \end{tabularx}
\end{table}

\subsection{Control}
\label{sec:4.control}

Through the control network, ARTS and Apertif are directed by a Central Control Unit (CCU) computer
that connects to the \ac{TABF} and the \ac{PL} via Ethernet.
The CCU is synchronized with the atomic clock of the WSRT using the standard network time protocol (NTP).
\section{Description of Time Domain System: II. Firmware and Software}
\label{sec:5}

\subsection{Firmware}
\label{sec:5b}

In the following subsections we describe how the ARTS \acp{TABF} in Fig.~\ref{fig:toplevel} 
are implemented in \ac{HDL} firmware on the \acp{FPGA} on \ac{UNB} and \ac{UNB2} (Sect.~\ref{sec:4d}, \ref{sec:4e}).
More detail is available in Appendix~\ref{sec:app:design:fw}.

Each observing mode (Table~\ref{tab:modes}) has a dedicated FPGA application design that either runs on all
128 FPGAs of the \ac{TABF} \acp{UNB} or on all 16 FPGAs of  \ac{UNB2}.
The designs contain functions for  the \ac{TABF} itself, 
for  monitoring
during operation, and for testing during commissioning.
Switching between modes occurs by changing the design image.
The flash memory of \ac{UNB} and \ac{UNB2} typically contains a default image and an application image.
Changing observing mode requires reprogramming the flash.
{ In practice this reprogramming takes $\sim$5 minutes for all FPGAs
  in parallel. This time
  includes steps to erase, write and read back to verify.
  Such reprogramming is part of the scripted mode switching that operators can initiate or schedule.
  Together with the required re-initialisation of the other subsystems, this switch takes $\sim$15 minutes. 
}

There are three different FPGA images for the 16 \acp{UNB}: search, timing and correlation.
The 4 \ac{UNB2} can run two images: the commensal search and the dedicated-search pass through.

The heart of the \ac{TABF} system is a voltage beamformer function %
that weights and sums the input dish voltages.
For the single voltage TAB (for timing or VLBI),
the VHDL component from the subband \ac{CB} beamformer in the Apertif \ac{FEBF} \citep{cho+20} was directly reused.
The  \ac{TABF} function for surveying contains further multi-beam parameters.

The subchannel filterbank and integrator functions are weaved into one.
This reduces aliasing losses.
On the one hand, samples should be integrated in time to 81.92\,$\mu$s to reduce data rate. 
On the other hand, the fine-channel \acl{PFB} \acused{PFB}
(\acs{PFB}: called F$_\mathrm{chan}$$_\mathrm{\_}$$_\mathrm{a}$ in Fig.~\ref{fig:app:XBF})
should separate the beamlet data into 4 channels.
But as channel 0 represents information from both the low and high edge of the subband, it is best discarded;
 implying 25\%  data loss. To mitigate this, 
the subchannel PFB  first forms 64 channels. Subchannel 0 is blanked.
The Stokes power data is then not integrated in time, but  over sets of 16 frequency subchannels in stead
(see Fig.~\ref{td-febf-tabbf-subchannels}).

Some functions in the FPGA application design require \acf{MAC}.
These act through a \ac{MM}
control interface
as visible in e.g. Figs.~\ref{td-fw-sc4-v01} and \ref{td-febf-map-v01}.
Some are set only at start of measurement (e.g. %
 \ac{PFB} coefficients);
others are updated regularly during the measurement (e.g. statistics, status monitoring).

\subsubsection{Multiple TABs on Uniboard}
\label{sec:5b.1}

\begin{figure}
  \centering
  \includegraphics[width=\columnwidth]{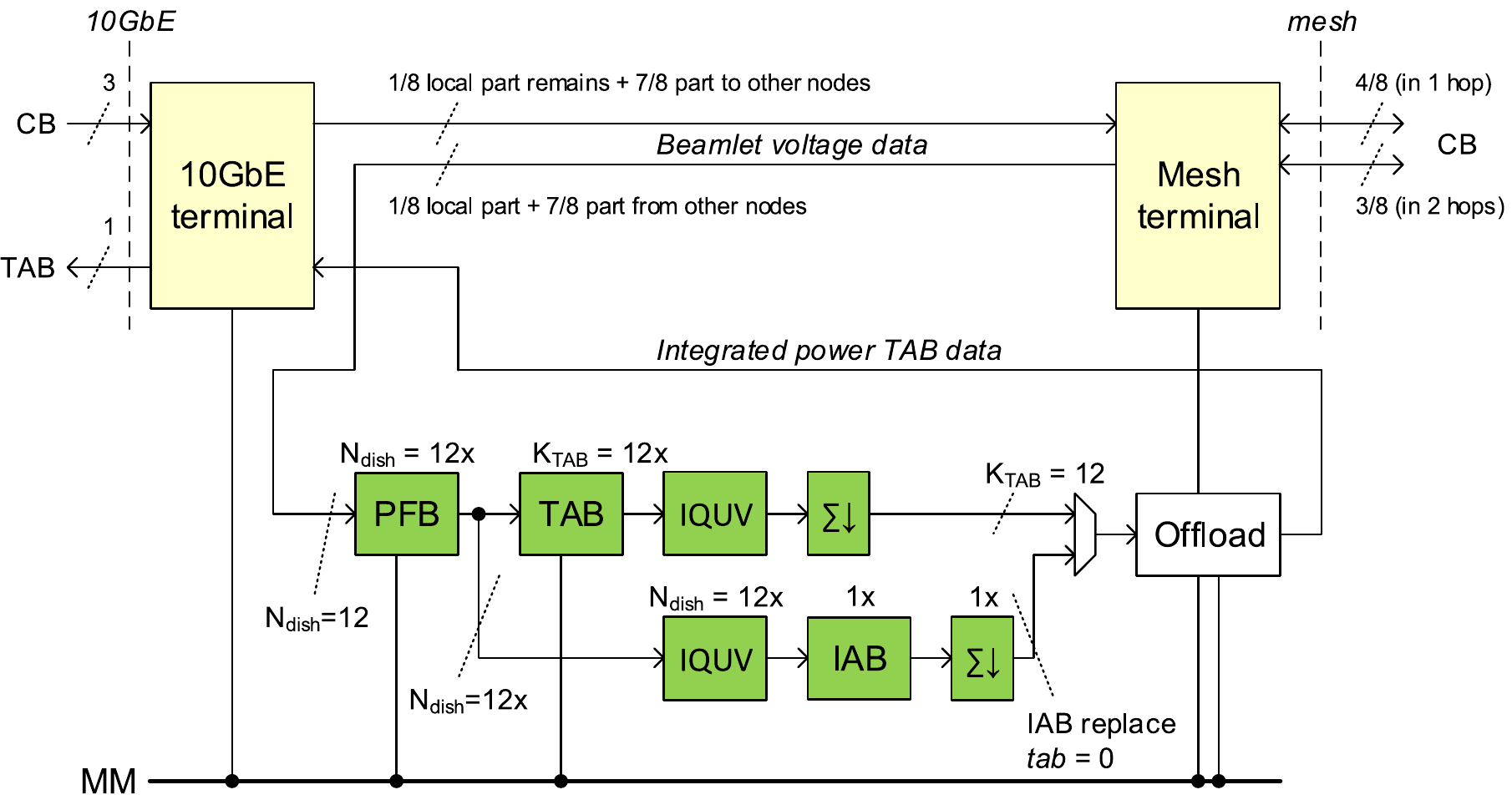}
  \caption{Top level block diagram for Dedicated Search FPGA application design on
    Uniboard.
    \label{td-fw-sc4-v01}
    }
\end{figure}  

The dedicated-search FPGA design %
(Fig.~\ref{td-fw-sc4-v01})
takes CB480 data as input.
Up stream, the \acp{FEBF} already compensated for the
time-varying geometrical delay due to the rotation of Earth.
That \ac{FS} was applied per \ac{CB} by
coarse, true sample \ac{DT}
plus fine-delay \acl{PT} (\acs{PT}; Sect.~\ref{sec:app:design:febf} and \citealt{cho+20}).
The  \acp{FEBF}  also already transposed the  CB480 data by T$_\mathrm{int\_x}$ (Fig.~\ref{td-febf-map-v01}).

To form and search \acp{TAB}, a series of transposes are required (Fig.~\ref{td-interconnect-map}).
The first is T$_\mathrm{array}$ = T$_\mathrm{dish}$ + T$_\mathrm{pol}$.
The wiring of the N$_\mathrm{link}$ = 384 10G input links to the \ac{TABF} implements the first stage of this
transpose, by bringing the data from N$_\mathrm{tp}$ = 24 telescope paths per band to a single UniBoard.
In the second part, these paths need to come together in a single \ac{FPGA}. 
This is achieved through the UniBoard mesh (Appendix~\ref{sec:app:design:hw:u}).
Of incoming beamlets, 1/8th stay at the receiving FPGA,  the other 7/8th are redistributed, to 
complete the transpose T$_\mathrm{array}$.

The 64 subchannels are created using a \ac{PFB} with an 8-tap  \ac{FIR} pre-filter.
On these subchannels, the coherent beamformer creates multiple intermediate single-polarization voltage TABs.
The phases of the 8-bit \ac{TABF} weights   $w_\mathrm{TAB}$ depend on pointing nor frequency, but only on
 dish index  $d$ and \ac{TAB} index $n$
(see Appendix~\ref{sec:app:beams}). They are simply
$\phi_\mathrm{TAB} = 2\uppi d \frac{n}{n_\mathrm{TAB}}$, 
where $n_\mathrm{TAB}$ is the total number of \acp{TAB}.
For the power TAB, the intermediate voltage TABs are converted into full Stokes power data.
Groups of 16 subchannels are  integrated to deliver %
a 81.92 $\mu$s sample time,
providing more than 10 time bins over an average FRB pulse.
The output data consists of separate I and IQUV data streams.
The available FPGA hardware allows this design to form up to 12 TABs (see Appendix \ref{sec:app:design:fpgares}
for the detailed usage). 

Each FPGA in the ARTS \ac{TABF} uses one 10GbE output link,
so in total there are M$_\mathrm{PN}$ = 128 output links  (Fig.~\ref{td-bf-map-v01}). 
Beamlets are mapped such that 5 CB directions end up at a FPGA.
The GPU cluster consist of 40 workstations, one workstation per CB.
Each FPGA thus only needs to output to 5 workstations in the GPU cluster. 
The rate per workstation is 1.8\,Gbps the Stokes I data, plus 
7.2\,Gbps for the Stokes IQUV data (cf. Table~\ref{tab:app:datarates}).

The design can also operate
to produce incoherent-array beams (cf.~Appendix~\ref{sec:app:design:fw:IAB}).

\subsubsection{Multiple TABs on \acl{UNB2}}
\label{sec:5b.2}

The CHAN320 input data for the commensal \ac{TABF} is already subchannelized in Apertif \ac{X} (Fig.~\ref{fig:app:XBF}),
and \mbox{offloaded} by its VHDL firmware.
One commensal \ac{UNB2} FPGA  \ac{TABF} design essentially includes 8 ported \ac{UNB} FPGA designs, processing the same as one entire
\ac{UNB}, without requiring a mesh interconnect.
 The voltage TAB360 processing and other functions
are similar
to the dedicated search (Sect.~\ref{sec:5b.1}). Background on the design can be found in
Appendix~\ref{sec:app:design:fw:UNB2}.
When forming 9 TABs the design uses about %
90\,\% of the multiplier elements in the FPGA.

\subsubsection{Single TAB}
\label{sec:5b.singleTAB}

The TAB1 \ac{TABF} for pulsar timing or VLBI
(Fig.~\ref{fig:app:XBF})
is a single coherent beamformer (Appendix~
\ref{sec:app:design:fw:TAB1})
that operates directly on the beamlet voltage
samples. 
The fringe stopping applied in the Apertif \ac{FEBF} means
TAB1 weights can be held fixed.
The output TAB1 is sent over 1GbE to the ARTS-0 machine.
That machine can interface the voltage TAB to a VLBI system (not in place at the time of writing)
or coherently fold and dedisperse it for pulsar timing (Sect.~\ref{sec:6c}).

\subsection{Software: Monitoring, Control, Operations} 
\label{sect:5c.1}
Control and monitoring of the overall time domain system is split over several 
controllers,
each described in the following subsections. 
In general, \ac{WSRT} and \ac{Apertif} are controlled by the \acl{MAC} system (\acs{MAC}; Sect.~\ref{sec:5c.mac}).
Observations are scheduled through \accom{ATDB}{Sect.~\ref{sec:5c.atdb}}.
Monitoring of the system health and component status is done through \acs{Artamis} (Sect.~\ref{sec:5c.artamis}).
Finally, the output data is moved into \accom{ALTA}{Sect.~\ref{sec:5c.alta}}.
These all interact with  the \ac{ARTS}-specific monitoring and control system (Sect.~\ref{sec:5c.artsmac}).
This interaction takes place through a message bus system.
The Apertif software is further detailed in \citet{cho+20} and Appendix~\ref{sec:app:design:sw}; the ARTS software in \citet{oostrum20}.

\subsubsection{\acf{MAC}}
\label{sec:5c.mac}

The Apertif \ac{MAC} \citep[][and Sect.~\ref{sec:app:design:sw}]{2016ADASS_mac}
comprises relatively independent software components that communicate through a message bus
(cf. Sect.~\ref{sec:app:design:sw:mb}) and are divided over three  layers: drivers, controllers and orchestration (Sect.~\ref{sec:app:design:layers}).
The drivers handle the I/O to all hardware components. %
The  high level controllers each cover one specific system function, such as control of a dish,  the
correlator, or  the beamformer.
Each controller typically interacts with several drivers. %
The orchestration layer sets different components through the entire system.

\subsubsection{\acf{ATDB}}
\label{sec:5c.atdb}

The \acl{ATDB} (\acs{ATDB}; \citealt{2019ADASS_atdb})
 workflow guides Apertif observations from specification to delivery into
\ac{ALTA}. %
ATDB is implemented as a microservices architecture.
The central database is wrapped inside a Django web application and can be
accessed by a range of (Python) ATDB services through a ReSTful API.
Each of the ATDB services in the automated workflow has a specific task
and is triggered by a specific observation status.
Together, this orchestra of services forms a very flexible, adaptable and lightweight workflow.

\subsubsection{ARTS Drivers and Controllers} 
\label{sec:5c.artsmac}

During system setup, a driver sets the static weights for the ARTS \ac{TABF}.
The rest of ARTS is configured every time an observation starts, when \ac{ATDB} sends a parameter-set file \citep[\emph{parset}, as in LOFAR;][]{sha+11}  %
over the Qpid messaging bus.
Two controllers direct the ARTS GPU cluster \citep[see Fig.~\ref{fig:controller} and][]{oostrum20}.
{\sc ARTSSurveyControl}, on the master node,  receives the parset, %
which lists  the \acp{CB} that need to be recorded. For each CB, %
pointing and processing settings
are sent 
to the {\sc ARTSSurveyNodeControl} controllers  on the worker nodes.

Processing on the nodes, as overseen by {\sc ARTSSurveyNodeControl}, revolves around the data memory buffers. 
Following  ARTS predecessor PuMa-II \citep{ksv+08}, these buffers use
\texttt{PSRDADA}\footnote{\url{http://psrdada.sourceforge.net/} \citep{2021ascl.soft10003V}}
formats and tools.
Four buffers are created: two for Stokes-I and two for Stokes-IQUV data; 
per set, the main buffer continuously holds all incoming data, while the trigger buffer uses \texttt{data\_dbevent} to store a selected amount
when an external trigger is received. Stokes-IQUV triggered data is written to disk using 
\texttt{dada\_dbdisk}. %
Multiple processes read from the main Stokes-I buffer.
First, {\sc ARTSSurveyNodeControl} %
starts three instances of GPU search pipeline {AMBER} (Sect.~\ref{sec:6b.2}) across three GPUs.
Then, it starts two data writers: 
\texttt{dadafilterbank} (for high-resolution {filterbank files}), and
 \texttt{dadafits} (for archive resolution FITS files).
Two instances of \texttt{fill\_ringbuffer} fill  the main Stokes I and IQUV buffers from the network.
Finally, {\sc ARTSSurveyNodeControl} starts  \ac{ARTS} processing pipeline DARC (Sect.~\ref{sec:6b.1}).

When an observation finishes,
each instance of {\sc ARTSSurveyNodeControl} reports back to ATDB which FITS data products were generated for
long-term archiving in ALTA.
\subsubsection{\acs{Artamis}}
\label{sec:5c.artamis}

The dish position,  subband statistics, %
 \ac{LNA} power,
and operational firmware image %
are monitored through \acl{Artamis} (\acs{Artamis}; \citealt{2019ADASS_artamis}). 
This real-time framework monitors the entire WSRT (and LOFAR), and is based on ETM/Siemens WinCC-OA.
Its web-based data is available to all users.
Adjustable alarms alert the telescope operators,
who have %
an instantaneous view on the whole system, including trends. %
\subsubsection{ALTA}
\label{sec:5c.alta}

The \acl{ALTA} (\acs{ALTA}\footnote{\url{https://alta.astron.nl}}; \citealt{2019ADASS_alta})
provides the long-term storage for raw and processed Apertif data.
It covers %
data management, discovery, and %
access.
The system combines %
2\,PB online disk-based storage, a PostgreSQL server,
and a web-based user portal (all hosted at \mbox{ASTRON}), with
tape-based scalable storage hosted at %
SURFsara.
Rule-based data management %
is provided through
iRODS\footnote{\url{https://irods.org}}.
Like all raw and processed data,  the \ac{ARTS} filterbank files are  ingested through an  automated workflow. 
Metadata %
is incorporated into the  database,
supporting full data provenance, %
compliant with \ac{VO} standards.

For public users, the primary entry point to access these ALTA data is through the
ASTRON Virtual Observatory services (Sect.~\ref{sec:12c}).
There, data collections including the one described in this
 paper are published.

\section{Description of Time Domain System: III. Science Pipelines}
\label{sec:6}

\subsection{FRB Searching}
\label{sec:6b}
Searching for FRBs is a computationally intensive process. This Section presents the parts constituting the ARTS search pipeline.

\subsubsection{AMBER}
\label{sec:6b.2}

The \acl{AMBER} ({\sc AMBER}; \citealt{2016A&C....14....1S})\acused{AMBER}
is  our real-time software pipeline for single pulse detection. %
\ac{AMBER}  identifies FRB candidates, i.e. dedispersed signals with high signal-to-noise ratio,
in the beamformed data stream (Sect.~\ref{sec:5b.1}). 
\ac{AMBER} is optimized for execution on highly parallel architectures
such as the ARTS \acp{GPU}.
\ac{AMBER} is  implemented in C++, with  GPU kernels  in OpenCL.
Its  design is modular: 
each component is separate, developed and maintained on its own.
This allows developers to add or modify modules, with minimal impact on the overall code base.
Moreover, modules can be reused in other projects (e.g.,  SKA) without a hard dependency on \ac{AMBER}.

In ARTS, the pipeline  sequentially applies all seven AMBER-provided processing stages
 to the data chunks:
(1) \acf{RFI} %
mitigation, (2) downsampling, (3) dedispersion, (4) integration, (5) signal-to-noise evaluation, (6) candidate selection, and (7) clustering.
We briefly describe each stage below. 

\newlength{\amberlistlengthspace}
\setlength{\amberlistlengthspace}{-3ex}
\vspace{\amberlistlengthspace}\paragraph{RFI mitigation (optional).}
\label{sec:4.rfi}
Optional stages can be enabled when starting AMBER.
Here, two distinct filters are applied to the input data to identify and remove -- in real time -- any bright, wide-band, low-DM RFI and bright narrowband RFI, respectively; this module is described in more detail in~\cite{sclocco2020}.

\vspace{\amberlistlengthspace}\paragraph{Downsampling (optional).}
This module reduces the time resolution of the input data. That is particularly useful when searching for
transients at high DM.
There, %
the intra-channel smearing %
 already exceeds the original sampling time.
 At high DMs, \ac{AMBER} also requires longer-duration chunks of data; at high time resolution these
 can %
 run into hardware memory limitations. %

\begin{figure*}
  \centering
  \includegraphics[width=0.75\textwidth]{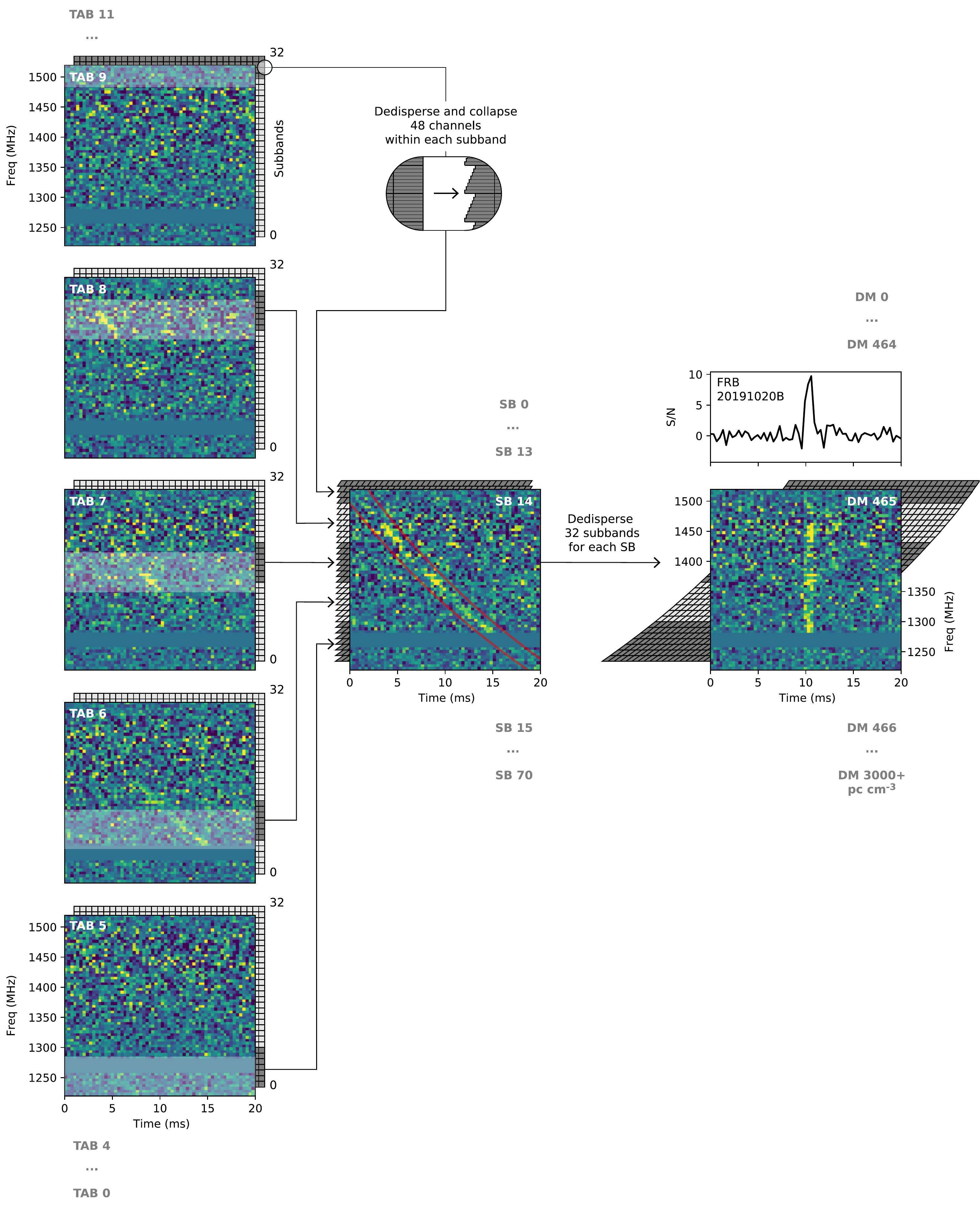}
  \caption{The formation of the SBs during subband dedispersion demonstrated through the detection of FRB~20191020B. 
  Combinations of 12 TABs (left column) produce 71 SBs.
  The TABs are combined in units of subbands, each consisting of 48 channels.
  Each subband is first dedispersed to a coarse DM.
  For SB\,14 the resulting intermediate time-frequency plot is shown in the center.
  We show this step here for clarity but in the
  production implementation it is optimized out. 
  For ease of visibility the FRB is shown here at a DM that is reduced by a factor 20.
  In the second step of dedispersion, the subbands are aligned over 32 trial DMs.
  FRB 20191020B was thus detected as shown on the right. 
  \label{fig:subband_dedisp}
  }
\end{figure*}

\vspace{\amberlistlengthspace}\paragraph{Dedispersion.}
\label{par:dedispersion}
This module is the base for the searching algorithm.
It implements dedispersion using an algorithm derived from the brute force approach.
Recognising that, in GPU dedispersion, optimizing for memory limitations is  more fruitful than for compute bounds,
the module is highly tuned for data reuse.
It can dedisperse  in both a single step, or using a more efficient two step process including subbanding.
During subbanding, the module creates the \acp{SB} described in Sect.~\ref{sec:3d.beam}, as
illustrated in Fig.~\ref{fig:subband_dedisp}.
A further overview of the module design and performance is available in~\citet{2016A&C....14....1S}.

\vspace{\amberlistlengthspace}\paragraph{Integration (optional).} 
The candidate \ac{S/N} is highest when
all burst emission is collected in a single time bin.
We thus search over a range of sampling times. %
During integration, the dedispersed time series are downsampled %
using a user-defined  set of trial pulse widths.
Each integration trial thus acts as a convolution kernel, smoothing the signal and approaching the maximum
intrinsic \ac{S/N} at the transient pulse width.

\vspace{\amberlistlengthspace}\paragraph{S/N evaluation.}
This module computes the \ac{S/N} of all peaks in the dedispersed, and optionally integrated, time series.
Different ways of computing the \ac{S/N} %
are implemented, and users can select the method most suitable for their data.

\vspace{\amberlistlengthspace}\paragraph{Candidate selection.}
Candidates whose \ac{S/N} exceeds a user-defined threshold are stored and made ready for output.

\vspace{\amberlistlengthspace}\paragraph{Clustering (optional).}
{If requested, AMBER will cluster results in the DM and integration (pulse-width) plane.  For groups of candidates
  with different pulse widths, but all occurring at the same time and the same DM, only the highest S/N candidate is
  preserved.  Next, groups of candidates that occur in neighboring DM trials are compacted, again keeping
  only the candidate with the highest S/N.  For each resulting DM-width cluster the single candidate with the highest
  \ac{S/N} is reported; reducing the total number of real-time output candidates.  Further per-observation clustering
  occurs off line (see next subsection).}

\vspace{1ex}
Eventually, all candidates found in the current input chunk are stored in a text file, and after the output is saved, \ac{AMBER} continues processing the next chunk of input data.

\ac{AMBER} %
is available on GitHub\footnote{\url{https://github.com/TRASAL/AMBER}; Apache License 2.0},
and portable in multiple ways.
\ac{AMBER} compiles and runs on different hardware platforms as it uses standardised and open languages:
C++ and OpenCL.
Through its  combination of run-time code
generation, user configurability, and auto-tuning, \ac{AMBER} furthermore
provides {\it performance} portability: it can be
automatically adapted to achieve high performance for different hardware platforms,
\citep[see e.g.][]{2018A&C....25..139M}, observational parameters and searching strategies.

\subsubsection{DARC}
\label{sec:6b.1}

In real time, {\sc AMBER} delivers metadata of FRB candidates.
Based on these,
the {\sc DARC}\footnote{\url{https://github.com/loostrum/darc}; Apache License 2.0}\acused{DARC} pipeline 
(the \acl{DARC}; \citealt{darc})
 decides whether  
to store the Stokes-IQUV data buffer to disk, and/or trigger a LOFAR observation.
\ac{DARC} also orchestrates the automated \emph{offline} processing
by the  deep neural network (Sect.~\ref{sec:6b.3}) of the best FRB candidates.
 \citet{oostrum20} describes \ac{DARC} in more detail.

\vspace{\amberlistlengthspace}\paragraph{Global design}
To ease development and improve readability, %
\ac{DARC}, like {\sc AMBER}, is split into several modules %
(Table~\ref{tab:darc_modules}).
These fall in three %
categories: monitoring and control, offline processing, and real-time processing.
Real-time processing is only active for the current observation, while offline processing may be
running for several observations simultaneously.
An overview of the \ac{DARC} modules and how \ac{DARC} connects to the
ARTS MAC (Sect.~\ref{sec:5c.artsmac}) is given in Fig.~\ref{fig:controller}.

\begin{table}
    \caption{{\sc DARC} modules and tasks, see also Fig.~\ref{fig:controller}.}
    \label{tab:darc_modules}
    \centering
    \begin{tabular}{lp{5.2cm}}
    \toprule
    Module & Task \\ 
    \midrule
    \multicolumn{2}{l}{\emph{Monitoring and control:}}\\
    \dm & Manages all other modules \\
    \sw & Generates a web page with the status of each module \\
    \midrule
    \multicolumn{2}{l}{\emph{Offline:}}\\
     \op & Full offline processing pipeline \\
    \midrule
    \multicolumn{2}{l}{\emph{Real-time:}}\\
     \al & Reads {\sc AMBER} candidates from disk \\
     \amc & Determines for which candidates to trigger a Stokes-IQUV dump and/or LOFAR TBB observation \\
     \dt & Executes Stokes-IQUV triggers through \psrdada \\
     \lt & Executes LOFAR triggers \\
     \vog & Sends VOEvents to outside world \\
    \bottomrule
    \end{tabular}
\end{table}

\paragraph{Monitoring and control}
Each node of the ARTS GPU cluster %
runs an instance of DARC.
The \dm module receives the control signals from the ARTS MAC
and manages all other DARC modules.
On the master node, the \sw module monitors the nodes, for a local status web page. 

\vspace{\amberlistlengthspace}\paragraph{Real-time analysis}
DARC runs real-time processing for every observation.
Every second, \al reads in the candidates from each {\sc AMBER} instance. %
Each candidate %
has a %
beam number, downsampling factor, arrival time, width, DM, and S/N.
A single transient may be detected in multiple points in this 6-dimensional parameter space. {\sc AMBER} already
clusters candidates that are close in time, DM, and %
downsampling factor. %
The \amc module clusters the candidates further over the DM/arrival time plane, and across all \acp{SB}
of the \ac{CB}. 
\begin{table}[b]
    \caption{Thresholds used during candidate clustering. }
    \label{tab:clustering_settings}
    \centering
    \begin{tabular}{lllll}
    \toprule
    
    Trigger      & Source type & S/N & DM range  & $D$ \\
                 &          &     &     (pc\,cm$^{-3}$)       &  \\
    \midrule
    \multirow{3}{*}{IQUV}
    & Known pulsar & >10 & DM$_\mathrm{src}\pm10$ & - \\
    & Known FRB    & >10 & DM$_\mathrm{src}\pm10$ & - \\ 
    & New source   & >10 & $>1.2 \times $DM$_\mathrm{YMW16}$ & 100 \\ 
    \midrule
    \multirow{2}{*}{LOFAR}
    & Known FRB  & >12 & DM$_\mathrm{src}\pm10$ & - \\
    & New source & >12 & $>2.0 \times $DM$_\mathrm{YMW16}$ & 80 \\
    \bottomrule%
    \end{tabular}
\tablefoot{$D$ is the maximum allowed downsampling factor for a candidate. \mbox{LOFAR} triggering is disabled for known pulsars.}
\end{table}

If the resulting highest S/N event  
exceeds the known or new source threshold (Table~\ref{tab:clustering_settings}),
the next pipeline step,  \dt, initiates
an IQUV dump and/or  LOFAR TBB trigger.

The IQUV dump consists of the set of 1.024\,s data chunks that encompass
 the dispersion delay
across the band with a minimum of 2\,s, plus an empty 2\,s
to determine noise statistics.
Overlapping  triggers are merged, %
eliminating %
dead time there; but disk-pool write speeds limit 
triggers to once per minute.

The \lt module on the master node collects the worker-node inputs, representing all CBs,  and selects the highest S/N
trigger to be sent to LOFAR.
 The \vog module can next format and 
 send an xml VOEvent to the broker, for 
 triggers to the outside world 
(Sect.~\ref{sec:12b}).

\vspace{\amberlistlengthspace}\paragraph{Offline analysis}
After recording finishes, the
offline processing is automatically started.
 \op performs a deeper search for FRB candidates, and informs the team of
the results via email. 
The module also takes care of system verification and calibration. %
This processing typically runs while the next observation is being analysed in real time.

The offline processing now clusters  {\it all} candidates
from the observation.  %
Any candidates with DM$\,<20\,\ppcc$ or S/N$\,<10$ are discarded.
{The real-time system already compacted candidates in the DM pulse-width plane.
Off line, we extend these clusters over the time dimension.
Within  a  0.5\,s time window,
 and in a number of DM windows that grow as the candidate DM increases,
 we keep only the highest-S/N candidate.}
For these remaining candidates, data from the
  \ac{TAB} filterbank files
are converted into the required \ac{SB} data using the {\sc DARC} multi-threaded \sbg tool,
and
 cleaned of RFI.
We then dedisperse to the DM found by {\sc AMBER} and downsample in time to the 
pulse width that maximises the S/N. %
That refines the initial, real-time {\sc AMBER} S/N estimate, which was based on only $1.024\,$s of signal and noise.
This re-calculation is especially important for wide pulses, detected at the highest downsampling factor of 250.
Offline, the method consists of a matched filter that 
tries many different box-car widths. %
For all candidates with %
post-processing  S/N$>$ 5, the
dynamic spectra, DM-time arrays, and metadata are saved to 
an HDF5 file, as input for  the  machine learning classifier (Sect.~\ref{sec:6b.3}). 

The resulting candidate plots from this classifier, together with 
overall statistics, %
are e-mailed  to the astronomer team and %
made available on a local website for human vetting and follow-up decision making.

In addition to FRB analysis, the \op module folds test pulsar data with {\sc prepfold} from {\sc PRESTO}\footnote{\url{https://github.com/scottransom/presto}} \citep{presto}. It also automatically processes 
drift scan data, which are used for sensitivity measurements (Sect.~\ref{sec:9b}).

\subsubsection{\acl{ML} Classifier}
\label{sec:6b.3}
Due to the real-time nature of the ARTS pipeline and the large number 
of false positives relative to true astrophysical transients, our candidate 
classification had to be automated. To this end, we built a binary classifier using 
deep neural networks %
to select true FRBs and discard false positives 
generated by \ac{RFI} and noise fluctuations \citep{connor-2018b}. The publicly-available package 
is called {\tt single\_pulse\_ml}\footnote{\url{https://github.com/liamconnor/single\_pulse\_ml}}
and uses {\tt Keras} \citep{chollet2015keras}
with a {\tt TensorFlow} \citep{tensorflow2015-whitepaper} backend 
for the construction, training, and execution of its convolutional neural networks 
\citep{singlepulseml}.
Our machine learning classifier was trained on tens of thousands 
of false positive triggers from Apertif, as well as an equal number of 
`true positives' that were generated either by injecting simulated FRBs into real 
telescope data or by detecting single pulses from Galactic pulsars. 

If the  classifier-assigned probability that an FRB candidate is  a real astrophysical transient exceeds 
a set threshold (currently 50\%), a diagnostic plot is generated showing the frequency-time intensity array (i.e. the
dynamic spectrum), a DM-time intensity array,  the pulse profile, plus metadata such as the beam number, S/N, classifier probability, and width.

\subsubsection{Localisation method}
\label{sec:6b.4}
To localise \acp{FRB} %
we make use of the multi-beam information provided by our setup.
We create a model of the telescope response on a grid spanned by RA and Dec,
and compare this to the measured \ac{S/N}.

The beam  model is constructed from the hierarchical beamforming techniques
described in Sect.~\ref{sec:3d.beam} and Appendix~\ref{sec:app:beams}, and is detailed in \citet{oostrum20}.
First, we create a model of the CBs, based on drift-scan data from both imaging and time-domain drift-scan data.
The performance of this model is described in Sect.~\ref{sec:9.FoV}. 
For each CB, we then simulate the tied-array beamformer, followed by the construction of the \acp{TAB} and \acp{SB}.
This final beam model predicts the relative sensitivity of the different \acp{SB} (see Sect.~\ref{sec:9.tabs}).

To localise a burst, we next compare its detection footprint against this model. 
We determine the detection \ac{S/N} of all \acp{SB} covering the \ac{CB} the burst was found in, and of all 
neighbouring \acp{CB}.
A beam model is next generated on a $40\arcmin\times40\arcmin$ grid with a resolution of $2\arcsec$, centred on the
highest-S/N \ac{CB}. %
The predicted relative S/Ns over all \acp{SB} are compared to the measured S/Ns, through a $\chi^2$ method.
The final 90\% confidence  localisation region is derived from the $\Delta\chi^2$ values.
As the region is generally close to elliptical, we often refer to it below as an ellipse.
The most likely localisation position is generally on the semi-major axis, but often away from the center.
The localisation  precision is evaluated in  Sect.~\ref{sec:9.loc}.

\subsubsection{Polarization calibration method}
\label{sec:6b.5}

{Although a Jones matrix was numerically determined for 
a model of the entire receiver chain of the Apertif prototype system \citep{6236092},
the polarimetric quality of the final system is known from on-sky calibration.

Both in imaging and in time domain, this calibration uses 
an unpolarized source %
to determine the relative gain of each polarization, and
remove the leakage, and a
known linearly polarized source %
to correct for the relative phase, and calibrate the polarization angle (PA).

In the raw autocorrelation CB data, the off-axis leakage is already $<$1\% \citep{Denes2022}.
After calibration, the 
leakage is also found to be $<$1\% over almost the entire imaging primary beam \citep{2022A&A...663A.103A}.
The on-axis leakage is even lower.
Tests in time domain find this same purity,
as we blindly reproduce the linear polarization fraction  
of 3C286 to 1\% accuracy.
Such precision allows for confidence in interpreting IQUV profiles \citep{Pastor-Marazuela_2022_Scattering};
the rotation measures we derive are even
more robust, as low leakage does not significantly impact the rotation of the PA with $\lambda^2$ \citep[see, e.g., ][]{clo+20}. 
}

\subsection{Pulsar Timing}
\label{sec:6c}

\ac{ARTS} can process a single tied-array beam
(Sect.~\ref{sec:5b.singleTAB}), to perform pulsar timing and to carry out general pulsar studies.
During such an observation,
the complex sampled voltages are processed in real-time on a dedicated server (Sect.~\ref{sec:4f}).
This machine, ARTS-0, has two classes of operation: a recording mode and a timing mode. 
In recording more,
the incoming TAB1 data is written to disk in \texttt{PSRDADA} format \citep{2021ascl.soft10003V}.
For timing, ARTS-0
predicts the pulsar periods and
phases for the duration of the observation using
\texttt{tempo2} \citep{2006MNRAS.369..655H, 2012ascl.soft10015H}.
Using these predictions, 16 instances of \texttt{dspsr} fold and coherently dedisperse
the data that come in through the input
ringbuffers (see Appendix~\ref{sec:app:design:sw:timing} for details).
Each \texttt{dspsr} instance processes a 18.75\,MHz subband
containing 24 channels of 0.781250\,MHz (as produced by the \ac{FEBF}, Sect.~\ref{sec:4b.1}).
The 16 instances are divided over 2 GPUs, which allows the system to reduce data in real time.
Subintegrations of the resulting folded profiles are stored in a
{\sc psrchive} \citep{hvm04} format filterbank file every 10 seconds.

\section{Commissioning results }
\label{sec:8}

\subsection{Pulsar detections}
\label{sec:8.pulsar}

During commissioning, we regularly observed the four pulsars
    B0329+54,
    B0531+21,
    B0950+08, and
    B1933+16.
Single pulses were detectable
from each.
As further detailed in  \citet{overview_arxiv_v1} and \citet{oostrum20} we used 
these detections and the radiometer equation 
to produce first, rough estimates of the \acf{SEFD} of the 8-dish system. %
Scintillation,  intrinsic brightness variations,
and the Crab Nebula background flux in the PSR~B0531+21 TAB
affect these measurements. 
The median of the four derived \ac{SEFD} values is 200\,Jy.
PSR~B1933+16 is the most stable pulsar in the set,
and the \ac{SEFD} its observations suggest (130\,Jy) 
is expected to be the most accurate.
Assuming an aperture efficiency of 70\%, a system temperature of 70\,K  \citep{2009wska.confE..70O,cho+20}, and
perfectly coherent beamforming, the theoretical SEFD of eight Apertif dishes is expected to be ${\sim}70\,$Jy. The
values derived here are a factor 2-3 higher. Fluctuations in pulsar brightness on that level are not unexpected.
A more accurate SEFD can be derived using scans of calibrator sources, as discussed in Sect.~\ref{sec:9b}.

\subsection{Coherence of the Tied-Array Beams}
\label{sec:8.coh}
\begin{figure}
\centering
\includegraphics[width=\columnwidth]{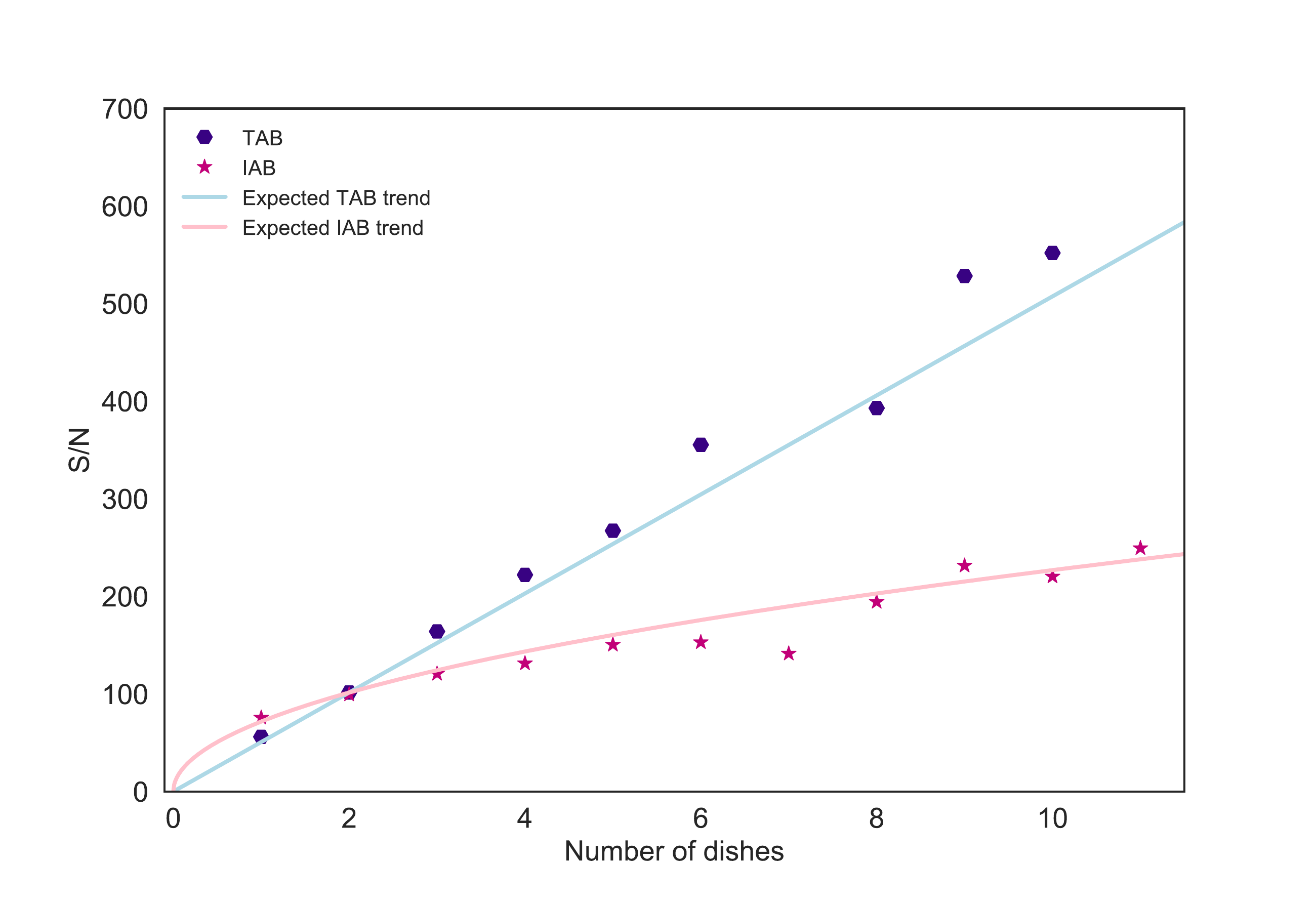}
\caption{\acl{S/N} of PSR B1933+16 versus number of dishes, for coherent and incoherent
  addition. The two trends are scaled to intersect at the 2-dish \ac{S/N} value.
\label{fig:snr_ndish}}
\end{figure}

Observing with an interferometric array is most profitable when full beam-forming coherence is obtained.
Then, the attained \ac{TAB} sensitivity evolves linearly with the 
number of dishes used.
For {incoherent} addition, on the other hand, the \ac{IAB} evolution follows a square-root law.
We measured the efficacy of our telescope addition by
observing pulsar B1933+16 over about two dozen telescope combinations
 \citep[for details, see][]{straal18},
 in both modes
 (Fig.~\ref{fig:snr_ndish}). 
The sensitivity increase is in good agreement with the expected relation
S/N $\propto \sqrt{n_{\rm{dish}}}$ for the \ac{IAB},
and
S/N $\propto n_{\rm{dish}}$ for the \ac{TAB},
demonstrating the coherence of the multidish
beam-forming system.

\subsection{Timing stability} 
\label{sec:8b.timing}

Early %
commissioning
flagged  that the synchronisation of the time stamps (the \acs{BSN}, \acl{BSN}, Sect.~\ref{sec:app:design:fw:timing})
to the \ac{PPS} signal %
did not occur completely as expected when the \acsp{ADC} (\aclp{ADC}; Sect.~\ref{sec:app:design:hw:febf})\acused{ADC}
were power cycled. Such power cycling (``cold starts'') %
severely reduced the effectiveness of
 the \ac{CB} calibration solution, as the time and phase information of the CBs jumped.
 Firmware switches under uninterrupted  power (``warm starts'') were found to keep calibration valid, and produce a
 system at adequate  sensitivity and stability for surveys.
 
Such timestamp jumps, however,  could impede %
 pulsar timing with Apertif (Sect.~\ref{sec:6c}).
Especially for long-term timing, intermediate system power cycling cannot be avoided.
We thus compared Apertif observations of a set of precisely timed millisecond pulsars,
taken over multiple, monthly epochs with intervening  power cycling,
against their 
European Pulsar Timing Array \citep{2013CQGra..30v4009K}  ephemeris. 
We find phase offsets that vary between epochs, of over 1\,ms in magnitude.
This shows the system currently %
does not straight-forwardly
support long-term timing. %

In multiple observations taken of PSR~J1022+1001
without power cycling in between, the pulse phase does line up to within
a $\sim$1\,$\mu$s. Short-term timing thus is phase coherent. %

\subsection{Perytons}
\label{sec:8.perytons}

\begin{figure}
  \centering
  \includegraphics[width=0.9\columnwidth]{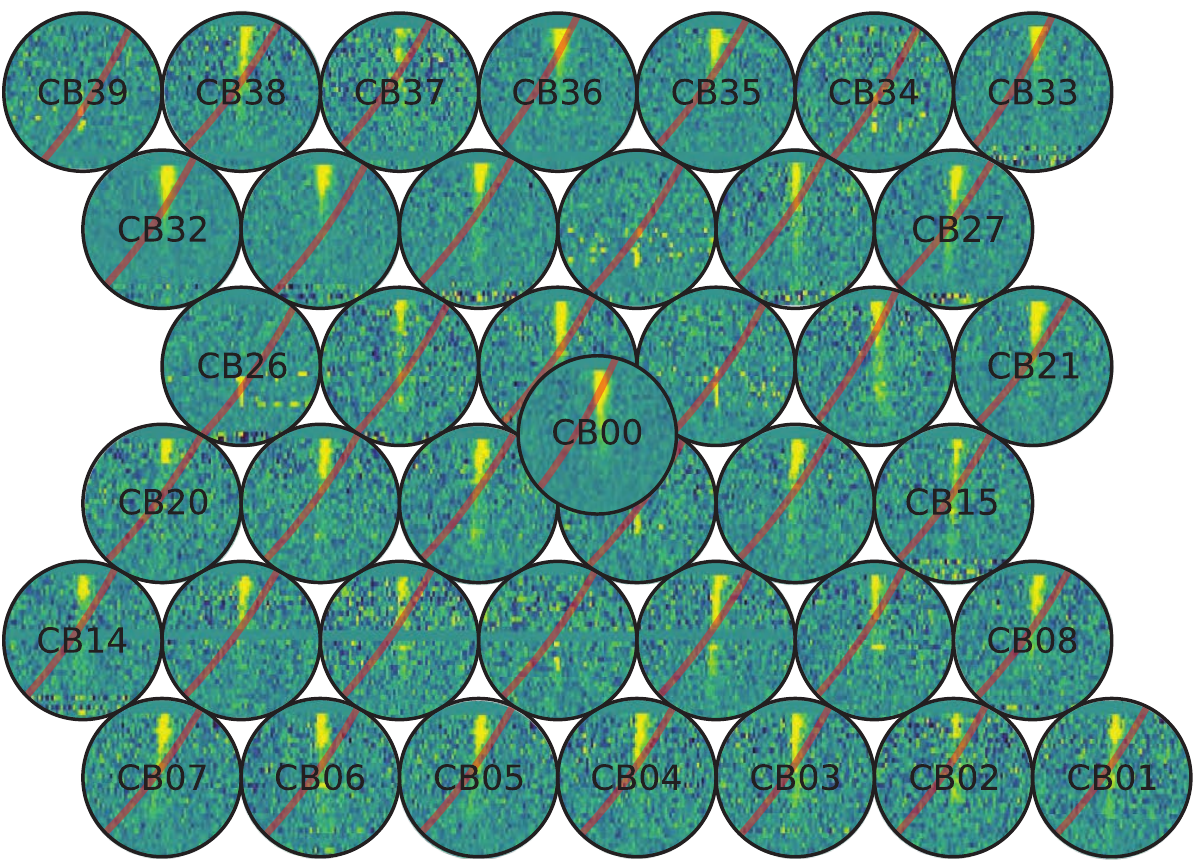}
  \caption{Peryton, detected in all 40 CBs (the circles).
    In each, the respective time-frequency plot is displayed (similar to, e.g., the bottom panel of
    Fig.~\ref{fig:frb:20191109A}), dedispersed to 395\,\ppcc. 
    The vertical axis spans 300\,MHz of bandwidth, the horizontal axis 500\,ms of time.
    The red curve indicates the track a signal with DM\,=\,0\,\ppcc\ would have followed. 
  \label{fig:perytons}
  }
\end{figure}

The ARTS real-time transient detection system was tested using perytons.
Perytons are a type of \ac{RFI} first discovered at Parkes \citep{2015MNRAS.451.3933P},
generated when microwave-oven doors are opened while the cavity magnetron still operates.
The frequency structure of the interference  
strongly resembles a %
high-DM  source.
During the Apertif Science Verification Campaign (SVC; Sect.~\ref{sec:12d}) we 
emitted 8 perytons using 2 different ovens, from outside the \ac{WSRT} control building.
AMBER detected the perytons in real time at a DM of nearly 400\,\ppcc\ (Fig.~\ref{fig:perytons}).

\section{Performance                                  }
\label{sec:9}

\subsection{Compound-beam Sensitivity}
\label{sec:9b}
The ARTS sensitivity depends on the sensitivity of
the individual \acp{CB} and on the calibration accuracy of the delay
and phase offset between the dishes. 
This CB and array-delay calibration determines if the phasing is coherent, as
the beam weights for \ac{TABF} itself are purely geometric (Sect.~\ref{sec:5b.1}).
The system is stable for 2$-$3 weeks \citep{cho+20}, and is re-calibrated weekly.
Following this weekly phase/delay calibration, we use quasars
    3C48,  
    3C147, or
    3C286 \citep{pb17}
to determine the sensitivity of the \acp{CB}, following a drift-scan method  \citep{oostrum20}  
summarised below. 

\begin{figure}[t]
\centering
    \includegraphics[width=\columnwidth]{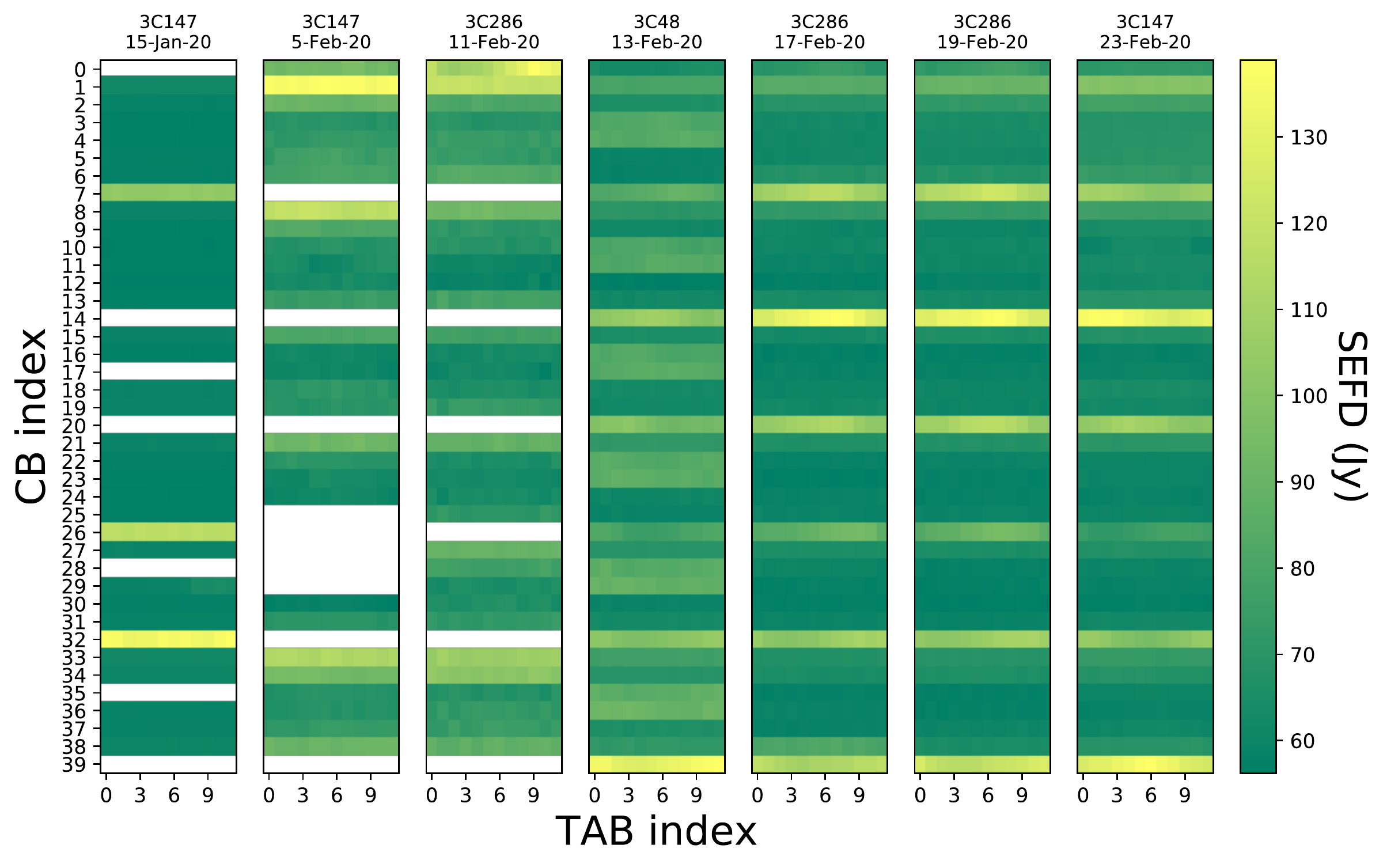}
    \caption{SEFD for all TABs of all CBs on seven different days. White regions indicate absence of data. Above each
      figure, the calibrator source and date are indicated.
    }
    \label{fig:sefd_allcb}
\end{figure}

For each TAB of each CB, we compare the
on-source and off-source values to determine the \ac{SEFD} from
\begin{equation}
\label{eq:onoff}
\frac{\mathrm{on-off}}{\mathrm{off}} = \frac{S_\mathrm{calibrator}}{\mathrm{SEFD}},
\end{equation}
where $S_\mathrm{calibrator}$ is the  calibrator flux density.
Fig.~\ref{fig:sefd_allcb} shows the derived SEFD for all 480 TABs, on seven different days.
The reported SEFDs are the median values in the frequency band not affected by strong RFI.
The TAB sensitivity within a CB is relatively constant.
The CBs themselves vary, most likely due to variations in the weekly calibration,
which can be affected by \ac{RFI} or the calibrator hour angle.

The CB sensitivity is expected to depend on its position within the \ac{PAF},
with central elements illuminated best.
Indeed we find that beyond ${\sim}1\degr$ from the PAF centre, the SEFD increases, by
${\sim}$30\% for the outermost beams
(Fig.~14 of \citealt{overview_arxiv_v1}; confirmed independently in Fig.~38 of \citealt{cho+20}).
The median \ac{SEFD} of the system is 85\,Jy.
\subsection{Field of View}
\label{sec:9.FoV}

\begin{figure}[b]
    \includegraphics[width=\columnwidth]{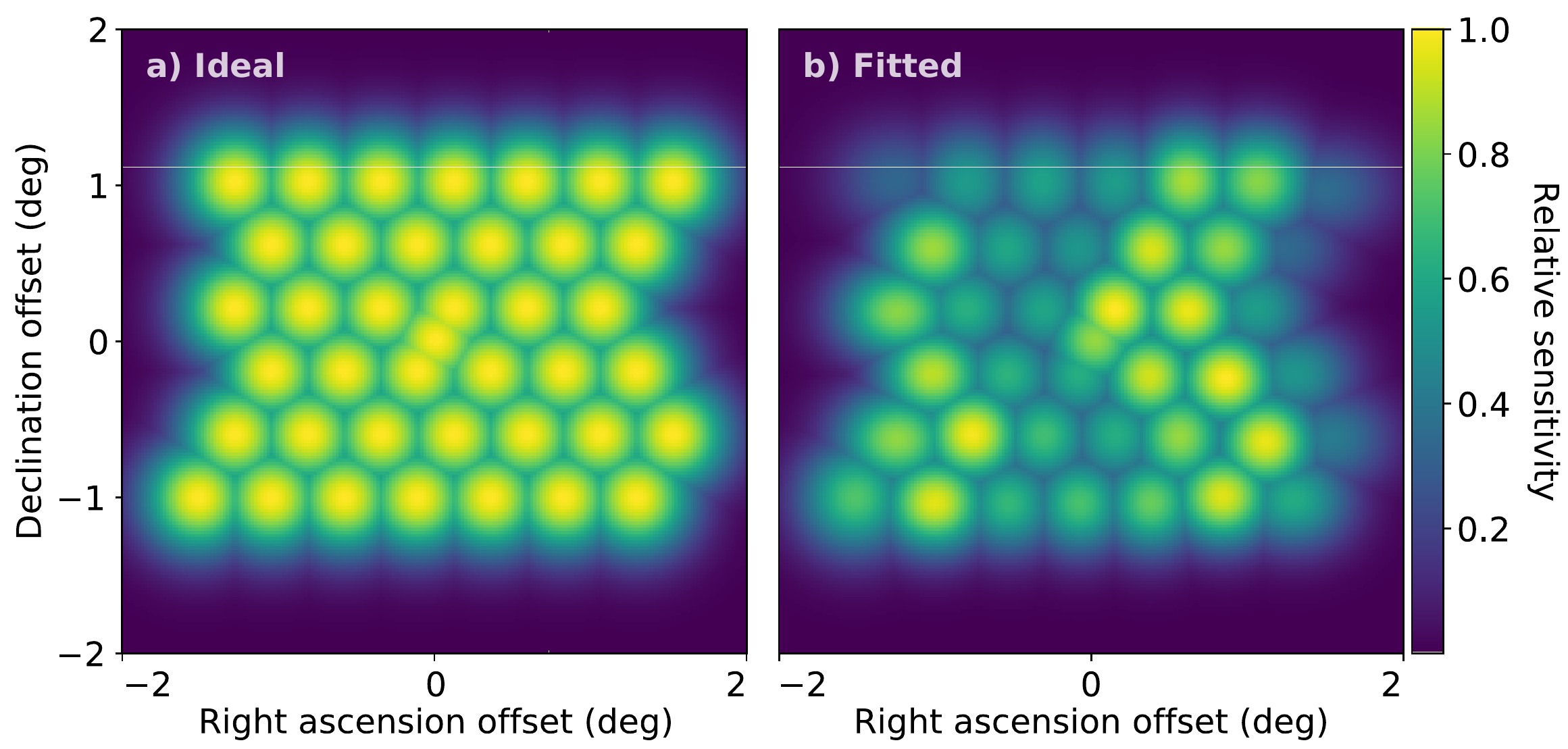}%
    \caption{Model of the 40 \acp{CB} of Apertif at 1370\,MHz using a) rotationally symmetric Gaussian beams with equal
    peak sensitivity and b)  2D Gaussian fits to each beam, scaled to their measured peak sensitivity in November 2019.}
    \label{fig:cb_models}
\end{figure}

Based on the beamforming models described in Sect.~\ref{sec:3d.beam} and Appendix~\ref{sec:app:beams}, 
we determine the theoretical \ac{CB} sensitivity pattern.
As the WSRT dishes have equatorial mounts, this pattern does not rotate on the sky.
As the \ac{PAF} itself is planar,
the CB offsets must first be
transformed onto the spherical (RA, Dec)
coordinates using a Gnomonic projection \citep[see][]{oostrum20}.
The resulting theoretical model at 1370\,MHz of the 40 \acp{CB} is as shown in Fig.~\ref{fig:cb_models}a.
This assumes each \ac{CB} reaches the same peak sensitivity. In reality, both the CB sensitivity and 
 shape depend on their position within the \ac{PAF}, and change over time as the system is
 recalibrated weekly (Sect.~\ref{sec:9b}). 
 Using drift scans in the image plane, the shape of the \acp{CB} was measured in June 2019
 \citep{Denes2022,Kutkin2022}. As long as none of the main PAF elements contributing to a CB fail, the shape of the \acp{CB}
should remain fairly constant. A 2D Gaussian was then fit to each CB averaged over all dishes. The size of the \acp{CB}
is assumed to scale linearly with wavelength.

To account for temporal variations in sensitivity (generally caused by \ac{RFI} during calibration),
a more accurate model is constructed for each observing session with an FRB detection.
There, the peak sensitivity of each CB is determined from the nearest time-domain drift scan;
leading to a model of CB sensitivity at each (RA, Dec) point on
the localisation grid.
The November 2019 model, for example, is shown as Fig.~\ref{fig:cb_models}b.

From these models,
we determine the Apertif FoV, out to the half-power contour, 
to be  8.2\,sq.\,deg. at 1370\,MHz. The FoV of a single CB is ${\sim}0.3\,$sq.\,deg. at 1370\,MHz, so Apertif increased the WSRT \ac{FoV} by a factor ${\sim}30$. At higher frequencies, the relative FoV increase is even higher.

\subsection{Tied-array beamforming performance}
\label{sec:9.tabs}

\begin{figure}
    \centering
    \includegraphics[width=0.4\textwidth]{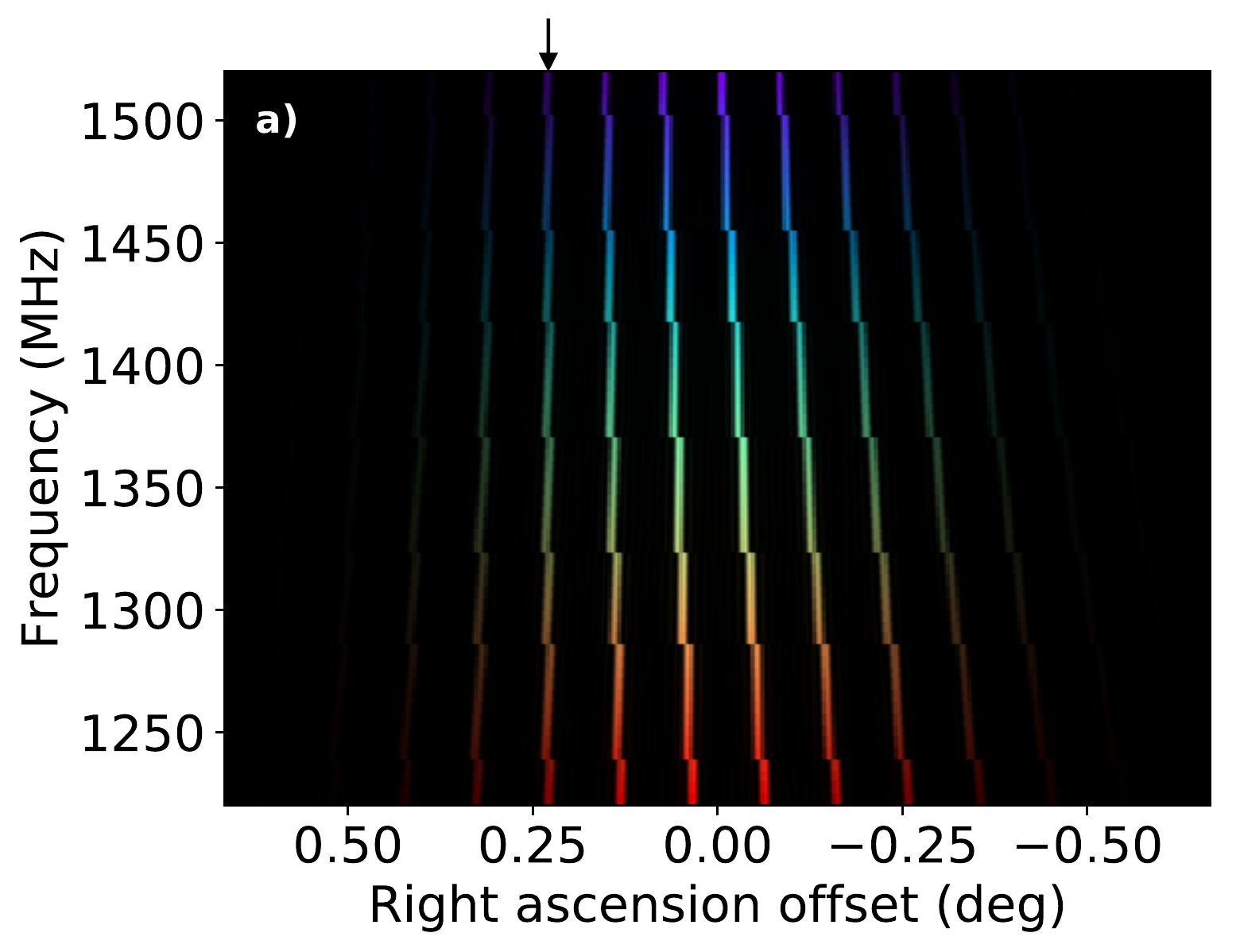}%
    \hfill
    \includegraphics[width=0.4\textwidth]{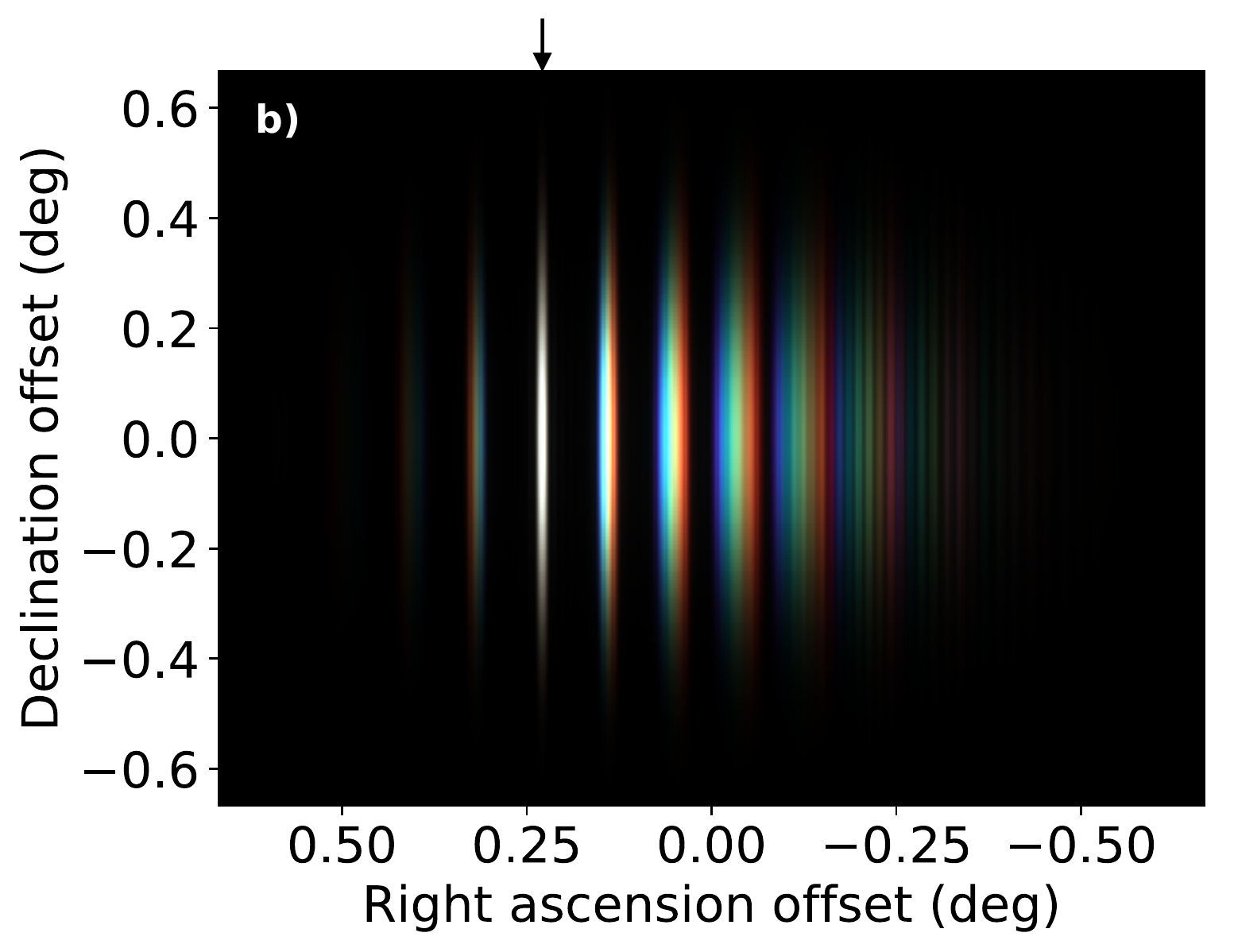}%
    \caption{Model of outermost \acl{SB} 70:
in a), as a function of RA and frequency. The intended main beam is indicated with an arrow. The 
    discontinuities in the sidelobes occur when the beamforming switches to a different TAB.
                    In b), as function of RA and Dec.
    The colours indicate frequency, where white
    means an SB is sensitive over the full frequency range. Only the main beam in (b) is broadband, the
    sidelobes are not.
    }
    \label{fig:sb_models}
\end{figure}

The \ac{FoV} spanned by the \acp{CB} is tiled out with \acp{TAB} and \acp{SB}
(Sect.~\ref{sec:3d.beam}).
To find the relative sensitivity of the different \acp{SB},
we simulate the ARTS beamformer (Sect.~\ref{sec:4b.1.tabf}), recreating its pointing through
the application of the
 geometric phase offset of the CB centre  plus the  additional TAB
offsets \citep[as detailed in][]{oostrum20}.
We take the projected baseline length and orientation into account.
The sensitive TAB sidelobes are %
attenuated by the CB response visible in Fig.~\ref{fig:cb_models}.
We finally simulate the mapping of the TABs to SBs, 
where for each of 32 subbands
the most sensitive TAB is used given the required SB pointing  (Appendix~\ref{sec:app:beams}).
Only the main SB beam contains signal of the full frequency
range, whereas the SB sidelobes do not.
This is best illustrated for  a beam on the edge of the CB, as shown in 
    Fig.~\ref{fig:sb_models}. For a main beam response, see e.g. Fig.~\ref{fig:20190709A_loc}.

Based on the drift scans of the calibrator quasars through the TABs of each CB (Sect.~\ref{sec:9b}), 
we determine the TAB beamforming efficiency.
Combining the aforementioned
\ac{SEFD} of 85\,Jy of the CB plus TAB system
with the 75\% aperture efficiency
and the  ${\sim}70\,$K system temperature
leads to a beamforming efficiency $\beta$ of 0.8.
The deviation from 1.0 is likely the result of \ac{RFI}, especially through its effects
on the weekly CB calibration.

\subsection{Localisation Accuracy}
\label{sec:9.loc}

\begin{figure}
    \includegraphics[width=0.9\columnwidth]{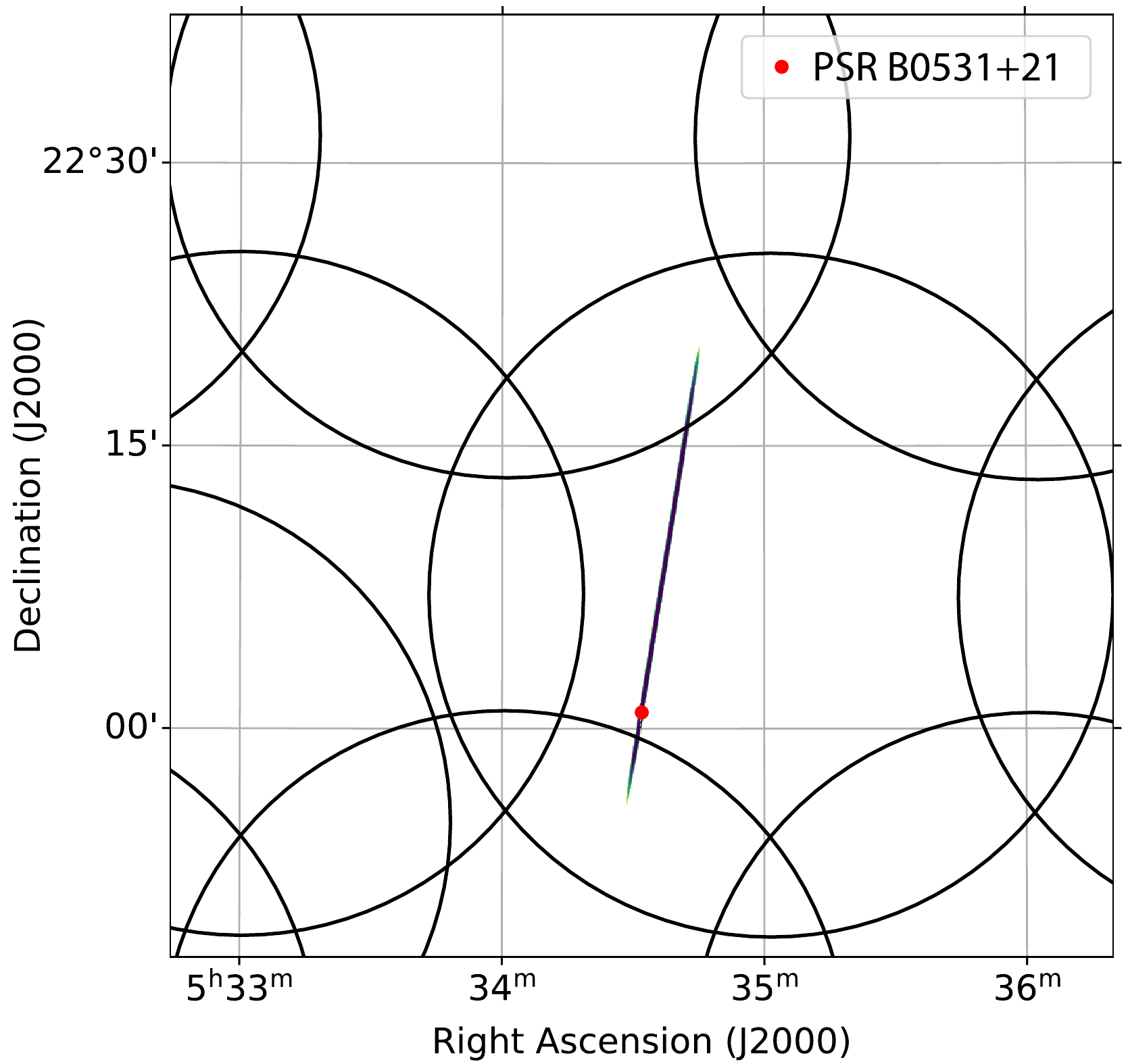}
    \caption{Localisation of the Crab pulsar using a single pulse.  The black circles indicate the size of the \acp{CB}
      at 1370\,MHz. The true location falls well within the 90\% confidence ellipse that is displayed.
    \label{fig:loc_B0525}
    }
\end{figure}

To validate 
 the accuracy of the localisation described in Sect.~\ref{sec:6b.4},
 we observed a field that includes
 both
 the Crab pulsar (PSR~B0531+21) and PSR~B0525+21.
An example of the localisation of
the Crab pulsar, based on the non-/detection footprint over the \acp{CB} and \acp{TAB} of a single pulse,
is shown in Fig.~\ref{fig:loc_B0525}.
The real position of the pulsar falls within the final 90\% localisation region.
As the likeliness is quite flat over this region, the fact that the source is away from the center of the region does not
signify any systematic offsets. 
Similar validation on PSR~B0525+21 corroborates this outcome \citep{oostrum20}.

\begin{figure}[t]
    \includegraphics[width=\columnwidth]{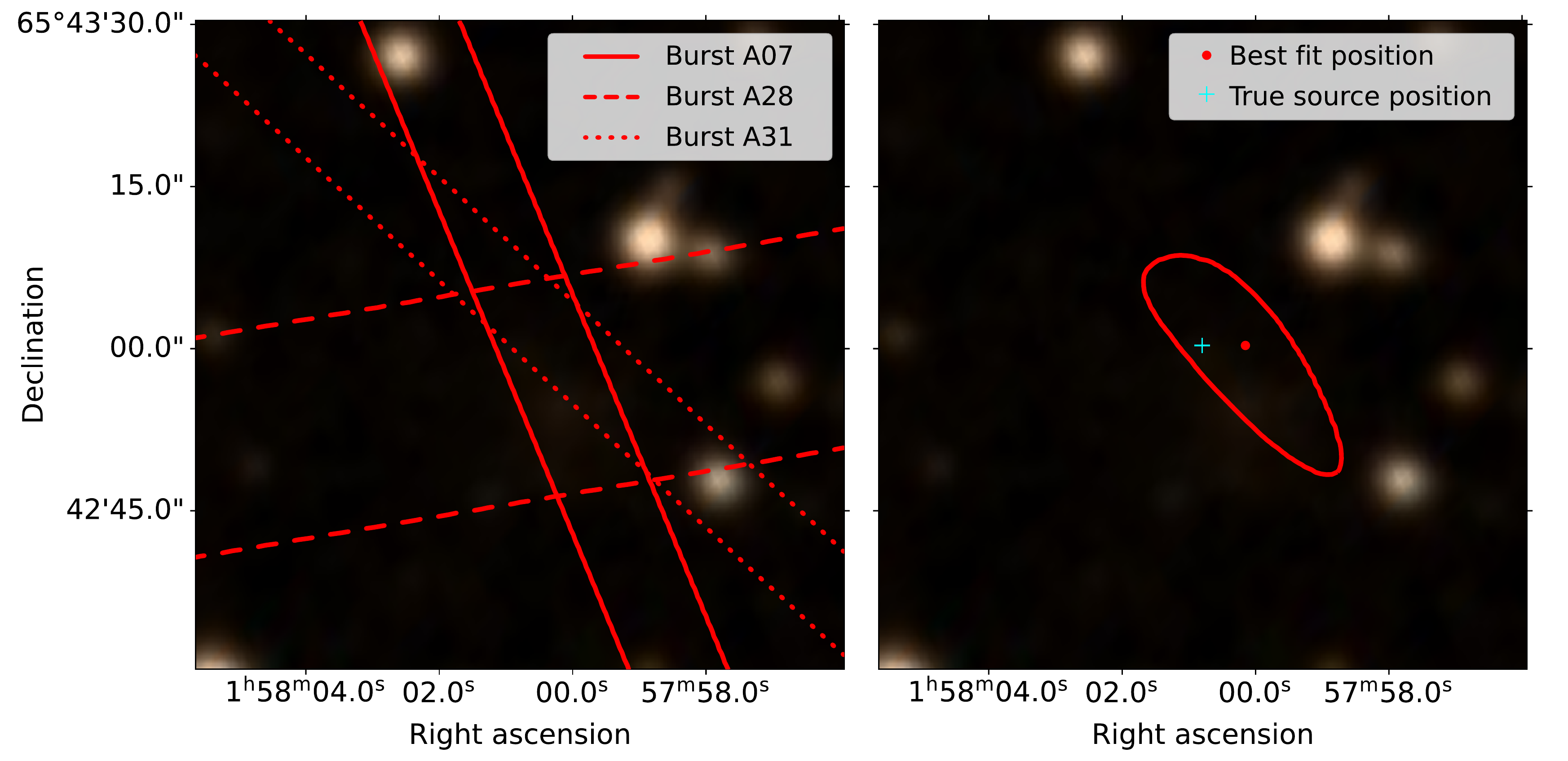}
    \caption{Localisation of FRB~20180916B from three individual bursts.
    Left: a zoom-in on the overlap  of the localisation regions for
    bursts A07, A28 and A31 (labels as in  \citealt{2020arXiv201208348P}).
           Right: the combined, final localisation region.
    \label{fig:R3}}
\end{figure}

For sources that repeat, multiple burst at different hour angles next intersect to further pinpoint the source
(Sect.~\ref{sec:3c.regular}, Fig.~\ref{fig:highlevelbeams}).
Two or three bursts generally suffice.
In Fig.~\ref{fig:R3} we demonstrate how the independent localisation of three separate bursts of FRB~20180916B
\citep{2020arXiv201208348P} combines to a final 
 90\%-confidence-level localisation area of 26\arcsec\ by 9\arcsec,
 in which  the true position is contained.

\subsection{RFI Identification and Mitigation}
\label{sec:9.rfi}

WSRT is located in a radio-quiet zone %
and its \ac{RFI} environment is good overall.
A number of known RFI sources do, however, remain \citep[as listed in][]{oostrum20}. These are generally  stronger than astrophysical signals
and can cause false-positive detections (i.e. non-astrophysical pulses erroneously classified as FRBs).
False positives hamper the search pipeline,  rapidly increasing the size of the
single-pulse candidates list: requiring both more processing and more human verification.
To reduce the impact of RFI, two mitigation  strategies are currently used in \ac{ARTS}:
on-line RFI removal %
using local statistics \citep[Sect.~\ref{sec:4.rfi} and][]{sclocco2020},
and an off-line deep-learning classifier (Sect.~\ref{sec:6b.3}). 

\subsubsection{Direction dependence of RFI}
\label{sec:9.rfidir}
We assessed the variability of RFI against sky direction 
in the {\sc AMBER} results of 448 observations taken March-Sept 2019,
when online RFI mitigation was not implemented yet in {\sc AMBER}.
Any trigger with DM=0 and S/N$>$10 was assumed to be RFI.
On a sky map of the resulting RFI trigger rates (Fig.~19 of \citealt{overview_arxiv_v1}),
a peak at az=$260\degr$ and alt=$65\degr$ stands out from the otherwise uniform RFI distribution on the sky.
The Smilde radio mast is located at this azimuth and its UHF antenna transmits in the
Apertif frequency band.
We likely observe a reflection on the troposphere. Mast emission that is reflected once, halfway between Smilde and
WSRT, at a typical troposhere height of 17\,km, is expected at an apparent altitude is $68\degr$, which matches our
observations. 
\subsubsection{Fraction of data lost to RFI}
\label{sec:9.rfifract}

\begin{figure}
  \centering
  \includegraphics[width=\columnwidth]{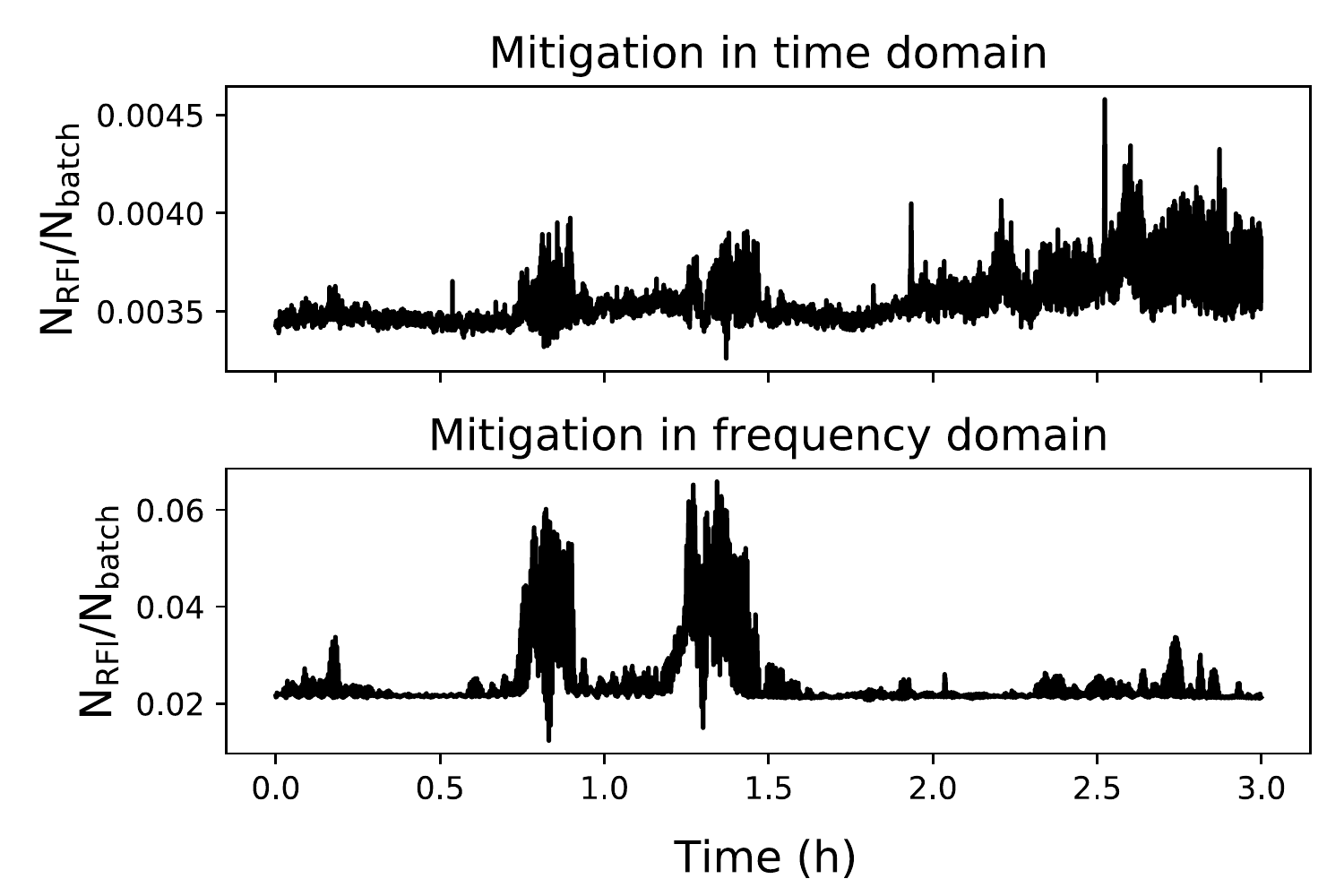}%
  \vspace{-1ex}\caption{Fraction of samples %
    cleaned of RFI by the two mitigation strategies, throughout a
    standard 3-hour observing session. %
  }
  \label{fig:rfi_3hrs_obs}
\end{figure}

The on-line RFI mitigation in {\sc AMBER} (Sect.~\ref{sec:6b.2}) employs two methods:
mitigation in time domain targets
bright low-DM broad-band signals, while the frequency-domain
method removes spurious narrowband RFI.
Both methods are applied in an iterative manner, where each
consecutive step applies mitigation to an ever-cleaner set of samples.
Three iterations of each method are
applied.
We evaluated the overall impact of RFI on our time and frequency samples, in a
representative 3-hour observation.
We find that at most about 7\% of the total bandwidth is affected by RFI for limited time periods (< 0.5\,hr).
In general, less than 3\% of data is lost to RFI.
Fig.~\ref{fig:rfi_3hrs_obs} highlights that the majority of RFI sources in our band consist of spurious narrow-band emission.

\subsection{Observing efficiency}
\label{sec:9e}

We express the end-to-end efficiency of the ARTS system
by the fraction of on-source time for which valid data reaches the archive (Sect.~\ref{sec:12c}).
Such an outcome is prevented by e.g., dead computing nodes and pipeline failures.
We compare the successful time against the total expected on-source time in dedicated mode
(so, not counting against e.g. slew time, setup time, or calibration time).
The ratio is prorated if a subset of beams fails.
This fraction  is a lower limit to our discovery efficiency though,
as detections happen in real time. These can and
have occurred when, for example, network issues prevented archiving, while real-time processing continued.
For 2019, this efficiency was $\sim$90\%, for 2020 it was 94\%. 

\subsection{Injection pipeline}
\label{sec:9g}
Throughout commissioning, each step in the ARTS pipeline (Sect.~\ref{sec:6b}) was tested via simulated FRB injection
tests and single-pulse observations of Galactic pulsars. For our simulated bursts,
we developed an injection package, {\tt injectfrb}\footnote{\url{https://github.com/liamconnor/injectfrb}}, as well as evaluation code for assessing completeness\footnote{\url{https://github.com/EYRA-Benchmark/frb-benchmark}}. The 
{\tt injectfrb} package accounts for 
physically realistic effects such as 
intra-channel dispersion smearing and sample 
smearing. 

Synthesized FRBs were simulated with parameters drawn 
from a wide distribution of DMs, widths, scattering timescales, 
and spectral index. Fluences were drawn from a Euclidean 
distribution, in line with current observations. Pulse durations 
were drawn from a log-Normal distribution with mean 1\,ms, 
allowing us to test our completeness over a large range, for bursts between 
0.01--100\,ms. Spectral index ranged from $-$4 to +4, and 
DMs were drawn from a uniform distribution between 
50$-$3000\,pc\,cm$^{-3}$.
The bursts were injected off-line into filterbank data from the ARTS observing system.
They thus include the same sky noise, RFI, instrument variability, and other limitations as the live data.
The simulated bursts were 
run through the full pipeline:
first searched 
by AMBER, next 
following the
post-processing steps, such as 
candidate clustering and machine learning 
classification.
Injection comparison against Heimdall \citep{2012MNRAS.422..379B}
helped improve, e.g., the AMBER real-time \ac{S/N} computation method (Sect.~\ref{sec:6b.2}). 
Early on, it also helped discover significant incompleteness above 
800\,pc\,cm$^{-3}$, identifying a bug in AMBER that could rapidly be fixed.
Afterwards, the sensitivity of the real-life survey was found to 
correspond to the expected fall off in S/N with DM due to intrachannel dispersion smearing. 
We find that in the relevant search 
parameter space (Sect.~\ref{sec:10.search}), the pipeline is over 90$\%$ complete.
Selection effects and biases qualitatively become more apparent at the edges of this search space:
candidates at low DM are more easily misidentified as RFI while at very high DM they are smeared out.
Broadband FRBs could potentially be more easily detected than strongly frequency-limited bursts. 
\subsection{Computing performance}
\label{sec:9h}

\begin{figure}[t]
    \centering
    \includegraphics[width=1.0\columnwidth]{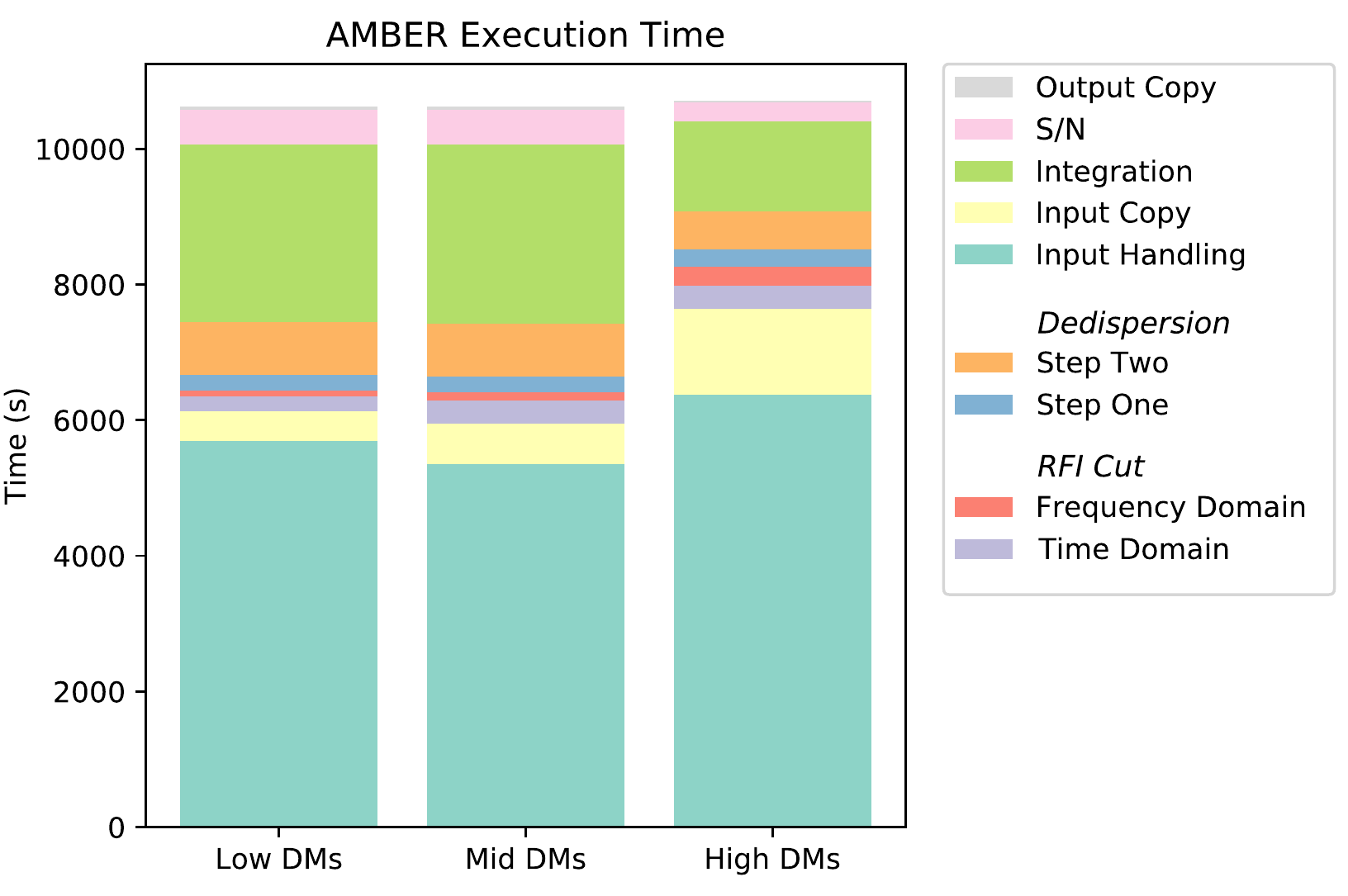}
    \caption{The break down of the functions for the three AMBER instances
    that run on each CB/compute node.
    Shown are the various total execution times per function,
    that together ran real-time on the 3\,hr observation. 
}
    \label{fig:amber_execution_time}
\end{figure}

The largest fraction of the computing requirements for ARTS is taken up by the FRB search.  On each
of the 40 nodes, 3 GPUs (Sect.~\ref{sec:5c.artsmac}) run AMBER to carry out this search.
These three
instances search the low, medium, and high DMs independently.  Fig.~\ref{fig:amber_execution_time}
shows the amount of time spent in each AMBER function (Sect.~\ref{sec:6b.2})
for a three-hour observation.
After the very effective optimisation of, first, the dedispersion
kernel \citep{2016A&C....14....1S}, and next, the \ac{S/N} kernel,
most execution time is taken up by input handling and by integration
(downsampling).

Due to the low arithmetic intensity, the dedispersion kernel and most other functions are memory bound.  Performance is
thus best understood by comparing our throughput against the memory bandwidth of the 1080 Ti of 484\,GB/s.
The high-DM instance achieves a throughput, averaged over all kernels, close to this value.
Given the delays at these high DMs, most data is used only once,
 meaning there is little opportunity for data re-use in the dedispersion kernel.
For the low and medium DM instances, however,
 such  data re-use \emph{can} be exploited, through the use of user-managed caching,
 allowing us to process twice as many  DMs as the high-DM instance, in the same time.
It is therefore by means of optimizing the memory accesses that we can process observations in real time.
  
\section{Survey and observation strategy }
\label{sec:10}

\ac{ARTS} enables an all sky survey for FRBs and pulsars to pursue
the science goals outlined in Sect.~\ref{sec:3}. 
This survey, ALERT\footnote{\url{https://alert.eu/}}, the 
\acl{ALERT}\acused{ALERT}, is described below.
Its first results are presented in the next Section.

\subsection{Frequency coverage}
\label{sec:10.freq}

During dedicated searching
ALERT uses a central frequency (cf. Sect.~\ref{sec:app:design:rf})
of 1370\,MHz to avoid the \ac{RFI} present at lower frequencies.
During commensal observing (SC3 in Table~\ref{tab:modes}),
the transient search shares
the Apertif imaging survey settings. These use a
 lower central frequency of
1280\,MHz, such that redshifted HI is observable throughout a larger cosmic volume. 
In the period covered here, 2019, commensal mode was not commissioned and all FRB observations used the 1370\,MHz band.

\subsection{Pointings and scheduling}
\label{sec:10b}

In the 2019 Apertif schedule, time domain observations were allotted a 1/3rd share of the science
time. These generally occurred as 2-week blocks every 6 weeks.
Pointing definition followed the same field  pattern used for the imaging surveys \citep{Hess2020}.
To provide a relatively uniform coverage of the Northern sky in the planned survey time, we
limit ourselves to pointings with declination $>$10{\degr};
and to maximize the number of FRB detections we avoid the increased foreground dispersion in fields
closer to the Galactic Plane than 5{\degr}.

The pointing selection was guided by our triple aims of
 detecting, localizing, and characterizing 
FRBs.
We thus target fields that
a) produced earlier \ac{ALERT} detections, such that in follow-up we can determine if or how these
repeat, or
b) contain known repeater FRBs such that these can be better localized and/or studied, or
c) are blank, to search for new FRBs.
In the known-source fields (a) and (b), new detections were expected at the same rate as in blank
fields, which indeed we found to be the case (Sect.~\ref{sec:11}). Blank fields are scheduled using
Apersched\footnote{\url{https://github.com/kmhess/apersched}} \citep{kelley_m_hess_2022_6988465}.
Each pointing lasts 3\,hr.
Generally, preference is given to fields at hour angles close to zenith to minimize terrestrial RFI,
which is more prevalent near the horizon.

\begin{table}[b]
    \centering
    \caption{The dedispersion setup for ALERT.}
    \label{tab:dedisp_plan}
    \begin{tabular}{llll}
        \toprule \toprule
        GPU    &    DM range       & DM step  & Subbanding DM step \\
               & (\ppcc)        & (\ppcc)  & (\ppcc) \\ 
        \midrule
        1 & 0--409.6       & 0.2      &  6.4 \\
        2 & 409.6--819.2   & 0.2      &  6.4 \\ 
        3 & 819.2--3379.2  & 2.5      & 40.0 \\
        \bottomrule
    \end{tabular}%
\tablefoot{The three GPUs cover subsequent parts of \ac{DM}-trial parameter space.
Each uses two-step subband dedispersion (Sect.~\ref{par:dedispersion}) with 32 subbands
and has optimal dispersion settings.}
\end{table}

\subsection{Search parameter space}
\label{sec:10.search}

The survey is optimally sensitive to FRBs characterized by certain sets of parameters.
Sources outside this space may still be detectable, but generally at reduced \ac{S/N}.
Our dedispersion plan (Table~\ref{tab:dedisp_plan})
starts at 0\,\ppcc. Sources with DM below \mbox{100\,\ppcc} (similar to e.g.,
FRB~20200120E; \citealt{2021ApJ...910L..18B}), however, may easily
be misidentified as RFI in the human vetting,
as they show little dispersion sweep over our 1.2$-$1.5\,GHz band. 
That is especially the case for wide bursts. So for these low-DM bursts, and for 
bursts above our maximum DM of $\sim$3400\,{\ppcc}, ALERT is not complete. 

AMBER searches through the incoming data at native time resolution of 81.92\,$\mu$s,
and at resolutions downsampled by factors 5, 10, 25, 50, 100 and 250.
Any bursts narrower than the original time samples, or wider than 
the last step, equaling 20\,ms, will only be detected at reduced \ac{S/N}.

\section{First results}
\label{sec:11}

\begin{table*}[bpt]
    \centering
    \caption{The ALERT FRBs up to 2019 Dec 31.}
    \label{tab:frb_overview}
    \setlength{\tabcolsep}{4pt} %
    \begin{tabular}{*{11}l}
    \toprule
    FRB      & Internal & MJD & S/N & DM      & DM$_\mathrm{MW}$  & $z_\mathrm{max}$ & Width & Fluence   &  RA      & Dec     \\
             & name &      &     & ($\pccm$) & ($\pccm$) & & (ms)    & (Jy\,ms)    &  (J2000) & (J2000) \\
    \midrule
    20190709A &190709   & 58673.21792057 & 26   &    663.1$\pm$0.1  & 52 / 45   & 0.65 & 0.49$\pm$0.05  & 7.0$\pm$1.4   &  01:39:19s & +32:03:13   \\
              &190903$^*$   & 58729.02228880 & 8    &    664$\pm$10     & 53 / 46   & 0.65 & 23$\pm$2       & 98$\pm$20 &  01:32:47 & +33:03:43   \\
    20190926B &190925   & 58752.03093855 & 15   &    957.3$\pm$0.5  & 51 / 44   & 0.97 & 2.2$\pm$0.2    & 7.6$\pm$1.5   &  01:42:06  & +30:58:05   \\
    20191020B &191020   & 58776.78080366 & 17   &    465.0$\pm$0.2  & 102 / 101 & 0.38 & 1.04$\pm$0.10  & 9.1$\pm$1.8   &  20:30:39  & +62:17:43   \\
    20191108A$^{**}$ &191108   & 58795.83082563 & 103  &    588.1$\pm$0.1  & 43 / 52   & 0.54 & 0.34$\pm$0.03  & 8.2$\pm$1.6  &  01:33:47  & +31:51:30   \\
    20191109A &191109   & 58796.54885262 & 22   &    531.2$\pm$0.1  & 108 / 108 & 0.44 & 0.48$\pm$0.05  & 4.5$\pm$0.9  &  20:35:15  & +61:49:02   \\
               \midrule
    \end{tabular}
    \tablefoot{First column lists TNS name, second lists previously used name
      \citep[cf.][]{oml+20, clo+20, oostrum20}. MJD refers to the arrival time at the solar system barycentre at
      infinite frequency. DMs and widths were measured with {\sc pdmp} from {\sc psrchive}
      \citep{hvm04}. DM$_\mathrm{MW}$ are the Milky Way DMs predicted using the NE2001 \citep{cl02} and YMW16
      \citep{ymw16} models, respectively. We assume a 10\% uncertainty on the observed pulse widths and 20\% on the
      derived fluences. The position error regions %
      are strongly elongated and presented individually. The best position is not always at the center of the
      elliptical region.
       The derivation of the redshift upper limit  $z_\mathrm{max}$ is described in Sect.~\ref{sec:disc:counterparts}.\\
    \tablefoottext{$*$}{Candidate}
    \tablefoottext{$**$}{The arrival time published in \citet{clo+20} contained an error in the barycentric
      correction.  In this table that is rectified.}
}
\end{table*}

We here present the first set of \ac{ARTS} results, based on observations up to 2019 Dec 31.
We discovered five new FRBs,
a significant addition to the $\sim$100 published at the time \citep[see][]{2016PASA...33...45P},
most of which were only very roughly localised.
The cut-off date. is aligned with the adjustment of the Apertif time-domain survey strategy
at the start of every calendar year.
Results from following years will appear in the very near future \citep[e.g.][]{Pastor-Marazuela_2022_Scattering}.
A summary of the observed and derived properties of the 2019 FRBs is given in Table~\ref{tab:frb_overview}.
It lists both the
{\tt FRB\_19MMDD} format names that we used in earlier
publications \citep{clo+20, oml+20}, and the TNS-network designations
for these bursts, in {\tt FRB\_2019MMDD[A-Z]} format \citep{2020TNSAN..70....1Y}, that we use in this paper.
In the following subsections, we discuss the characteristics and localisation of each FRB separately.
Where required we describe analysis methods  where they are used for the first time. 
The analysis of the ensemble is presented in Sect.~\ref{sec:disc}.

\subsection{FRB 20190709A}
\label{sec:frb20190709A}
\begin{figure}[b]
  \centering
  \vspace{-1ex}
  \includegraphics[width=0.98\columnwidth, trim=60 50 110 90, clip]{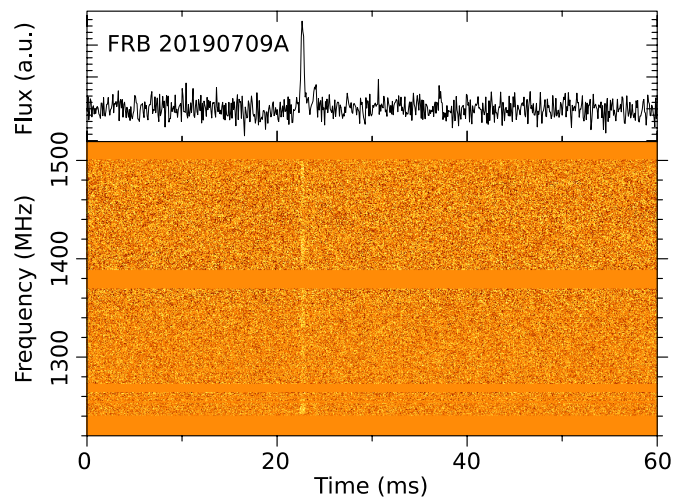}
  \caption{Dynamic spectrum (bottom) and pulse profile
    (top; flux density scale in relative, arbitrary units) of FRB 20190709A.
  }
  \label{fig:frb:20190709A}
\end{figure}
\subsubsection{Detection}

FRB~20190709A, the first ALERT FRB, was discovered
in first week of the survey. 
It was detected in a 3C48 calibration drift scan observation.
Its fluence of 7\,Jy\,ms puts it at about the 50th percentile of the known population
at the time (cf.~Fig.~\ref{fig:F_vs_z}).

The FRB fluences we report are calculated from the modified radiometer equation
\citep[e.g.,][]{cm03}
using Eq.~2 of \citet{overview_arxiv_v1},
the SEFD from the weekly calibration (per CB; Sect.~\ref{sec:9.FoV}) and 
the relative sensitivity of the \ac{SB} the FRB was found in (Sect.~\ref{sec:9.tabs}), at the best-fit position.  

At 663.1$\,\pccm$, the DM of FRB~20190709A far exceeds
 the   Galactic electron density
predicted in this direction, in either NE2001 \citep{cl02} or YMW16
      \citep{ymw16}, as listed in Table~\ref{tab:frb_overview}.
As in \citet{clo+20} we roughly relate the  $\mathrm{DM_{IGM}}$ contributions, for this and subsequent detections, to redshift  $z$  as 
\begin{equation}
  \centering
  \frac{\mathrm{DM_{IGM}}}{\pccm} \approx 930\,z.
\label{eq:dm-z}
\end{equation}
Assuming no DM contribution from the host galaxy then gives a redshift upper limit for this FRB of
$z = 0.65$.

The FRB exhibits a strong,  $\sim$400\,$\upmu$s narrow component and potentially a second equally
narrow %
component (see Fig.~\ref{fig:frb:20190709A}).
These widths are comparable to the %
intra-channel dispersion smearing, %
suggesting the  components are intrinsically  extremely narrow.
The separation between the components is about 1.3\,ms.
At $z = 0.65$
the actual separation at the emission site would
be only about 0.8\,ms. 

FRB~20190709A  was detected before IQUV triggering (Sect.~\ref{sec:6b.1}) was in production.
  In the Stokes-I data,
the auto-correlation function of the burst spectrum indicates smooth structures
with characteristic bandwidth of  order 20--30\,MHz. The Galactic
scintillation bandwidth is expected to be only a few MHz. However, effects
of any scintillation caused by the host galaxy medium cannot be ruled out.

\subsubsection{Localisation}
\begin{figure*}[t]
    \centering
    \includegraphics[width=0.9\textwidth]{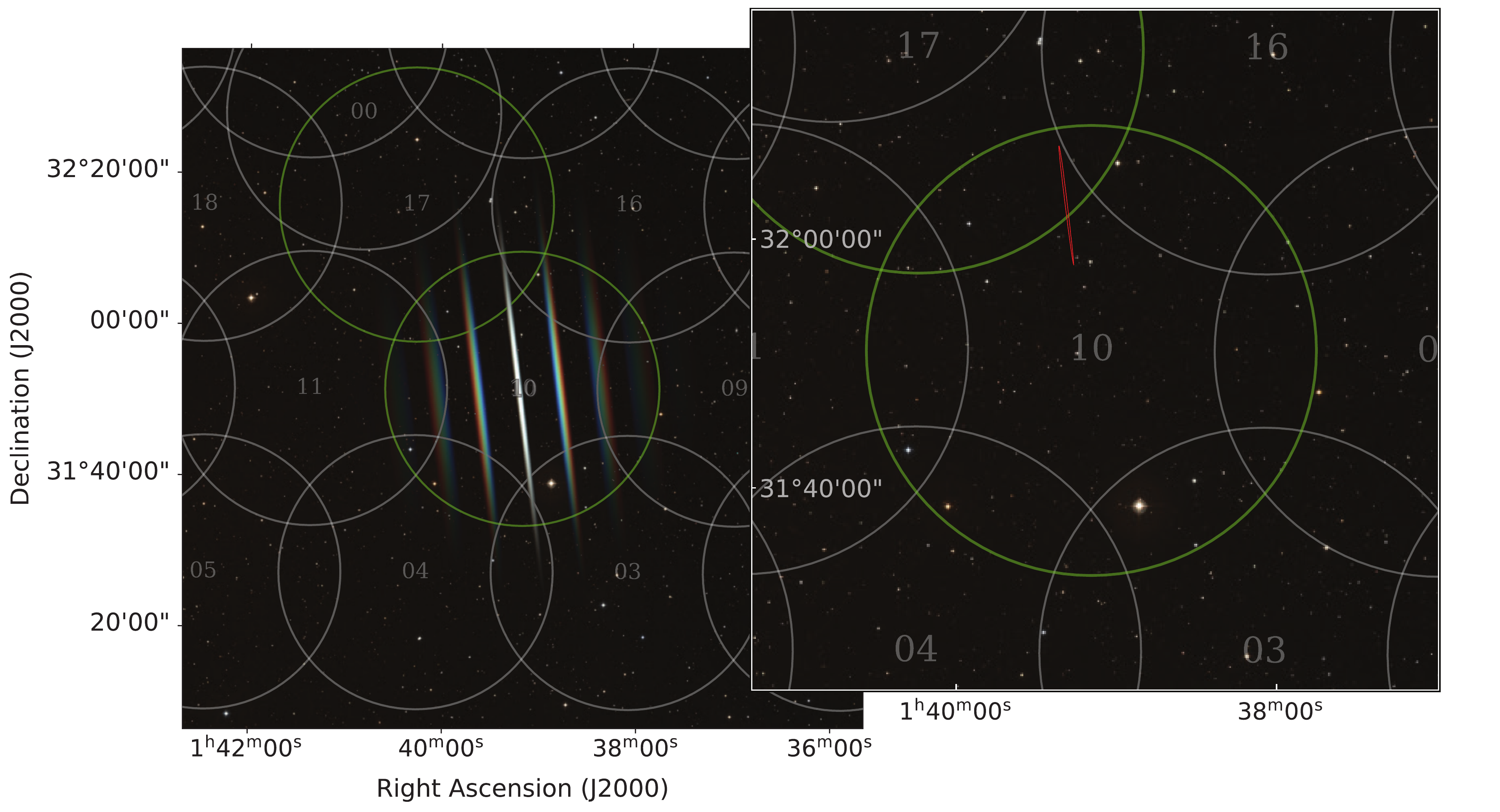}
    \caption{The localisation region of FRB~20190709A. 
      In both panels, and in the localisation region Figures that follow,
      the CBs at 1370\,MHz are shown as
    white (non-detection) and green (detection) circles. 
    In the left-hand side we show in the colour scale the simulated response of the SB in which the FRB was detected
    most strongly (cf.~Fig.~\ref{fig:sb_models}).
    On the right hand side, and in the follow Figures,
    the red, elongated and very narrow area indicates the 90\% confidence level localisation
    area that results from combining the SB detections and upper limits of all surrounding CBs. Background image here
    and in the following Figures are from the Sloan Digital Sky Survey \citep[SDSS;][]{SDSS}.}
    \label{fig:20190709A_loc}
\end{figure*}

FRB 20190709A was detected in two CBs, across a total of 23 SBs. The derived localisation region is well represented by an ellipse of
$10\arcmin \times 39\arcsec$, as shown in Fig.~\ref{fig:20190709A_loc}.
These FRB localisation regions are derived using statistical errors. %
The systematic offset that \citet{oostrum20} found for the localisations from drift scan observations
(relevant for this FRB 20190709A and for candidate FRB 190903; Sect.~\ref{sec:frb190903})
is solved here.

For all FRBs, we have searched for putative host galaxies using the GLADE galaxy catalogue \citep{dalya_glade_2018}. The catalogue was created by combining previously existing galaxy and quasar catalogues (GWGC, 2MPZ, 2MASS XSC, HyperLEDA and SDSS-DR12Q). Although its primary objective is to identify potential gravitational wave host galaxies, we used this database to look for potential FRB hosts within the error region of ARTS FRBs. The catalogue is complete up to a luminosity distance $d_L=37^{+3}_{-4}$\,Mpc and contains all of the brightest galaxies up to $d_L=91$\,Mpc. %
While most ARTS FRBs could originate from distant galaxies beyond the completeness limit of GLADE, this catalogue does allow us to check for the presence of possible nearby host galaxies which, if they are the host galaxy, could point to a large amount of DM local to the source. 

For FRB 20190709A, the GLADE catalogue contains no galaxies within the error region limits.

\subsection{Candidate FRB 190903}
\label{sec:frb190903}
\subsubsection{Detection}
\begin{figure}[h]
  \centering
  \includegraphics[width=\columnwidth, trim=60 50 100 80, clip]{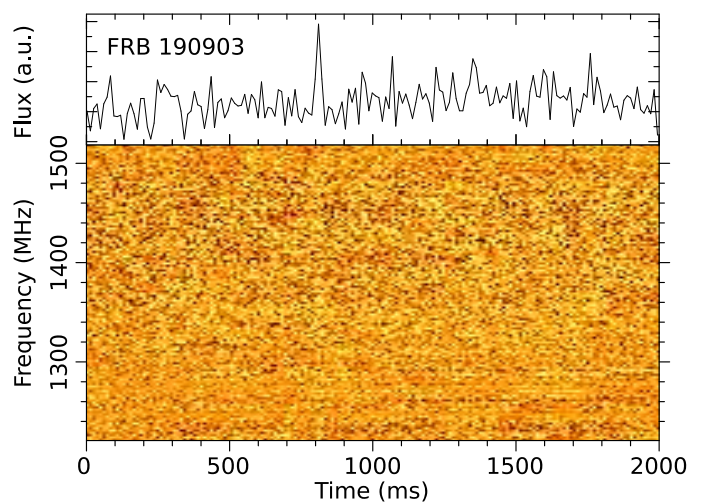}
  \caption{Dynamic spectrum (bottom) and pulse profile (top) of FRB 190903.}
  \label{fig:frb:190903}
\end{figure}

Candidate FRB~190903 (see Fig.~\ref{fig:frb:190903})
was detected just above our
detection threshold.
Given its low S/N it is of course quite possible that FRB 190903 is %
a non-astrophysical signal fluctuation.
The candidate did, however,  survive blind machine-learning and human vetting.
Only then was it realized the FRB was from the repeat follow-up field on FRB~20190709A, at the same DM.
No further repeat detections at this DM were, however, seen in
120\,hrs of follow-up of the field (see
Sect.~\ref{sec:probeM33}).
Its width of 23\,ms is an outlier compared to the generally $\sim$ms widths of the unambiguous detections (Table~\ref{tab:frb_overview}).

\begin{figure}[t]
    \centering
    \includegraphics[width=0.98\columnwidth]{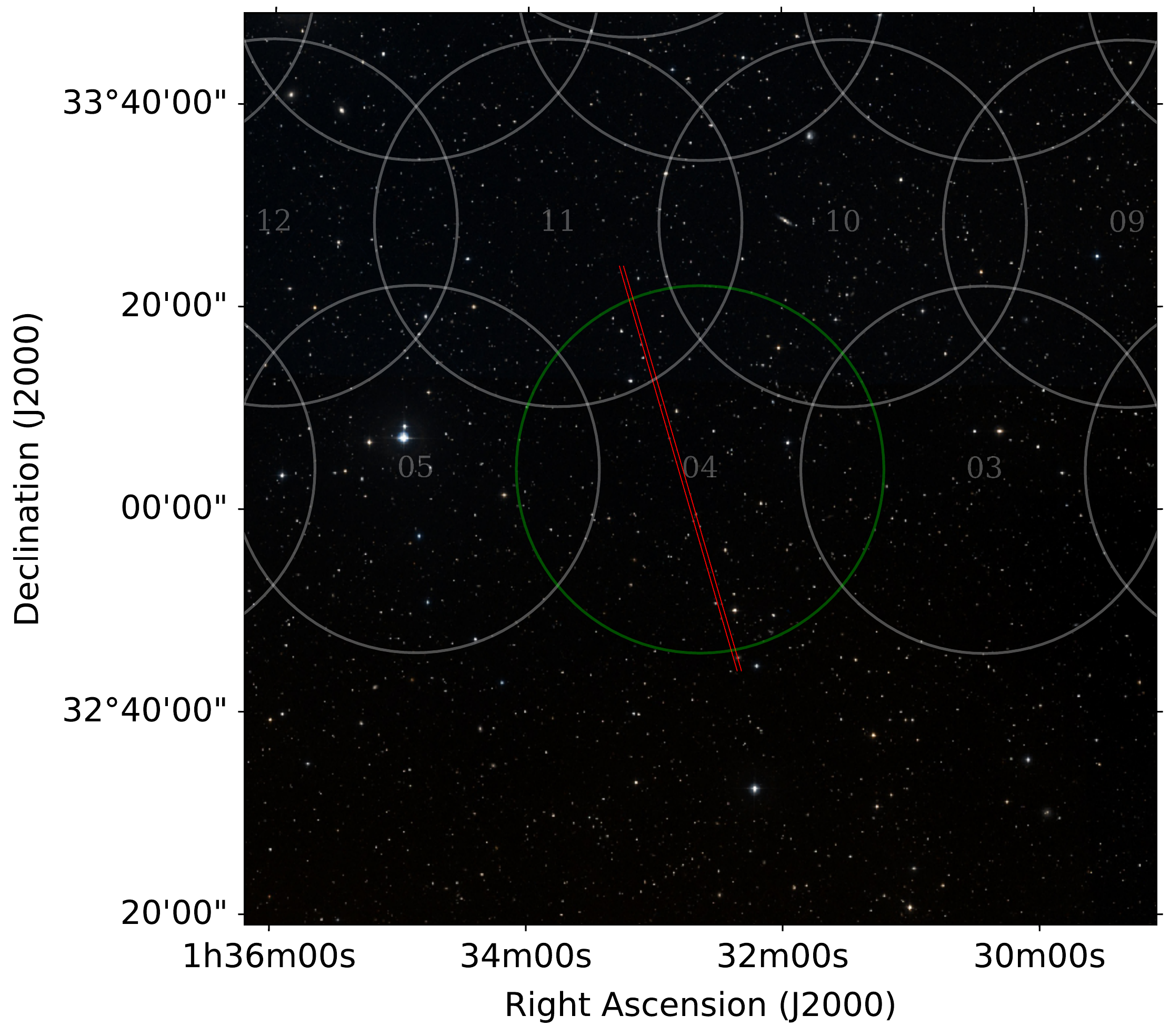}
    \caption{The localisation region of FRB~190903. The circles and red area are as described for
      Fig.~\ref{fig:20190709A_loc}. The localisation area is not constrained towards lower declinations.
      \vspace{-2ex}
    }
    \label{fig:190903_loc}
\end{figure}
\begin{figure}[b]
  \centering
  \includegraphics[width=\columnwidth, trim=60 50 110 90, clip]{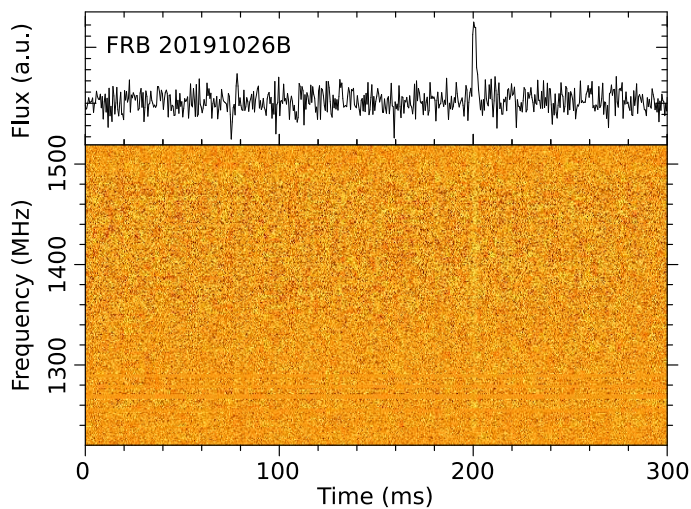}
  \caption{Dynamic spectrum (bottom) and pulse profile (top) of FRB 20190926B.}%
  \label{fig:frb:20190926B}%
\vspace{-5mm}%
\end{figure}%

\subsubsection{Localisation}
Because of its marginal detection in an outer beam, the localisation region of FRB 190903 is quite large: out to the SB
model limit of $40\arcmin$ from the CB centre, the region is roughly an ellipse of $43\arcmin \times 30\arcsec$
(Fig.~\ref{fig:190903_loc}).
However, we cannot exclude that FRB 190903 originated further away from main beam pattern. Given its similar DM and
approximate location to FRB 20190709A, we consider that FRB 190903 might in fact be a repeat pulse of FRB
20190709A. This would place FRB 190903 far outside the main beam pattern, assuming FRB 20190709A did occur within the
main beam pattern.
As a precise enough model of the CB sidelobes is not currently available, we cannot further constrain the localisation of FRB 190903 within the FRB 20190709A field and thus cannot prove whether or not these two bursts do in fact originate from the same source.
There are two galaxies within a 1$\arcmin$  distance from the localisation region in CB 04 at a redshift
$z<0.65$, one of which is located within the ellipse limits.

\subsection{FRB 20190926B}
\label{sec:frb20190926B}

\subsubsection{Detection}

\begin{figure}[t]
  \centering
  \includegraphics[width=\columnwidth]{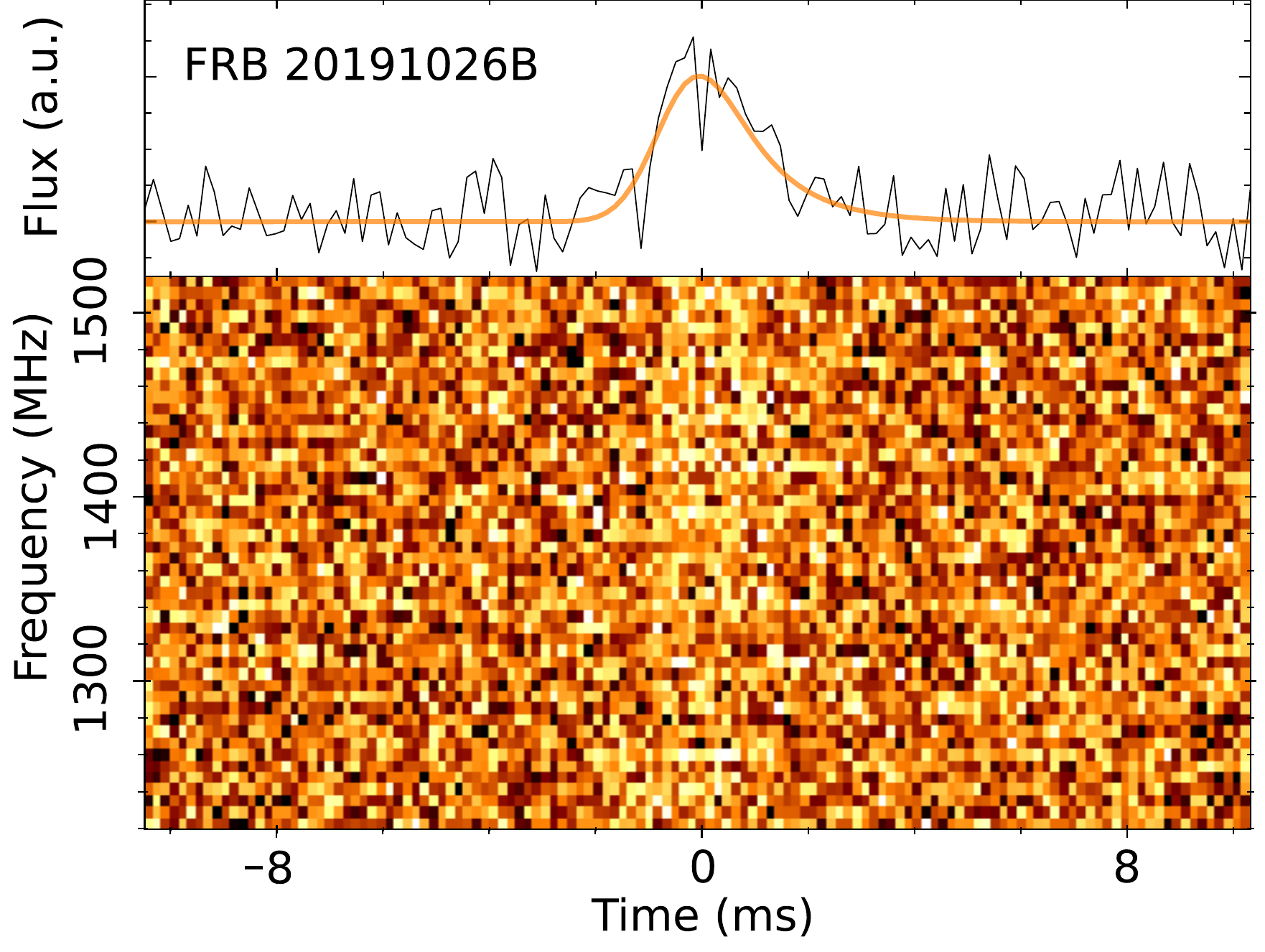}
  \caption{FRB 20190926B at 15$\times$ higher time resolution than Fig.~\ref{fig:frb:20190926B}.
  The orange line in the top panel is the best fit when including a scattering tail. }
  \label{fig:frb:20190926B_zoom}
\end{figure}

FRB~20190926B was detected with a single, 2-ms wide component at a
DM of $957.3\,\pccm$ (Fig.~\ref{fig:frb:20190926B}). As the intra-channel dispersive
smearing is %
0.6\,ms %
the burst is well resolved. The Galactic scintillation bandwidth
on this sight line is expected to be only a few MHz, but
we see no evidence of
such scintillation in this burst.

This is the only  burst in the  sample we present here that potentially shows  exponential
decay of the intensity, indicative of scattering (Fig.~\ref{fig:frb:20190926B_zoom}). 
Following the scattering characterisation described in \citet{Pastor-Marazuela_2022_Scattering}, we fit the dedispersed
pulse profile to a Gaussian function, with an optional convolution with a decreasing exponential.
The Bayesian information criterion %
values are the same with or without the  exponential tail;
we thus cannot prove its presence, or rule it out.
The best resulting scattering timescale $\tau_{\text{sc}}=1.0\pm0.3$\,ms is several orders of magnitude larger than the
expected contribution from the Milky Way. If present, it  has thus likely been produced  at the host galaxy or circumburst environment.

\subsubsection{Localisation}
\begin{figure}[t]
    \centering
    \includegraphics[width=\columnwidth]{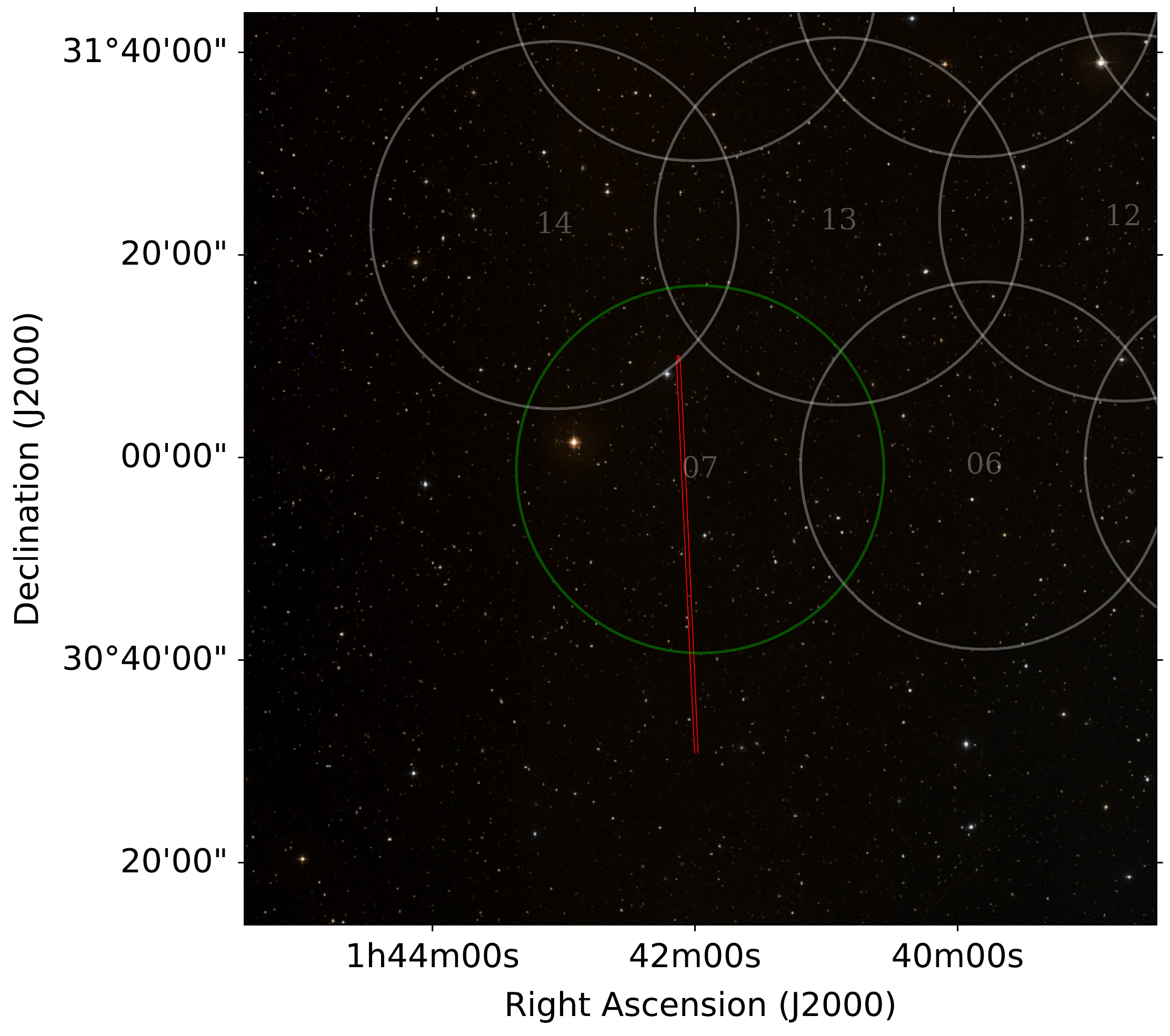}
    \caption{The localisation region of FRB~20190926B. The localisation is not constrained towards lower declinations. }
    \label{fig:20190926B_loc}
\end{figure}
FRB 20190926B was detected in one CB, across a total of 7 SBs. The derived localisation region is an ellipse of
$39\arcmin\times30\arcsec$, as shown in Fig.~\ref{fig:20190926B_loc}.
Because the FRB was found in a CB on the edge of the beam pattern, the localisation region
is not constrained towards lower declinations.
However, if the FRB was  detected in a sidelobe, it should 
intrinsically have been extremely bright.
We conclude  it more likely it occurred within the main beam of CB 07, as
indicated by the green circle in the figure. The GLADE catalogue contains one galaxy,
SDSS~01424.24+305214.4, redshift 0.68,
located at an angular distance of ${\sim}45\arcsec$ from the localisation region, within the main beam.

\subsection{FRB 20191020B}
\label{sec:frb20191020B}
\subsubsection{Detection}
\begin{figure}[h]
  \centering
  \includegraphics[width=\columnwidth, trim=60 50 110 90, clip]{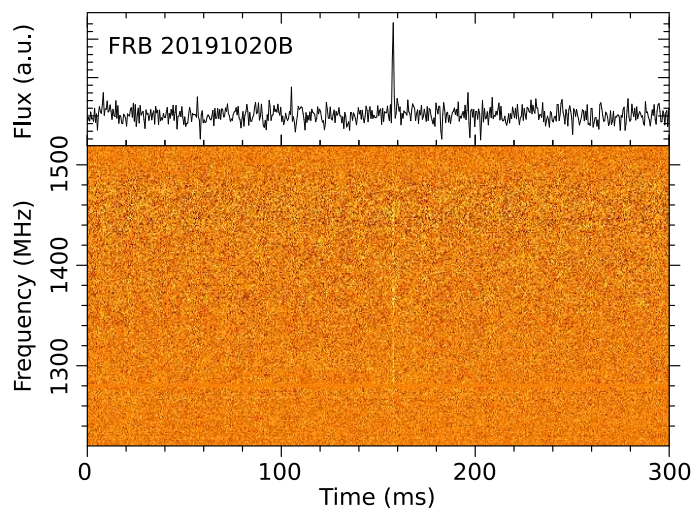}
  \caption{Dynamic spectrum (bottom) and pulse profile (top) of FRB 20191020B.}
  \label{fig:frb:20191020B}
\end{figure}

With a width of about 1\,ms, FRB~20191020B is also resolved
(Fig.~\ref{fig:frb:20191020B}). The burst is consistent with a single
Gaussian component. The burst spectrum is fairly uniform.
{This
  detection was the first for which IQUV data saving (Sect.~\ref{sec:6b.1})
  was successfully triggered in real time.
There is, however, no quasar data available in this epoch for polarization calibration, and the \ac{S/N} is
too low to perform the calibration method on the FRB data itself that was demonstrated in \citet{clo+20}. }

\begin{figure}[t]
    \centering
    \includegraphics[width=\columnwidth]{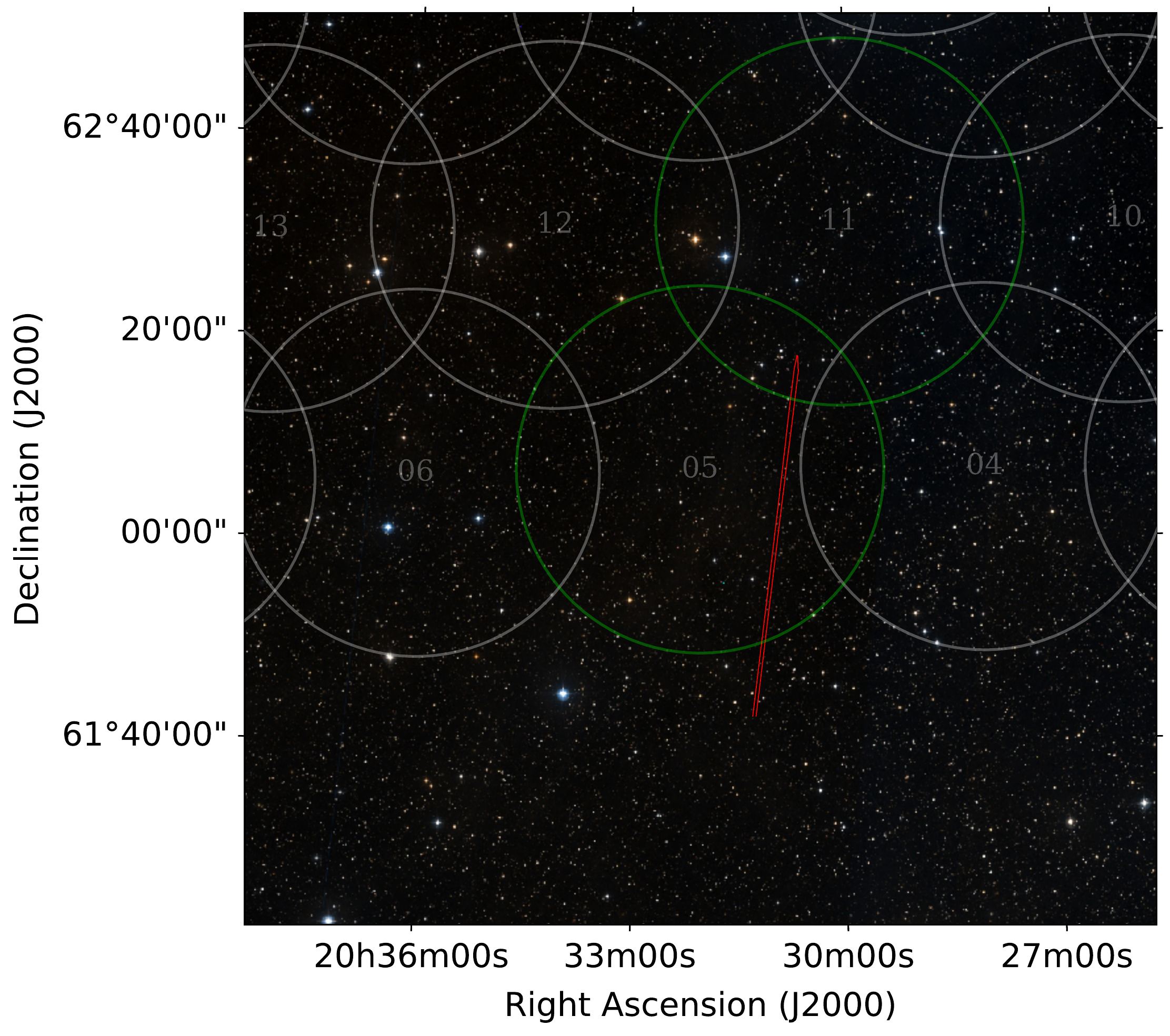}
    \caption{The localisation region of FRB~20191020B. Again, the localisation is not constrained towards lower declinations. }
    \label{fig:20191020B_loc}
\end{figure}

\subsubsection{Localisation}
FRB 20191020B was detected in two CBs, across a total of 9 SBs. The derived localisation region is an ellipse of $37\arcmin\times35\arcsec$, as shown in Fig.~\ref{fig:20191020B_loc}. Similar to FRB 20190926B, it was found in an outer CB and its localisation region is open-ended towards lower declinations. However, as for FRB 20190926B, we find it more likely that the FRB occurred in the primary beam of CB 05. There are no galaxies in the GLADE catalogue located at less than $1\arcmin$ from the localisation ellipse within CB 05.

\subsection{FRB 20191108A}
\begin{figure}[b]
  \centering
  \includegraphics[width=\columnwidth, trim=60 50 100 90, clip]{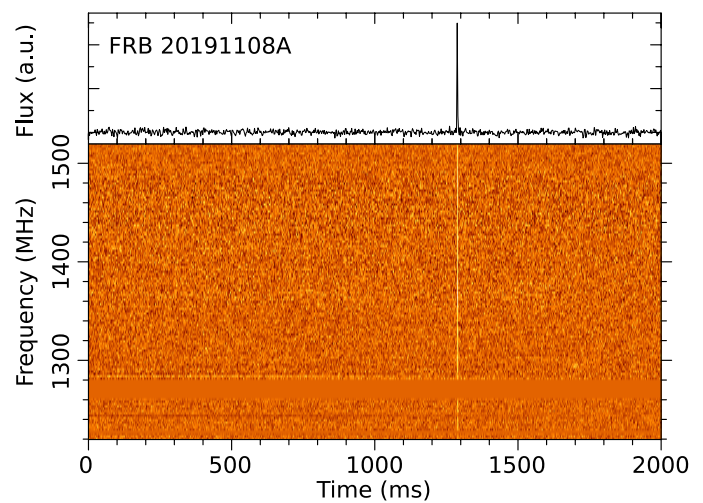}
  \caption{Dynamic spectrum (bottom) and pulse profile (top) of FRB 20191108A.}
  \label{fig:frb:20191108A}
\end{figure}
FRB~20191108A (Fig.~\ref{fig:frb:20191108A}) was detected in  48 \acp{SB} spread over three \acp{CB},
at a maximum S/N in the real-time pipeline of 60.
The discovery DM was 588\,pc\,cm$^{-3}$.
Full-stokes data was captured and calibrated.
The FRB exhibited a rotation measure %
of +474$\pm3$\,rad\,m$^{-2}$,
much higher than expected from the Milky Way foreground, and much higher than
seen in other sources close to this line of sight. 
The FRB is localised to a small, 5\,$\arcsec$ by 7\,$\arcmin$ region that is only $1.20\pm0.05\degr$ away
from the core of Local Group galaxy M33.
This discovery is described in detail in \citet{clo+20}.
The implications of  our various FRB lines of sight through the M33 halo are discussed in Sect.~\ref{sec:probeM33}.

\subsection{FRB 20191109A}
\subsubsection{Detection}
\begin{figure}[t]
  \centering
  \includegraphics[width=\columnwidth]{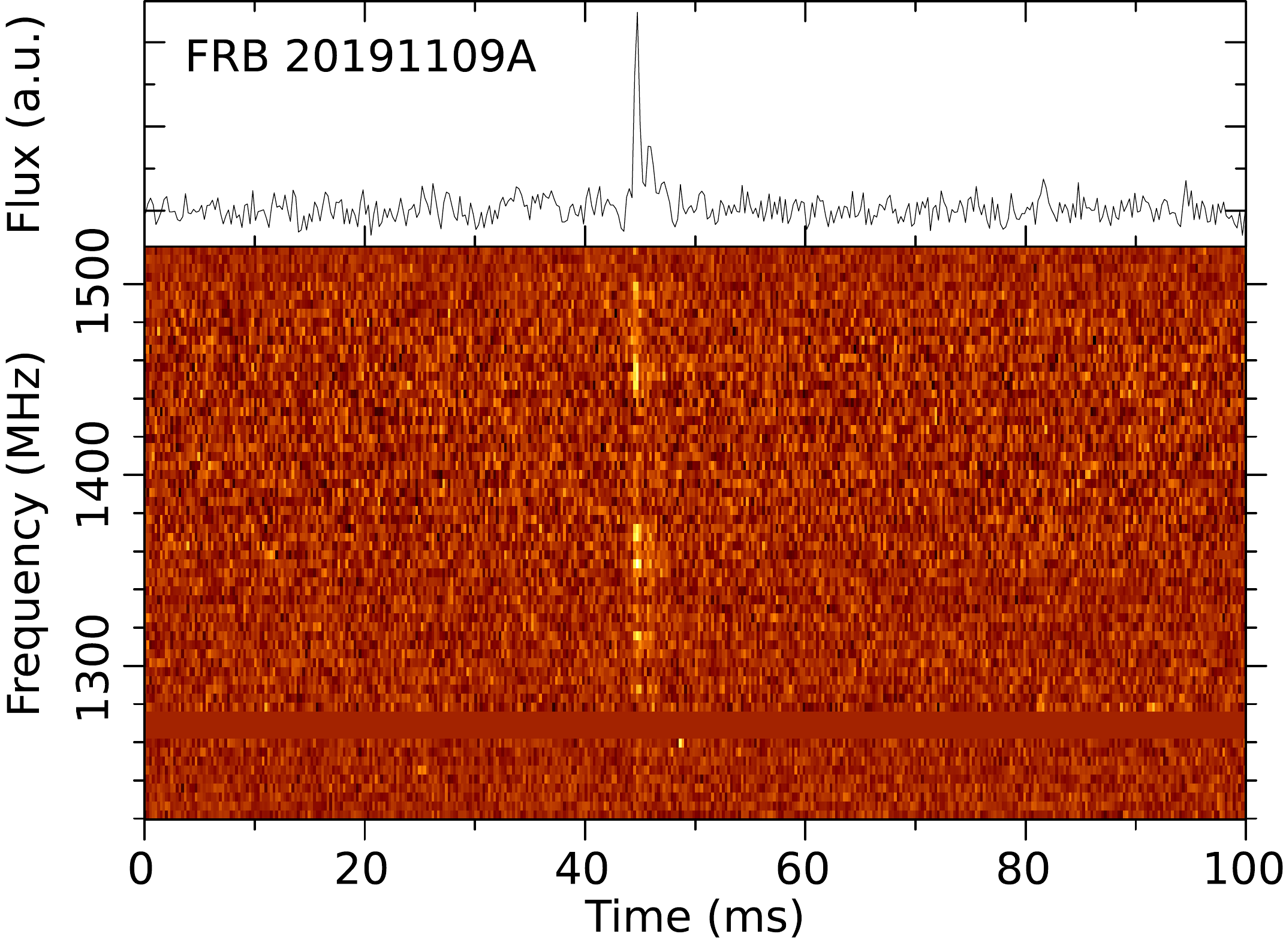}
  \caption{Dynamic spectrum (bottom) and pulse profile (top) of FRB 20191109A.}
  \label{fig:frb:20191109A}
\end{figure}
\begin{figure}[b]
    \centering
    \includegraphics[width=\columnwidth]{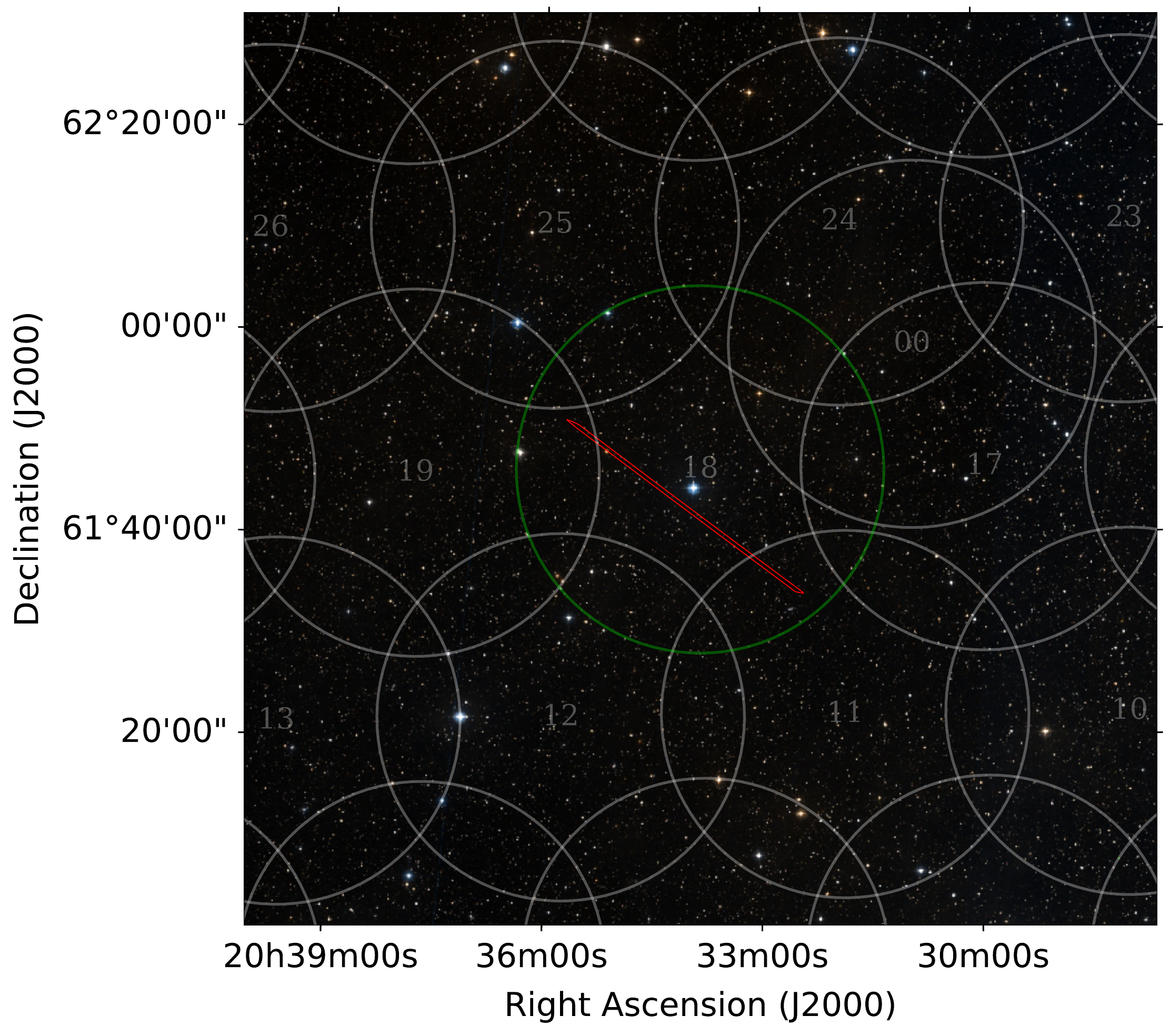}
    \caption{The localisation region of FRB~20191109A. The CBs at 1370\,MHz are shown in
    white (non-detection) and green (detection). The red, elongated and very narrow area indicates the 90\% confidence level localisation area. }
    \label{fig:20191109A_loc}
\end{figure}

FRB~20191109A exhibits multiple components --- 2 discernible bright components
and perhaps a faint third trailing component (Fig.~\ref{fig:frb:20191109A}).
The widths of the individual components are around 0.6\,ms, and at a DM of
531\,\ppcc  these are resolved over the intrachannel smearing.
The first two components are  separated by 1.2\,ms.
A one-off FRB model by
\citet{FR14}, which involves collapse of a supermassive star into a black
hole, predicts a leading precursor, a main burst and ringdown within 1\,ms
or more. While there is no sign of a precursor, the temporal structure of
FRB~20191109A might be reminiscent of the above predicted ringdown.
This multi-component profile, with its $\sim$1-ms burst separation, could also be the FRB equivalent of quasi-periodic
microstructure seen in pulsars.
In \citet{Pastor-Marazuela_2022_arXiv} we present additional FRBs with such structure,
and provide a more detailed discussion.

FRB~20191109A also exhibits frequency structures with bandwidths of the
order of 10$-$12\,MHz. The structures are significantly correlated between the first two components,
which means they are either intrinsic or are caused by scintillation affecting a broadband intrinsic pulse.
The NE2001 and YMW16  electron density models 
both predict Galactic scintillation
bandwidths to be much less than 1\,MHz. Hence, the observed frequency
structures are likely either intrinsic to the emission mechanism, or caused by the medium in the host galaxy.

Several repeating FRBs are known to exhibit frequency structures which
drift downwards with time. By cross-correlating the spectra at
the peaks of the first two components, we find an offset of
about $0.6\pm0.6$\,MHz between the frequency structures under the
corresponding components, consistent with zero.
For a non-zero offset, the best-fit value would imply a subburst
drift to lower frequencies with time at a rate of $\sim$0.5\,\mhzpms,
i.e., much lower than that observed for the repeating FRB~121102
\citep[$\sim$200\,\mhzpms;][]{Hessels19}.

\subsubsection{Localisation}
FRB 20191109A was detected in one CB, across a total of 15 SBs. The derived localisation region is an ellipse of $29\arcmin \times 5\arcsec$, as shown in Fig.~\ref{fig:20191109A_loc}. GLADE does not contain any galaxy located at less than $1\arcmin$ from the localisation region.

\section{Discussion}
\label{sec:disc}

\subsection{Characteristics of the discovered sample}

\begin{figure}[t]
  \centering
  \includegraphics[width=\columnwidth]{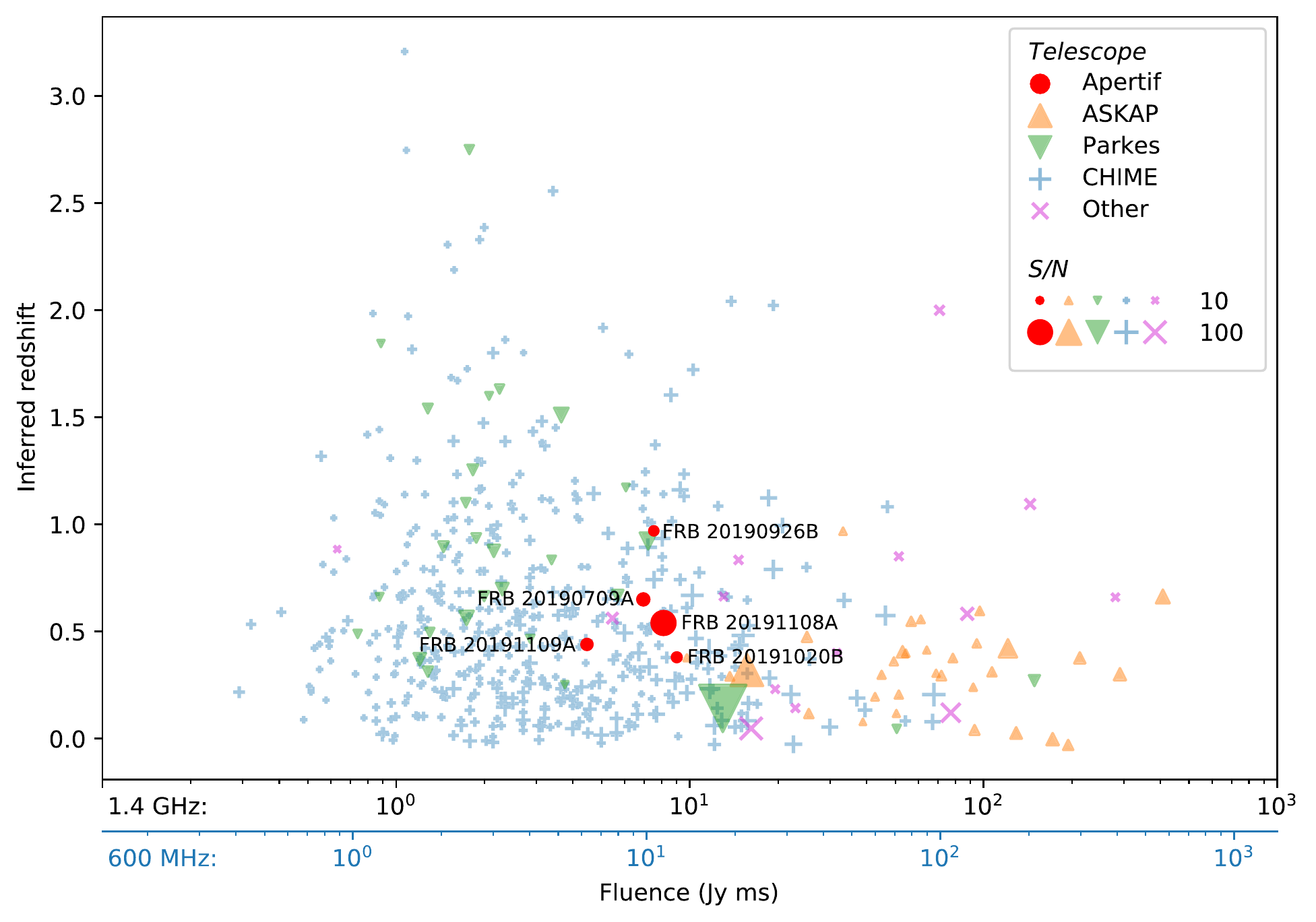}
  \caption{The redshift and fluence for the  FRBs presented here compared against those  in the
    TNS 
    also detected in     or before 2019. 
    In this semi-log plot, a spectral index of $-0.4$
    is used to
    compare the CHIME bursts (blue abscissa) with the other bursts (black abscissa). 
    The Apertif FRBs are positioned around the mid point of the 1.4\,GHz FRB fluence distribution.
  } %
  \label{fig:F_vs_z}
\end{figure}

The fluences of the Apertif-discovered bursts are around the median of the known fluence distribution
(Fig.~\ref{fig:F_vs_z}).
When comparing the 1.4\,GHz surveys,
Parkes detected bursts that were generally fainter, as its SEFD is lower \citep[30\,K;][]{mlc+01}.
In contrast, only brighter FRBs are generally detected by ASKAP, as that system searches
data from incoherently added dishes \citep{bannister19}.
As a result, following the relationship between the fluence limits
and the volume up to the equivalent redshift limit,
the inferred redshifts of our bursts are also higher than the ASKAP bursts, and lower than the Parkes
bursts.

Given the high time and frequency resolution of ARTS (detailed below), even our distant bursts
suffer less from dispersion-smearing related selection effects than the CHIME bursts (Fig.~\ref{fig:F_vs_z}).
Assuming a spectral index of $-0.4$, the best-fit value for the FRB population per
\citet{2021A&A...651A..63G}, to compare the fluences, the CHIME bursts are generally dimmer,
yet lower redshift.
The Apertif FRB sample is free of selection effects out to larger distances than CHIME
\citep[][]{Yuyang2022}.

\begin{figure}[b]
  \centering
  \includegraphics[width=\columnwidth]{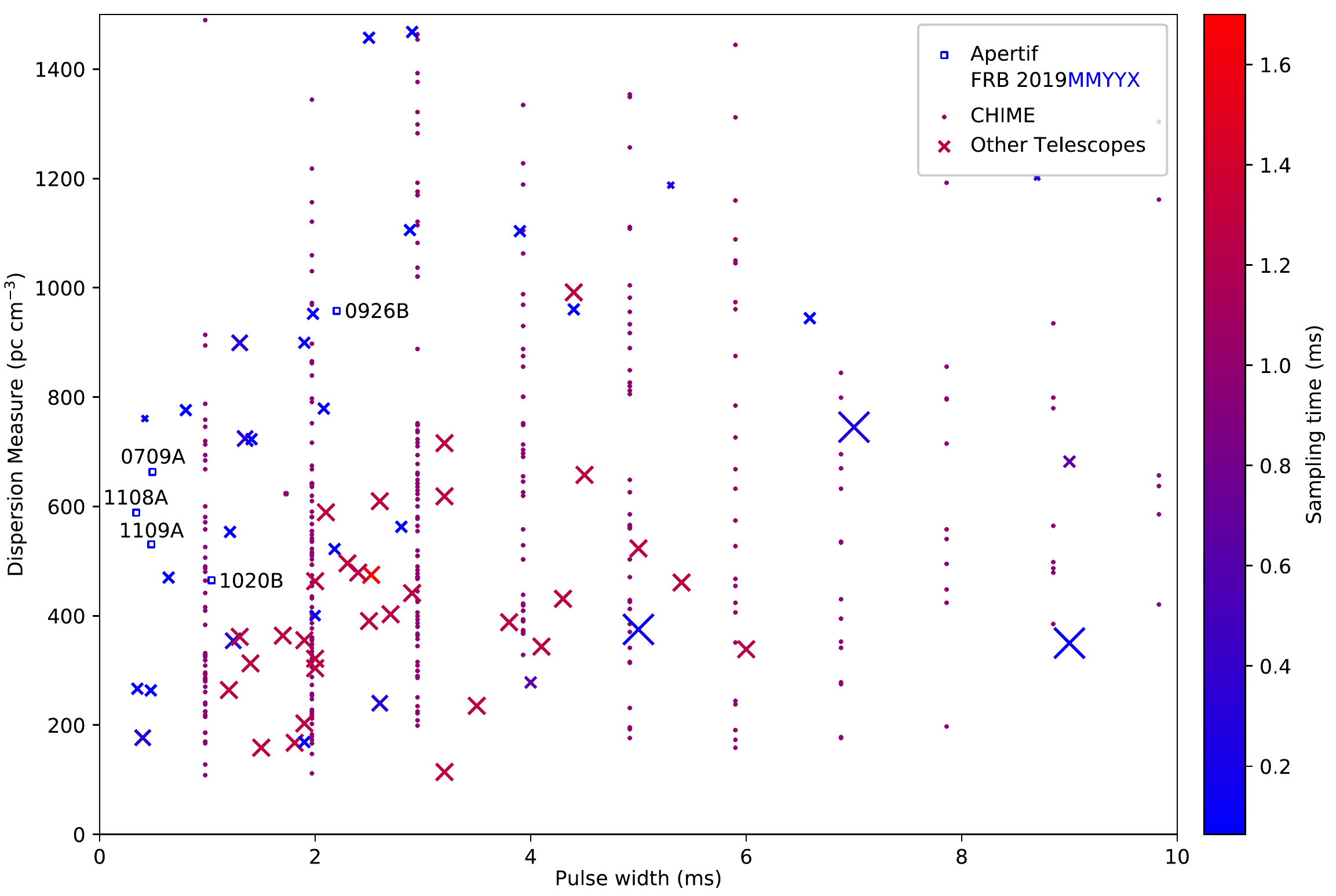}
  \caption{The pulse width and dispersion measure of the 5 FRBs presented in this work,
    contrasted against the same
    bursts as in Fig.~\ref{fig:F_vs_z}.
    The Apertif FRBs are among the narrowest known, and have high dispersion measure.
    Markers are sized and colored to reflect the frequency and time resolution, respectively.
    For Apertif, both are very high.
  }
  \label{fig:w_vs_DM}
\end{figure}

The stand-out characteristic of the Apertif bursts is
that they are detected at significantly narrower widths
than the FRB population that was known before.
Fig.~\ref{fig:w_vs_DM} shows that 3 of the Apertif FRBs are among the 6 narrowest overall.
The reason for this unique trait is two-fold.
The observing system has very high native time resolution,
and very good frequency resolution, such that instrument smearing
is low.
For comparison, this time and frequency resolution are 
10$\times$ and 5$\times$ higher, respectively,
than for the ASKAP incoherent-sum mode.
This allows us to determine the second, more fundamentally intrinsic characteristic:
the bursts display very little temporal scattering. 
This means we can discern individual components even if they are only a millisecond apart,
as for FRB 20191109A (and as analysed for Apertif FRB 20201020A in \citealt{Pastor-Marazuela_2022_arXiv}).

Notable, too, are the generally large dispersion measures.
There, the good frequency resolution limits the intra-channel dispersion smearing, in turn allowing short bursts to
reach high signal-to-noise ratios, and be detected;
whilst these same bursts might be washed out and missed by other telescopes.
That brings a larger cosmic volume of bursts within our S/N reach, increasing detections at high DMs.

The 5 bursts reported here were the first discovered with the system,
and IQUV polarization recording and calibration was only successful for
FRB~20191108A. For bursts discovered in 2020 and after, the availability of full-Stokes data is much higher
\citep{Pastor-Marazuela_2022_Scattering}.

Finally, all 5 bursts are broadband, covering the entire 300\,MHz Apertif band,
even if that coverage is modulated by frequency structure.
This spectral modulation and band coverage fraction are qualitatively similar to the burst sample
presented for ASKAP \citep{shannon_dispersionbrightness_2018}.
Of the four FRB archetypes proposed in \citet{2021ApJ...923....1P}, only two classes are represented in our sample.
Morphologically, FRBs~20190926B, 20191020B and 20191108A display straight, Gaussian
profiles in time, without much hint of a scattering tail, that are broadband (class I).
FRBs~20190709A and 20191109A are spectrally similar but temporarily  more complex (class
III).
We found no bursts that are narrowband (class II) or downward drifting (class IV). 
In principle the broad band nature of all bursts is
encouraging for simultaneous detections of the same bursts with Apertif and \mbox{LOFAR}
(Sect.~\ref{sec:3c.3})
although the separation of their observing frequencies is admittedly much larger than
the  bandwidth coverage achieved here.

\subsection{Survey detection rate and localisation}
\begin{figure}[t]
  \centering
  \includegraphics[width=\columnwidth]{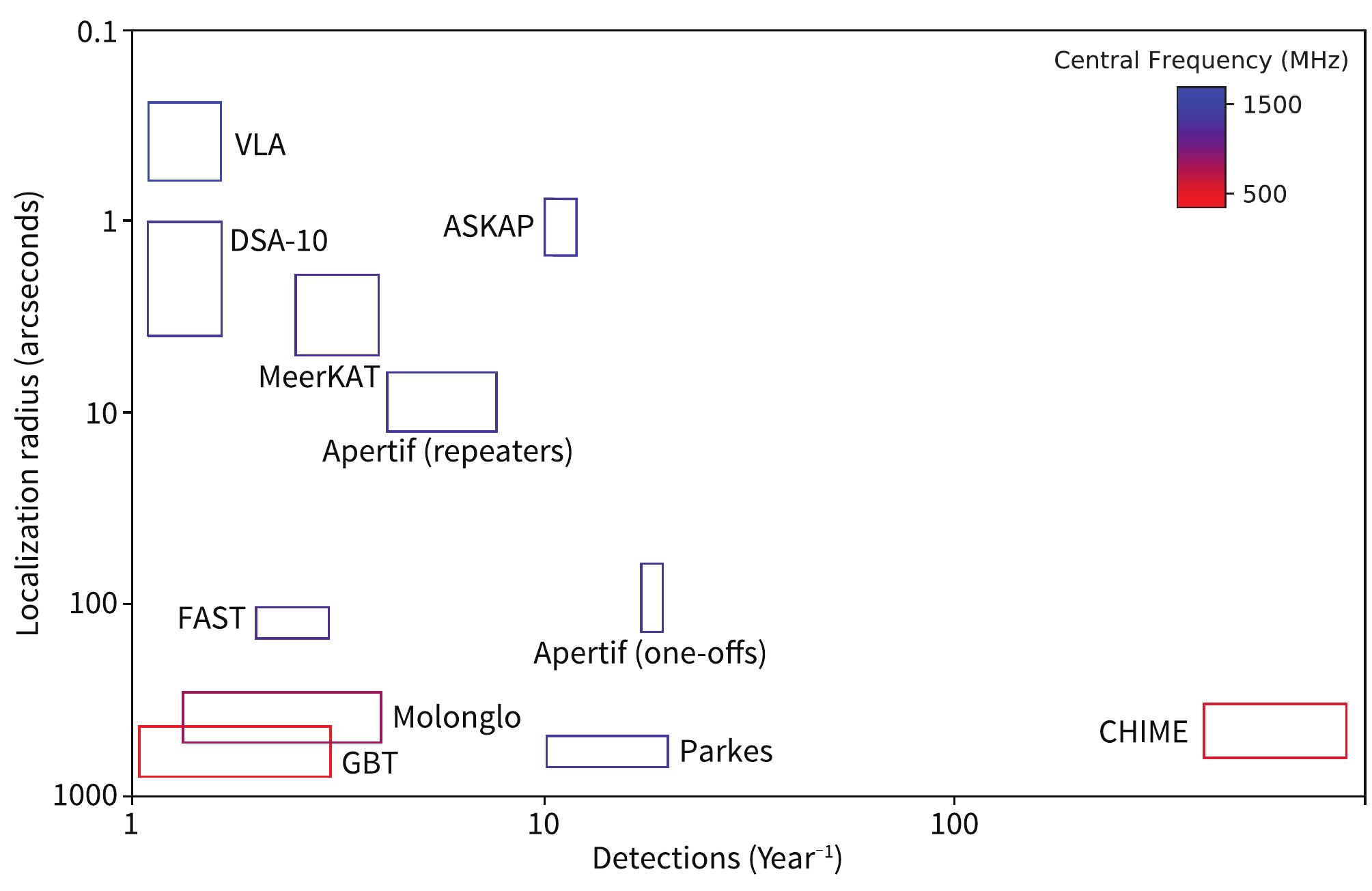}%
  \vspace{-1ex}%
  \caption{Rate per calendar year versus general localisation properties for the main FRB surveys to date.
    Shown is localisation radius (or equivalent for elongated regions) versus the rate per year given the observing
    time. 
    Apertif localisation numbers are
    provided for both single-burst detections (``one-offs'') , and for repeat dections at different hour angles
    (``repeaters''), 
    see Sect.~\ref{sec:9.loc}.
    Data for other surveys from \cite{sbs2018}, \citet{2019A&A...632A.125G} and \citet{2021ApJS..257...59C}.%
  }%
  \label{fig:SurveyComp}%
\vspace{-1ex}%
\end{figure}

Our  2019 results show the total number of FRB detections per calendar year is high for ALERT at Apertif.
Fig.~\ref{fig:SurveyComp} shows this large advance over
(earlier) FRB surveys as carried out with GBT, VLA, and Molonglo. 
The high rate is a results of  three main factors:
the large time ALERT covered the sky, of about 1/3rd of the available
observing hours;
the large field of view enabled by the \acp{PAF};
and the high sensitivity from the full-field coherent beamforming. 

As Apertif also offers 1-D interferometric FRB localisation,
ALERT can localise bursts significantly better than single-dish surveys do.
Fig.~\ref{fig:SurveyComp} shows the localisation radius
(or its equivalent in case localisation regions are elongated)
for the survey comparison. For Apertif we consider both the one-off case
($\sim$$100\arcsec$, Sect.~\ref{sec:11}), 
and the case where multiple bursts are detected
($\sim$$10\arcsec$, Sect.~\ref{sec:9.loc}).
For the rate of the latter, we assume a detected population comprised of Apertif discoveries that repeat,
plus localisations from other, known, repeater FRBs.
Overall, ASKAP, CHIME and Apertif each have their own tradeoff between rate
and localisation accuracy, with Apertif providing both.

\subsection{Detection of new FRBs versus field type}

The ALERT pointing strategy (Sect.~\ref{sec:10b}) contains three types of fields.
The priority order is for fields that contain a) earlier
\ac{ALERT} detections, b) known repeater FRBs, or c) are blank.
We find the approach works well in allowing both follow-up study and new detections.
The repeat visits to detection fields (a) and the new detection in there,
have allowed us to
chart multiple lines of sight within a field,  as demonstrated in Sect.~\ref{sec:probeM33}. 
Fields with other known FRBs (b) led to results detailed separately in e.g. \citet{2020arXiv201208348P} for FRB~20180916B.

\subsection{Prospects for counterpart identification}
\label{sec:disc:counterparts}
While for some FRBs we have already identified one or more galaxies in their localisation regions,
there are probably more,  too faint to be included in the  GLADE  catalogue.
Here we estimate the total number of galaxies we might expect in the localisation regions of our FRBs.

The number of potential host galaxies in a localisation region depends strongly on which types of galaxies are
considered to possibly host FRB progenitors: Dwarf galaxies are far more common than massive galaxies. As the first
repeating source, FRB 121102, was localised to a dwarf galaxy with a high specific \acl{SFR}
\citep[][]{Tendulkar2017}, it was thought that this type of galaxy might be related to the FRB progenitor
type. However, other FRBs --- both repeating and non-repeating --- have now been localised to a variety of galaxies
\citep{Chatterjee2017,bannister19,ravi2019,prochaska-2019b,2020Natur.577..190M}.
We here apply the same analysis as done
for FRB 110124 in \citet{2019MNRAS.482.3109P} and \citet{oostrum20}, and estimate the number of dwarf galaxies
($4\times10^7\,\Msun < M_\mathrm{stellar} < 10^{10}\,\Msun$, i.e.
between the mass of the host galaxy of FRB 121102 and the maximum dwarf-galaxy mass) and massive galaxies
($M_\mathrm{stellar} > 10^{11}\,\Msun$) in the FRB localisation volumes.
From the stellar mass function of \citet{Baldry2012},
and from converting the \citet{Haynes2011} H{\sc i} mass function at ratios between 1$-$10, we estimate a dwarf galaxy number density of $n = (0.02 - 0.06)\,\mathrm{Mpc^{-3}}$. For massive galaxies we use the luminosity function of \citet{faber2007}, and find $n = (1.5 - 2.0)\times10^{-3}\,\mathrm{Mpc^{-3}}$. 

The expected number of galaxies in an FRB localisation region is the galaxy number density multiplied by the comoving volume out to the redshift of the FRB, assuming that the mass functions do not evolve significantly up to the maximum redshift ($z \approx 1$) of our FRB sample. The redshift is estimated from the \ac{IGM} DM contribution (DM$_\mathrm{IGM}$) using Eq.~\ref{eq:dm-z}. DM$_\mathrm{IGM}$ can be related to other sources of DM as 
\begin{equation}
\label{eq:dms}
\mathrm{DM_{IGM}} = \mathrm{DM} - \mathrm{DM_{MW}} - \mathrm{DM_{halo}} - \frac{\mathrm{DM_{host}}}{1+z},
\end{equation}
where DM$_\mathrm{MW}$, DM$_\mathrm{halo}$, and DM$_\mathrm{host}$ are the DM contributions from the Milky Way, its
halo, and the host galaxy (which includes the environment local to the source), respectively. For the Milky Way
contribution, we take the lowest value predicted by the NE2001 and YMW16 models (see
Table~\ref{tab:frb_overview}). Based on the \citet{yamasaki-2019-halodm} model, we conservatively assume 10\,$\pccm$
from the Milky Way halo, and we set the host galaxy contribution to zero. The resulting redshift estimates thus are conservative upper limits.

\begin{figure}
    \centering
    \includegraphics[width=\columnwidth]{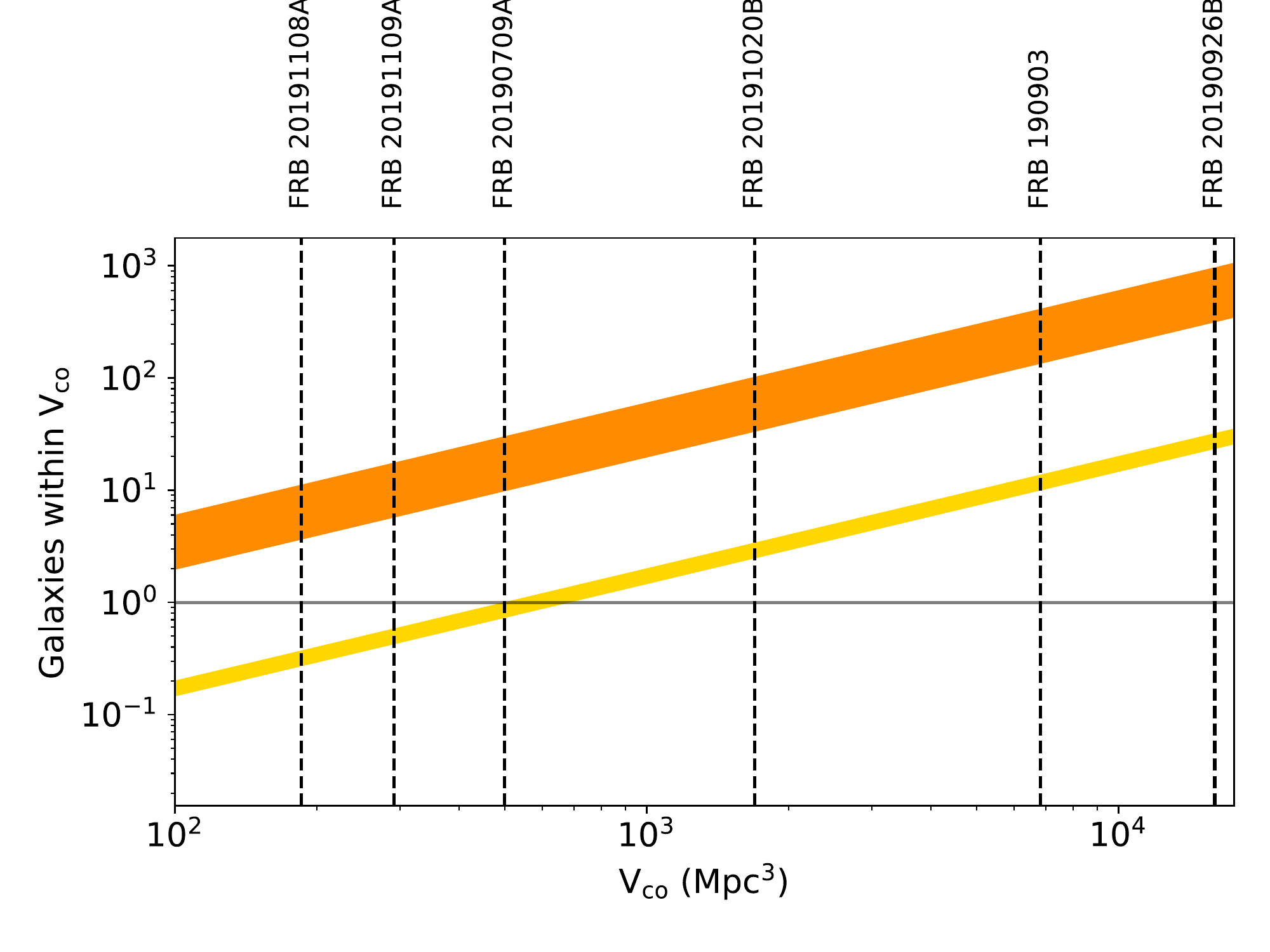}
    \caption{The expected number of galaxies in the FRB localisation areas for a range of dwarf galaxy (orange) and massive galaxy (yellow) number densities as function of comoving volume. The comoving volume depends on both the redshift of the FRB and the size of the localisation region on-sky, and should be regarded as an upper limit. The horizontal line indicates where the expected number of galaxies is one.}
    \label{fig:hostgalaxies}
\end{figure}

In Fig.~\ref{fig:hostgalaxies} we show the resulting number of expected galaxies as a function of
the comoving volume spanned by the FRB localisation region and 
the redshift limit. %
The number of expected dwarf galaxies is $\ge1$ for all our FRBs, up to several hundred for FRB 20190926B.
If a localisation region contains a known dwarf galaxy, one thus cannot straightforwardly
conclude  it is the host. %
In contrast, the number of expected massive galaxies is less than one for three of our FRBs.
For the FRB with the smallest
comoving volume associated with its localisation region, FRB 20191108A, the expected number
is ${\sim}0.3$. Assuming Poissonian error bars, the probability of finding a massive galaxy in the region by chance
is then $27\%$.
It is therefore not possible to definitively associate a massive galaxy with any of our FRBs.

In order to rule out spatial coincidence at the 95\% level, we find that the comoving volume should be limited to
${\sim}30\,\mathrm{Mpc^3}$ for association with a massive galaxy, and ${\sim 2}\,\mathrm{Mpc^3}$ for association with a
dwarf galaxy. Such volumes could be reached by either better FRB localisation, or by finding an FRB at a relatively
small distance.
For example, an FRB with the same localisation region as our best-localised burst, FRB 20191108A, but at
redshift of $z=0.28$, as opposed to the current upper limit of $z=0.54$ (see Table~\ref{tab:frb_overview}), would
be
localised to the required comoving volume to rule out spatial coincidence with a massive galaxy.
For a dwarf galaxy
association, the redshift upper limit is $z = 0.11$.
Such redshifts are not unreasonable within the overall FRB
population; about a quarter of FRBs localised to
hosts %
are within that
redshift range (we note, though, that localisable FRBs will generally be relatively close by, which biased the quoted fraction).
In \citet{Pastor-Marazuela_2022_Scattering}, we will present such bursts.

Another class of potentially interesting FRB counterparts
are the persistent radio sources (PRSs). These were discovered for
the first repeater FRB 121102 \citep{Chatterjee2017,Marcote17},
and for repeating FRB~20190520B \citep{2021arXiv211007418N}.
As also discussed in \citet{clo+20}, radio point sources are sparser than optical galaxies, hence ARTS localisation regions might be small enough to identify radio sources associated with our FRBs. 

For each FRB, we set a lower limit to the flux density of a persistent radio source that is ruled out to be in the
localisation region by chance at the 10\% level, following \citet{2018ApJ...860...73E}. The resulting flux densities are
listed in Table~\ref{tab:persistent_source_limits}. Given the $5\sigma$ sensitivity limit for the Apertif imaging
 of $350\,\mathrm{\upmu Jy}$ as used in \citet{clo+20}, we note that Apertif imaging could identify persistent radio sources that
 are unlikely to be in the localisation region by chance for the majority of our FRB sample.
 In the first Apertif imaging data release \citep{Adams2022} the sensitivity limit is even lower.
 Apertif thus not only has the
 capability to discover FRBs; it can identify the  potentially   associated persistent radio sources as well.

\begin{table}
    \centering
    \caption{Flux densities above which no persistent radio sources are expected to be found in the FRB localisation
      region by chance at the 10\% level, for a $5\sigma$ sensitivity limit of $350\,\mathrm{\upmu Jy}$.}
    \label{tab:persistent_source_limits}
    \begin{tabular}{*2l}
        \toprule
        FRB & S   \\
               & (mJy) \\
        \midrule
         20190709A & 0.21 \\
         190903 & 5.4 \\
         20190926B & 5.3 \\
         20191020B & 5.4 \\
         20191108A & 0.14 \\
         20191109A & 0.34 \\
         \bottomrule
    \end{tabular}
\end{table}

\subsection{Probing the M33 halo}
\label{sec:probeM33}
\begin{figure}[tb]
    \centering
    \includegraphics[width=\columnwidth]{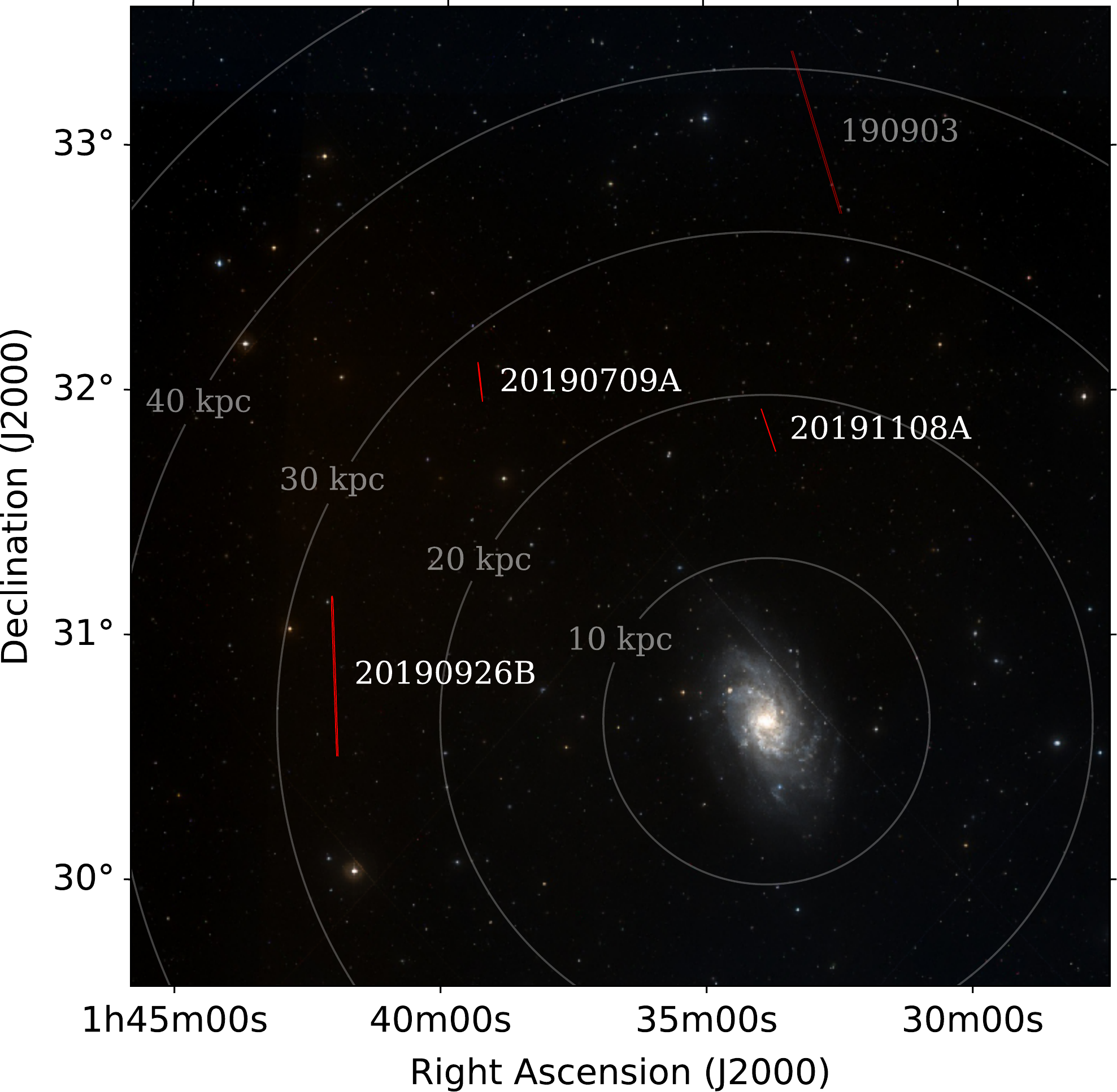}
    \caption{The location of the three FRBs and one FRB candidate that cut within 50\,kpc of M33.
    }
    \label{fig:M33_multi}
\end{figure}

As shown in Fig.~\ref{fig:M33_multi}, three of our first four detections and one unverified FRB candidate were found in
the angular vicinity of Local Group galaxy M33 (the Triangulum galaxy).
This is because our first detection,
FRB~20190709A, was discovered during a calibration drift scan observation of the quasar 3C48,
which is in the Triangulum constellation.
FRB~20190926B and FRB~20191108A were later detected during follow up observations of our first discovery.
The dispersion and scattering of this set of FRBs might help investigate the M33 halo.
FRB~20191108A has the lowest angular separation from the core of M33 and has the smallest localisation region,
but all three sources cut well through the M33 halo.
They next also intersect the halo of the much larger, M31 (Andromeda) galaxy that is close to M33;
and finally, the halo and disk of our own Milky Way.
Interactions between these three Local Group galaxies produce connecting gas bridges
that the FRBs also skewer.
The hot gas bridge that \citet{2021ApJ...907...14Q} identified  between the Milky Way and M31,
  for example, is in the FRB line of sight.
Combining the $\sim$15$\degr$ angular distance between the FRBs and M31
with the bridge model of  \citet{2021ApJ...907...14Q}
suggests the baryon bridge  disperses the bursts  by an additional
$\sim$40$-$200\,{\ppcc}. 
All the FRB DMs thus ought to have  components
that are attributable to the plasma of M33 and M31,  and to the Local Group bridges; 
and that amount can be no larger than the minimum extragalactic DM of the three FRBs.
As the lowest extragalactic DM of the three is $\sim$540\,pc\,cm$^{-3}$,  %
 the electron column density in the \acl{CGM} of M33 and M31 must be less than $\sim$500$-$340\,{\ppcc}.
A larger sample of FRBs from Apertif and other surveys may better
establish this bridge floor, or even a spatial DM gradient, in the direction of M31 in the future.
In that sense it is unfortunate that dedicated surveys with LOFAR did not find further FRBs in either M31
\citep{2020A&A...634A...3V} or M33 \citep{2016A&A...593A..21M}.
None of these three Apertif bursts show any evidence of temporal scattering.
As described in \citet{clo+20}, FRB~20191108A shows some frequency structure, but the broad, $\sim$\,40\,MHz fluctuations are likely not due to propagation effects in the M33  halo.

\subsection{All-sky burst rate}
\label{sec:burst_rate}
With ALERT we have discovered five FRBs in the 800 hours observed during 2019:
one FRB every ${\sim}6.9$
days.
To convert this to an all-sky rate,
we use the FoV derived in Sect.~\ref{sec:9.FoV}, of 8.2\,sq.\,deg. 
Using a Poissonian 90\% confidence interval \citep{geh86}, the inferred all-sky rate is $700^{+800}_{-400}$\,bursts\,sky$^{-1}$\,day$^{-1}$.
We derive the fluence completeness threshold for this rate using
the system SEFD. %
At the CB centres the SEFD is best, typically 85\,Jy (Sect.~\ref{sec:9b}).
The most conservative completeness threshold would be based on 2 $\times$ this SEFD,
if we consider our FoV out to half-power.
Most FRBs, however, will be found in the most sensitive part of the FoV. %
In 85\% of the FoV, the sensitivity is at least 70\% of the maximum value. %
We thus take that 70\% %
as the realistic sensitivity threshold. Using the radiometer equation
we then find a fluence completeness threshold of $1.6 \sqrt{{w}/{\mathrm{ms}}}$\,Jy\,ms for pulses of width ${w}$. 
The derived burst rate is in agreement with earlier values at 1400\,MHz from surveys with similar fluence thresholds \citep{Champion2016,Bhandari2018,2016MNRAS.455.2207R}.

\section{Data and Code Releases}
\label{sec:12}

\ac{ARTS} data are distributed as real-time alerts, and as a time-series archive.
The \ac{ARTS} codebase is publicly accessible.

\subsection{Real time: VOEvents}
\label{sec:12b}

\ac{ARTS} sends out real-time FRB detection alerts using
the International Virtual Observatory Alliance (IVOA) event
messaging format, VOEvent.
Detections that meet a pre-defined threshold for viability are communicated using the FRB-specific
VOEvent class, developed and standardized  by \citet{Petroff_2017_arXiv}.
Events are generated locally on the ARTS cluster and broadcast via the ASTRON
network. An internal link between ARTS and LOFAR allows for rapid triggering of the LOFAR system \citep[cf.][]{2020arXiv201208348P}; currently this system is limited to triggers on previously discovered FRBs where the approximate DM and expected signal-to-noise ratio are known.
ARTS VOEvent alerts for new detections are distributed publicly through a COMET broker \citep{2014A&C.....7...12S}.

\subsection{Archive: \acs{ALTA} and the \acs{VO}}
\label{sec:12c}

The public access to Apertif archived data
is organized through data releases in the \acf{ALTA}
and in the \acf{VO}.
These contain products for imaging \citep[e.g.,][]{Adams2022} and for time domain.
The \ac{ALTA} web interface\footnote{\url{https://alta.astron.nl/}, includes manual}
provides querying, including an Aladin Lite based sky view, and download capabilities.
ASTRONs VO service\footnote{\url{https://vo.astron.nl}} is published in the VO Registry \citep{demleitner2015},
making it accessible by VO applications by default. The data sets are accessible 
using the Table Access Protocol (TAP; \citealp{dowler2019}).

The time-domain data are in PSRFITS format, 
with 1-bit sampling. The data are total intensity (Stokes I).
For each pointing, all \acp{TAB} are available. 
Data from before 2020 May 1
supply 384 channels of 0.8\,MHz each, with a  time resolution instead of 2.048\,ms.
Later data are archived using higher temporal and spectral resolution,
and provide 768 channels of 0.4\,MHz each, every 0.8\,ms.

\begin{figure*}
    \centering
    \includegraphics[width=\textwidth]{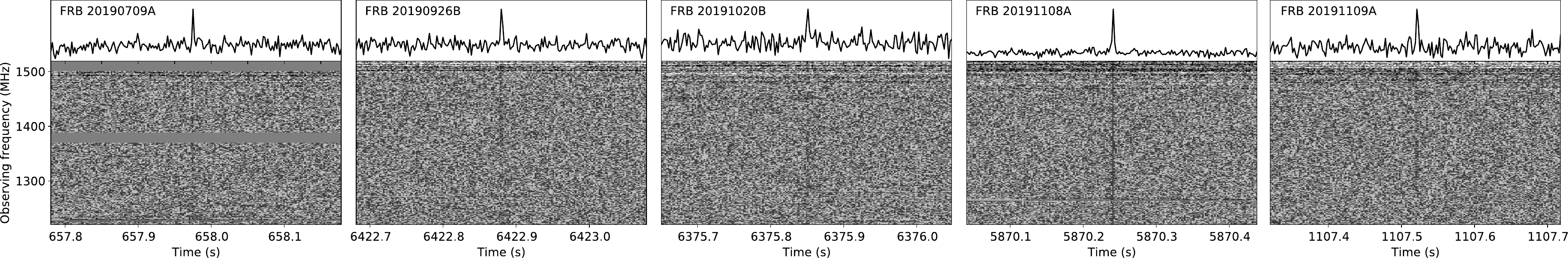}
    \caption{Detection of the five 2019 FRBs in the archival FITS data. Top: pulse profile, bottom: dynamic spectrum.}
    \label{fig:frbs_alta}
\end{figure*}

\subsection{Data releases}
\label{sec:12d}

Generally, data are released in batches per calendar year, together with an associated interpretation paper.
Specific additional releases  accompanied 
\citet{oml+20,clo+20} and \citet{2020arXiv201208348P}.

Furthermore, the data specifically recorded during the Apertif Science Verification Campaign
(SVC; cf.~Sect.~\ref{sec:8.perytons}) are in a central
data release on \ac{ALTA}.
These data include all \acp{TAB} for 47 survey field from that campaign,
and are publicly available\footnote{\url{https://alta.astron.nl/science/dataproducts/release__release_id=SVC_2019_TimeDomain}}.
They are in a similar \mbox{PSRFITS} format as the final data (Sect.~\ref{sec:12c}), but 
recorded at a central frequency of 1400\,MHz.

Together with this paper covering our first, 2019 discoveries, we have released the full set of 2019
survey data through
ALTA\footnote{\url{https://alta.astron.nl/science/dataproducts/release__release_id=APERTIF_DR1_TimeDomain}}
and the \ac{VO}.
The data releases for subsequent calendar years will follow on much shorter timescales.

\label{sec:12c.frb}
The five FRBs presented in the current paper (Sect.~\ref{sec:11})
were discovered in raw, high-resolution data,
but are also visible in this archived data.
The resulting detection plots are displayed as Fig.~\ref{fig:frbs_alta}.
All archive data for these five pointings are directly accessible through the VO\footnote{\url{https://science.astron.nl/sdc/astron-data-explorer/data-releases/apertif-time-domain-dr1/}}.
For other pointings the metadata is directly available, but the data itself first needs to be staged from tape by the
ASTRON helpdesk.

\subsection{Code releases}
\label{sec:12e}

The codes that are used to operate ARTS and ALERT are open source.
They are free for re-use and open to contributions through %
the TRAnsients Software ALliance  (TRASAL\footnote{\url{https://github.com/TRASAL}}).
The realtime search system code is released as
AMBER \citep{2016A&C....14....1S, sclocco_amber_2020} and RFIm \citep{sclocco2020}.
The accompanying real-time data handling is released as \citet{dadafilterbank,dadafits}.
The pipeline orchestration code suite is freely available
\citep{darc,arts-tools,arts-tracking-beams,arts-localisation}.
Finally, the machine-learning post processing is released as \citet{singlepulseml}.

Custom code used to generate published Apertif FRB results are released through Zenodo, 
for both the current paper\footnote{\url{https://doi.org/10.5281/zenodo.6415175}}
\citep{vanleeuwen_artsrdm_2022}
and for previous publications \citep[e.g.][]{pastor_alertr3_2020}.

\section{Future}
\label{sec:13}

\subsection{Expected detection rates}
\label{sec:13b}
Detection rate simulations help  measure
the operational performance of the system,
help set survey strategies,  and even help select the science priorities. 
To these ends, we use the FRB population synthesis code base \texttt{frbpoppy} \citep{2019A&A...632A.125G} to derive the expected detection rate for ALERT.
We adopt the `complex' intrinsic Euclidean FRB population, for an ALERT survey
with an SEFD of $\sim100$~Jy and the beam sensitivity pattern as presented in Sect.~\ref{sec:9}.
Based on these inputs, and scaling with respect to the HTRU detection rates,
 ALERT should detect 1 FRB every  $4.75^{+3.33}_{-2.10}$ days (where the quoted  margin is the  1$\sigma$
Poissonian error).
The current detection rate found in Sec.~\ref{sec:burst_rate} falls within this range, implying the system to be functioning as expected.

\subsection{Operations time scales}
\label{sec:13c}

The Apertif surveys are operated from 1 July 2019 until 28 Feb 2022.
Per that end date, the ALERT survey for pulsars
and FRBs ceases.
Down-sampled data from the pointings that were covered will continue to be available through the archive.

\subsection{Possible upgrade to 2-D, baseband recording system}
\label{sec:13d}
{
The current system is optimized for FRB detection and localisation with a 1-D array.
As the \ac{TABF} already coherently forms beams at  array angular resolution, over the full \ac{FoV},
the benefit of a complex-voltage buffer and dump system for offline analysis \citep[such as in ASKAP;][]{bannister19} is limited.

One upgrade under consideration is to extend the WSRT to a 2-D array, 
and add  such a baseband buffer system.
ARTS would detect FRBs over RT2-RT9/B in real time, as now.
Real-time freeze plus off-line analysis of complex-voltage buffer data from the new and existing dishes
then provides 2-D localisation.
A new 3\,km North-South arm, of similar length as the existing 12-dish East-West arm, 
would provide localisations of $\sim$1\arcsec\,$\times$\,1\arcsec, a good match for optical follow-up.
Given the \ac{S/N} produced by the highly sensitive E-W arm, the N-S arm could be more sparsely populated, using
smaller dishes to match the existing \ac{FoV}.

Each central \ac{UNB} of the \ac{TABF} (Sect.~\ref{sec:4d}) can hold up to 4\,TB of DDR3 RAM,
enough to store 10\,sec of CB480 voltage data (Table~\ref{tab:app:datarates}).
Such a circular buffer, covering the E-W arm, is sufficiently long for our real-time triggering.
The required complex-voltage buffers for the new N-S arm could also be implemented on  \ac{UNB}; the absence of
beam-forming requirements, however, may also allow for a simpler hardware solution.
After a trigger, the voltages are combined
in software correlation and imaging, providing improved FRB localisation.  
}

\section{Conclusions }
\label{sec:14}
The Apertif Radio Transient System achieves full coherent-addition sensitivity over the entire
field of the view of Apertif.
ARTS powers the ALERT survey, and we detect 1 FRB every 7 days of observing.
The fact that the interferometer consists of steerable dishes, allows for follow-up
campaigns on known and newly detected FRBs.
While none of the discoveries reported here were seen to repeat,
other known repeater FRBs were detected and studied.
All the while, new FRBs were found in these fields at the same rate as for blank fields.
Five one-off FRBs were discovered in 2019, during the first 6 months of overall operation,
and each was interferometrically localized.
This combination of solid detection rates and good localisation
is essential for mapping the magneto-ionic material in the Universe
along different, well-defined lines of sight.

\begin{acknowledgements}
  We thank 
  Richard Blaauw, 
Raymond van den Brink, 
  Chris Broekema, 
  Lute van de Bult, 
John Bunton, 
Adam Deller,
  Paul Demorest, 
Roy de Goei, 
Peter Gruppen, 
Jason Hessels, 
Gemma Janssen,
Gert Kruithof, 
  Hans van der Marel, 
Rob van Nieuwpoort, 
Jan David Mol, 
Gijs Molenaar,
  Kaustubh Rajwade,
  John Romein, 
Gijs Schoonderbeek, 
  Mike Sipior, 
Jurjen Sluman,
  Laura Spitler,
Ben Stappers, 
  Marc Verheijen, 
  Nico Vermaas,
  and
  Ren{\'e} Vermeulen
  for contributions to the realisation and operation of Apertif, ARTS and ALERT,
  and for comments and discussions.
We thank 
Vlad Kondratiev, Joe Callingham, %
Sarrvesh Sridhar,  Emma Tiggelaar,
and Matthijs van der Wiel for suggesting, providing and operating the microwave oven used in Sect.~\ref{sec:8.perytons}.
This research was supported by 
the European Research Council (ERC) under the European Union's Seventh Framework Programme
(FP/2007-2013)/ERC Grant Agreement No. 617199 (`ALERT'); 
by Vici research programme `ARGO' with project number
639.043.815, financed by the Dutch Research Council (NWO);
by the Netherlands eScience Center under the project `AA-ALERT' (027.015.G09, grant ASDI.15.406);
by the
Netherlands Research School for Astronomy, NOVA, under `NOVA-NW3', and `NOVA5-NW3-10.3.5.14';
and through  CORTEX (NWA.1160.18.316), under the research programme NWA-ORC, financed by NWO.
Instrumentation development was supported 
by NWO (grant 614.061.613 `ARTS') and
NOVA (`NOVA4-ARTS'). PI of aforementioned grants is JvL.
EP acknowledges funding from an NWO Veni Fellowship.
The contributions of SMS were supported by NASA grant NNX17AL74G issued through the NNH16ZDA001N Astrophysics Data
Analysis Program (ADAP).
EAKA is supported by the WISE research programme, which is financed by NWO.
BA acknowledges funding from the German Science Foundation DFG, within the Collaborative Research Center SFB1491
''Cosmic Interacting Matters - From Source to Signal''.
KMH acknowledges financial support from the State Agency for Research of the Spanish Ministry of Science, Innovation and
Universities through the "Center of Excellence Severo Ochoa" awarded to the Instituto de Astrof{\'i}sica de Andaluc{\'i}a
(SEV-2017-0709), from the coordination of the participation in SKA-SPAIN, funded by the Ministry of Science and
Innovation (MCIN).
KMH and TvdH acknowledge funding from the ERC under the European Union's Seventh Framework Programme
(FP/2007-2013)/ERC Grant Agreement No. 291531 ('HIStoryNU').
RM acknowledges support from the same programme under ERC Advanced Grant RADIOLIFE-320745.
This work makes use of data from the Apertif system installed at the Westerbork Synthesis Radio Telescope owned by ASTRON. ASTRON, the Netherlands Institute for Radio Astronomy, is an institute of NWO.

\end{acknowledgements}

\bibliographystyle{Bib/yahapj}
\begin{appendix}

\section{Detailed Design I: Functions and Data}
\label{sec:app:design:i}

Three appendices with more detail on this \ac{ARTS} design follow:
(\ref{sec:app:design:i}) covers function and data;
(\ref{sec:app:design:hw}) describes hardware; and
(\ref{sec:app:design:fwsw}) presents software and firmware.

\begin{figure}[h]
  \centering
  \includegraphics[width=\columnwidth]{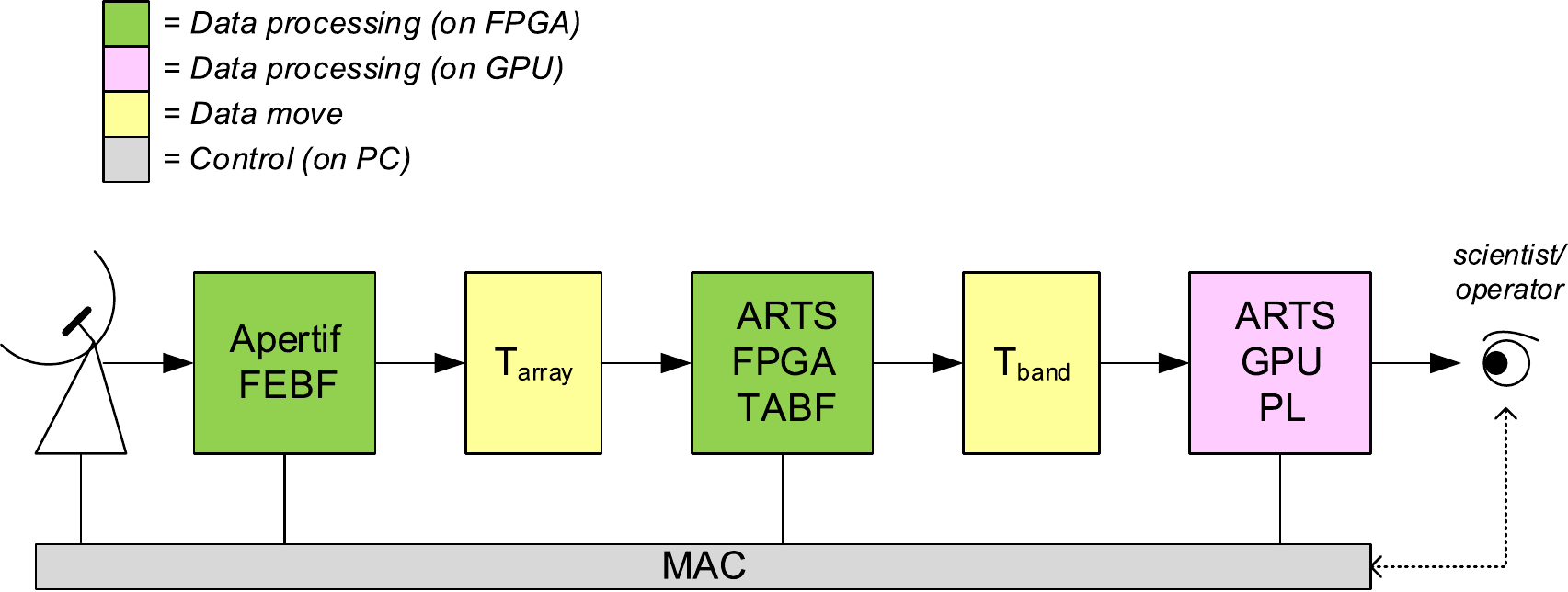}
  \caption{ARTS top level data path, with main interfaces.
    A user controls ARTS via the \ac{MAC} and accesses the data products via the ARTS \acf{PL}.
    \label{td-app-toplevel-v01}
    }
\end{figure}

We move through the system following the end-to-end data path shown in Fig.~\ref{td-app-toplevel-v01}.
Functions on this path either \emph{process} data (e.g. fringe stopping, filterbank, and beamformer), or
 \emph{move} data (e.g., reordering, selection, packetizing, and physical I/O).

\subsection{Relevant general subsystem: \acl{FEBF}}
\label{sec:app:design:febf}

\begin{figure}[b]
  \centering
  \includegraphics[width=\columnwidth]{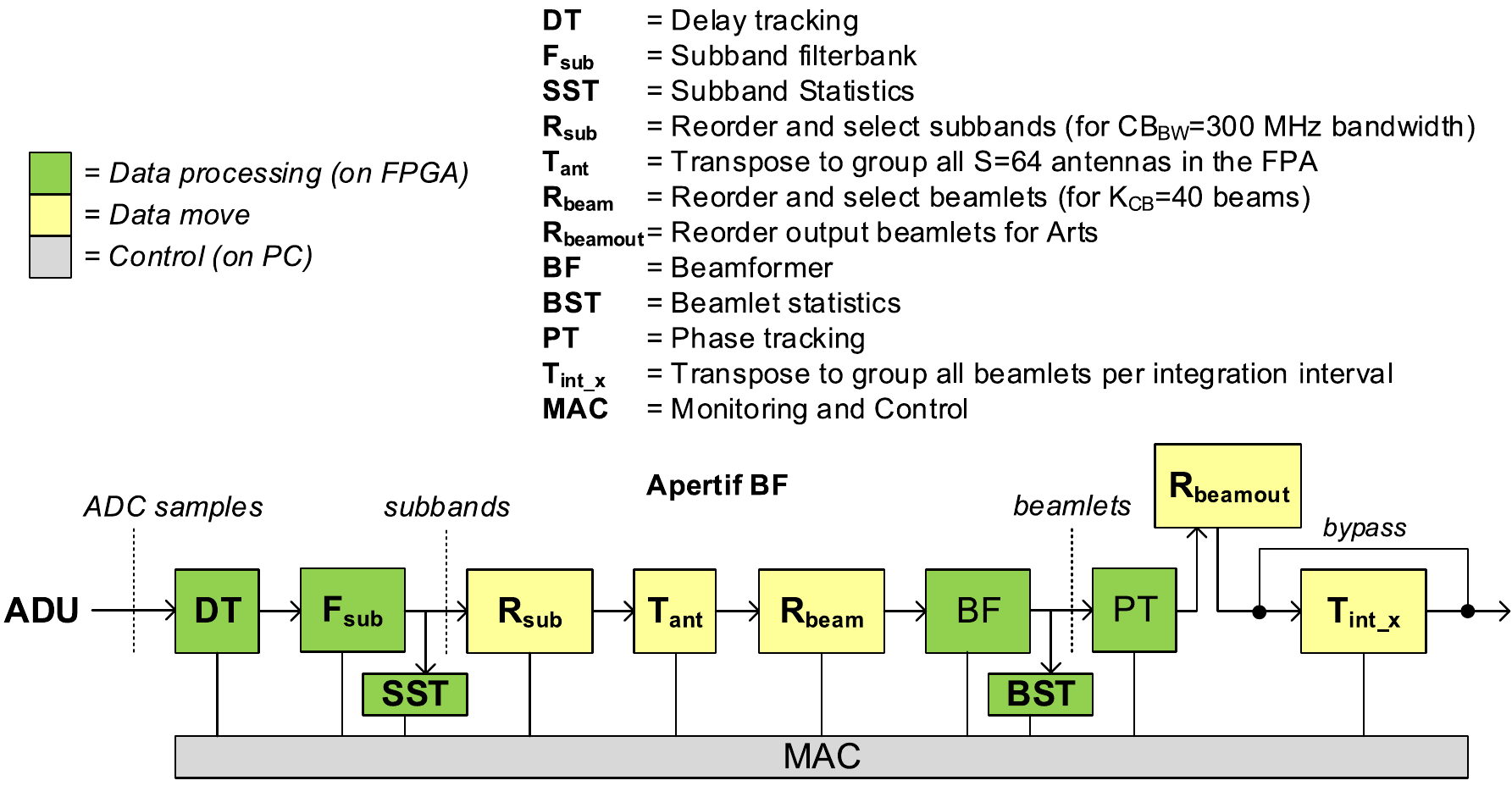}
  \caption{The Apertif \ac{FEBF} subsystem that is input to \ac{ARTS}.
The T$_\mathrm{int\_x}$ transpose is bypassed for ARTS Timing (TAB1).
  }
  \label{td-febf-map-v01}
\end{figure}  

The \aclp{FEBF} (\acsp{FEBF}; \citealt{cho+20}) provide the CB480  data to  \ac{ARTS}.
The \acp{FEBF} first sample the  data of all antenna elements at 800\,MHz.
This produces 800\,MHz $\times$ 8\,bits/sample $\times$ 128\,ADCs/dish $\times$ 12\,dishes = 9.8\,Tbps.
  Of these ADCs, 121 are connected to a PAF receiver element.
  The two polarizations are processed independently.
The sampled data is next separated into 512
  subbands by means of a  \ac{PFB} that uses a 1024-point \ac{FFT}.
  The subband filterbank
(F$_\mathrm{sub}$; see Fig.~\ref{td-febf-map-v01})
  is critically sampled, %
  at 781250\,Hz. 
The \ac{PAF} beamformer then forms \emph{beamlets}, %
a beam formed for one subband. %
This requires a transpose T$_\mathrm{ant}$ that groups subbands from all 64 \ac{PAF} antenna elements per polarization. %
A \ac{CB} is formed from 384 beamlets with identical direction, together
spanning 300 MHz bandwidth (BW). 
With N$_\mathrm{dish}$ = 12 dishes $\times$  K$_\mathrm{CB}$ = 40
compound beams the total output of the \ac{FEBF}
is referred to as CB480.
The CB480 data flows at 
 12\,dishes $\times$ 2\,polarizations $\times$ 40\,CBs $\times$ 300\,MHz $\times$ 2\,complex $\times$ 6\,bit =
3.456\,Tb/s, at 9 Gbps per 10G link (Table~\ref{tab:app:datarates}).

\acf{FS} compensates for the change in geometrical delay between dishes due to Earth  rotation.
In the \ac{FEBF},
it consists of true sample \acf{DT} operating on the \ac{ADC} samples, followed by a residual \ac{PT}
on the beamlets per compound beam.
\ac{FS} is thus valid for the center of each CB.

\subsection{Frequency and bandwidth}
\label{sec:app:design:rf}

\begin{figure}
  \centering
  \includegraphics[width=\columnwidth]{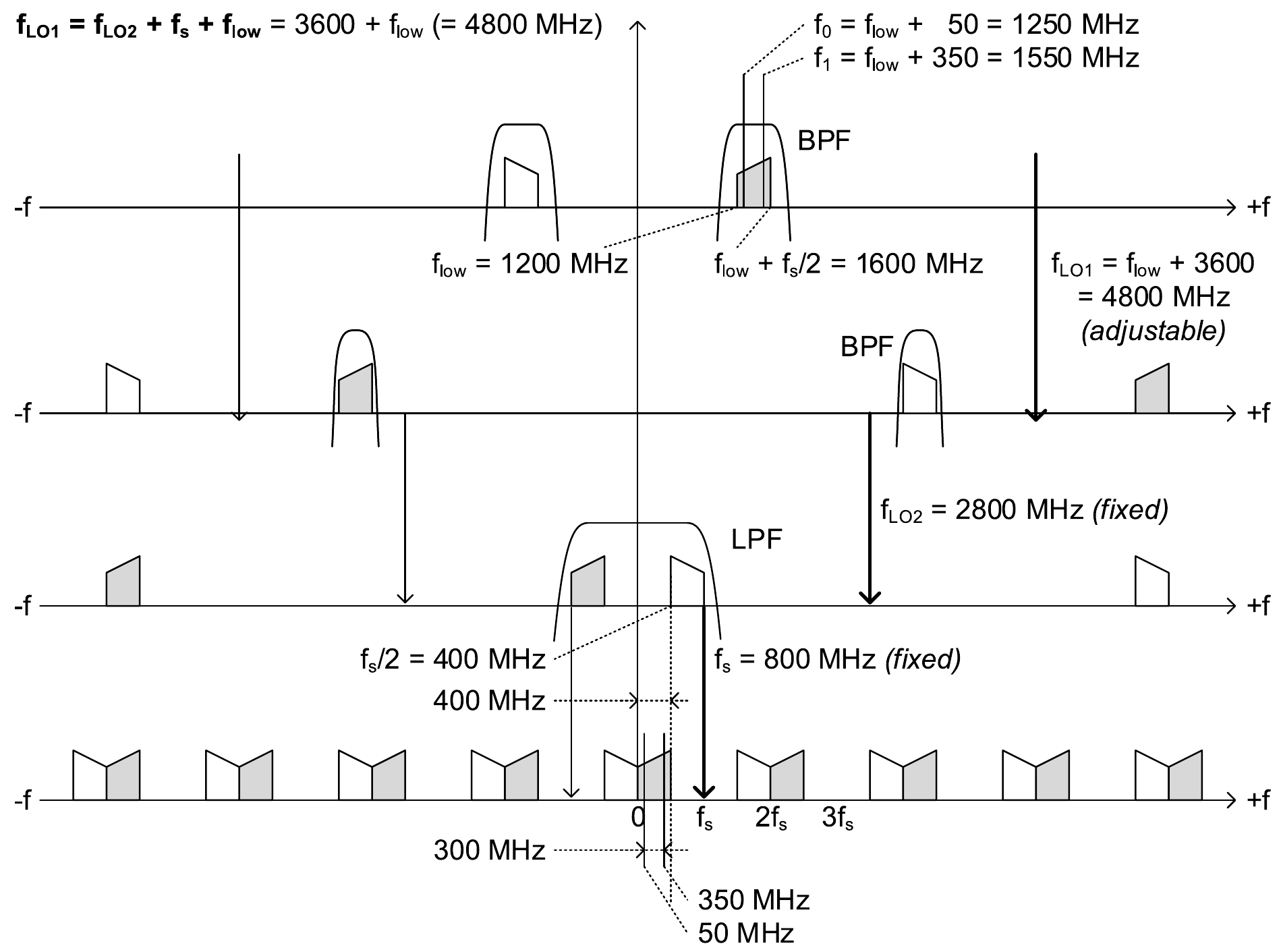}
  \caption{Apertif \ac{FEBF} mixer and subsampling scheme for 300\,MHz beams, from e.g. f$_\mathrm{0}$ = 1250\,MHz to
    f$_\mathrm{1}$ = 1550\,MHz. 
The RF$_\mathrm{BW}$ = 400\,MHz.
The f$_\mathrm{LO2}$ = 2800\,MHz and is fixed.
 By setting f$_\mathrm{LO1}$ appropriately, 
 the f$_\mathrm{low}$ maps to 0\,Hz and f$_\mathrm{0}$ to 50\,MHz.
\label{td-febf-subbands-v01}
    }
\end{figure}

The \ac{FEBF} operates in the 1130$-$1720\,MHz \acf{RF} range \citep{cho+20}.
Fig.~\ref{td-febf-subbands-v01} shows how 400\,MHz is down converted and subsampled, 
and how the central frequency  can be adjusted in steps of 10 MHz.

\begin{figure}
  \centering
  \includegraphics[width=\columnwidth]{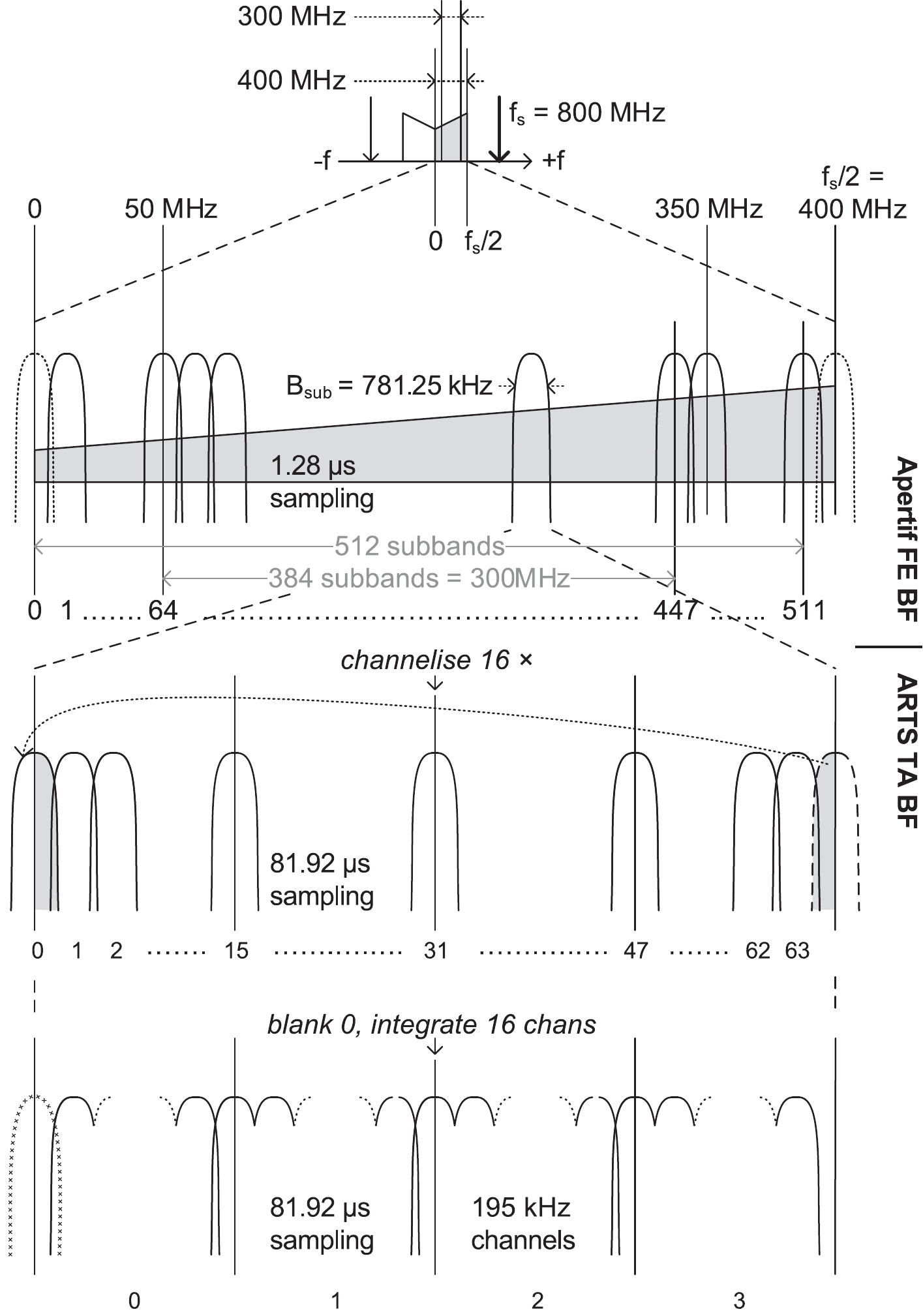}
  \caption{ \acp{PFB} in the Apertif \ac{FEBF}, with \mbox{N$_\mathrm{sub}$ = 512} subbands,
    followed by a second filterbank, plus an integration step, in the \ac{TABF}.
    \label{td-febf-tabbf-subchannels}
    }
\end{figure}

Typically the middle subbands 64 to 447 (50 to 350\,MHz) are selected
(cf. Fig.~\ref{td-febf-subbands-v01} and \ref{td-febf-tabbf-subchannels}).
Down stream, these subbands are split into 4 sub-channels in the \ac{TABF}
(Fig.~\ref{td-febf-tabbf-subchannels}).
By implementing this split as a 64-channel separation followed by a 16-channel integration,
the impact of the  channel leakage in the FFT is reduced, while effectuating the intended downsampling in time
(Sect.~\ref{sec:4b.1.tabf}). 

{The \ac{PFB} for the 64 channels uses a prototype \ac{FIR} filter with 8 taps $\times$ 64 points = 512
  coefficients. The FIR filter
  coefficients\footnote{These
  can be recreated using \texttt{run\_pfir\_coeff.m} with \texttt{application = 'apertif\_channel'} 
  from \url{{https://doi.org/10.5281/zenodo.4748331}} \citep{kooistra_e_2021_4748331}, the code that defines the
  subband and channel characteristics.} 
  are 9 bit real values. The half power bandwidth of the channels is equal to the
  channel frequency (12.2\,kHz) and the stop band attenuation is
  $>$55\,dB.}

The early design choice for 400/512\,MHz = 781250 Hz channels means frequency channels need to be converted to tie
Apertif into e.g. LEAP for pulsar timing \citep{2016MNRAS.456.2196B}  and into standard VLBI.
Both use (multiples of) 1\,MHz channels.

\subsection{Data formats and interface data loads}
\label{sec:app:design:dataformats}

\begin{figure}[b]
  \centering
  \includegraphics[width=\columnwidth]{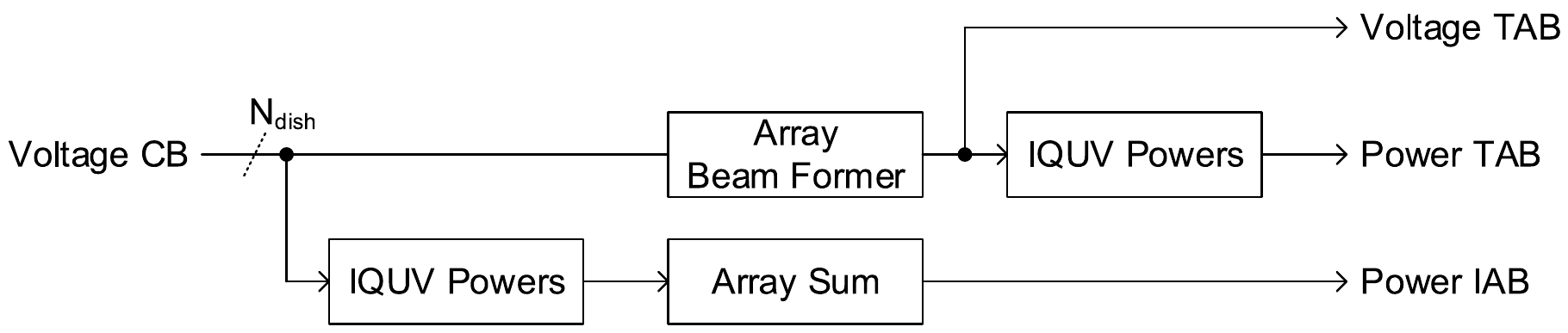}
  \caption{Coherent and incoherent array beamforming in ARTS.
    \label{fig:app:coh_bf}
  }
\end{figure}

The ARTS beamformers (Sect~\ref{sec:app:design:bfalgos}) produce data in voltage and/or power data formats
(Fig.~\ref{fig:app:coh_bf}).

For voltage beams, such as the CB and the TAB1 for Timing, %
the  data vector %
consists of the complex
samples for both polarizations.
The search modes use integrated Stokes power data beams, 
such that samples can be integrated to achieve data reduction (see Sect.~\ref{sec:5b.1}).
\label{sec:app:design:dataload}

Table~\ref{tab:app:datarates}  summarizes the data loads on the ARTS \ac{BF} interfaces.
The default input data is CB480 for a dedicated search (Sect.~\ref{sec:app:design:febf}).
For  commensal searching (Sect.~\ref{sec:4e}), the input data load in CHAN320 mode is
12\,dishes $\times$ 2\,polarizations  $\times$ 40\,CB  $\times$ 300\,MHz  $\times$ 2\,complex  $\times$ 9\,bit = 3.456\,Tbps.

The default search output is 12 TABs per \ac{CB} for TAB480.
Optionally each first TAB can be replaced by an \ac{IAB} for IAB40.  
The search mode internally first creates TAB480 voltage streams,
totaling
12\,dishes $\times$ 2\,polarizations $\times$ 40\,CB $\times$ 300\,MHz $\times$ 2\,complex $\times$ 8\,bit = 4.608\,Tbps.
For each TAB the output consists of both a Stokes-I and a Stokes-IQUV data stream.
As the Stokes I data (for real time detection) and IQUV data (for data buffering)
have different purpose and format, the Stokes I data is included (again) in the  IQUV data stream.
For downstream performance the Stokes-I streams are
demultiplexed and re-ordered on the \ac{TABF} such that the time index is fastest changing index. 
The  Stokes-IQUV streams remain in the original multiplexed format.

\begin{table}
  \caption{Data-rate loads for ARTS \ac{TABF}     interfaces.}
    \centering
    \begin{tabularx}{\columnwidth}{llllllr}
    \toprule
    \bf{Mode}  & \bf{Data} & \bf{Dish} & \bf{CBs} & \bf{TABs} & \bf{N$_{int}$} & \bf{Load}  \\
       & Type & N   & N     &  N    &    &  Gbps\\ 
     \toprule
       CB480   & CV & 12 & 40 &  - &  - & 3456  \\ 
       CHAN320 & CV &  8 & 40 &  - &  - & 3456  \\ 
       TAB1    &CV & 12 &  1 &  1 &  - & 9.6    \\ 
       TAB480  &CV   &           12 & 40 & 12 & -  & 4608  \\
       TAB480  &IQUV &           12 & 40 & 12 & 16 & 288   \\
       TAB480  &I    &           12 & 40 & 12 & 16 & 72    \\
       \multicolumn{2}{l}{~~~~ I+IQUV/node}   &    12 & 40 & 12 & 16 & 9     \\
       IAB40  & IQUV &            12 & 40 &  - & 16 & 24    \\
    \bottomrule
    \end{tabularx}
    \tablefoot{The number of samples per integration is N$_{int}$ = 16, for 81.92\,$\mu$s sampling.
    In CV modes the data are complex voltages. I and IQUV modes are detected Stokes data.
    I+IQUV/node denotes the load of the combine Stokes I and IQUV data per GPU workstation, 
    at (288\,+\,72)\,/\,40 = 9\,Gbps.
  }
    \label{tab:app:datarates}
\end{table}

\subsection{Beamforming Algorithms}
\label{sec:app:design:bfalgos}

The coherent beamformer weights and sums the input voltage data from dishes $d$.
The weights $w_\mathrm{TAB}$ are complex values that are purely geometric and generally static (Sect.~\ref{sec:5b.1}).
The coherent beamformer operates separately per polarization $p$.
The %
\emph{single} voltage output data $vTAB$, 
that directly operates on the input beamlets $b$, is
  \mbox{$vTAB(p, b) =  \sum^{N_\mathrm{dish}-1}_{d=0} w_\mathrm{TAB}(d, b) ~\times~ CB(p, d, b)$}.

For the intermediate voltage \acp{TAB}  used  in the \emph{survey} modes,
the same weight %
is used for all subchannels:
  \mbox{$vTAB(p, s) =  \sum^{N_\mathrm{dish}-1}_{d=0} w_\mathrm{TAB}(d, b) ~\times~ CB(p, d, s)$}.
These vTABs are first converted into full Stokes power data and next integrated over
N$_\mathrm{int\_a}$ = 16 subchannels into each channel $c$, resulting in the integrated power $pTAB$ stream to the search \ac{PL}:
  \mbox{$pTAB(c) =  \sum^{N_\mathrm{int\_a}-1}_{s=0} \mathrm{Stokes}(vTAB(p, s))$}
In contrast to the TABs, the incoherent beamformer already operates on  power data (see Fig.~\ref{fig:app:coh_bf}).
Weights are generally 0 or 1.
It, too, next integrates over the subchannels: %
  \mbox{$IAB(c) =  \sum^{N_\mathrm{int\_a}-1}_{s=0} \,\, \sum^{N_\mathrm{dish}-1}_{d=0} w_\mathrm{IAB}(d)~\times~\mathrm{Stokes}(CB(p, d, s))$}.

\subsection{FPGA Usage}
\label{sec:app:design:fpgares}

Each  \ac{UNB} uses 8 Stratix IV FPGAs.
Each  \ac{UNB2} uses 4 Arria10 FPGAs.
Both are clocked at 200\,MHz.

The FPGA usage of the \emph{connectivity} is defined by the required optical links.
Each entire \ac{UNB} (with \ac{OEB}, see Sect.\ref{sec:app:design:hw:u}) and
each individual FPGA on \ac{UNB2} can connect 24 optical 10G
links. The 384 10G links that come in from the \ac{FEBF}
thus determine that the \ac{TABF} requires at least 16 UniBoards or 4 \ac{UNB2}.
The usage of the \emph{logic} in the FPGA is split over the utilization of the flip flops (FF) and the lookup tables
(LUT). Each Stratix IV  has 182,400 FF, each Arria10 has 1,708,800 FF. 
The \emph{memory} in the FPGA is organized in \acp{BRAM},
that can also be used as FIFOs.
The \emph{multiplier} usage follows from
Sect.~\ref{sec:app:design:bfalgos} and the number of TABs/CB.
Two telescope paths are multiplexed %
to approach 100\% %
\ac{DSP} efficiency. %
The 16 \acp{UNB} can thus form up to 14 TABs/CB, where only 12 are required for the dedicated search
(cf.~Eq.~\ref{eq:NTAB}, Sect.~\ref{sec:app:beams:TAB}). 
The 9 TABs/CB required for commensal search 
can exactly fit 4 \ac{UNB2}, but only if channel filterbanking is done in the Apertif \acl{X} (Sect.~\ref{sec:4e}).
Table~\ref{tab:app:fpgaresources} lists the %
FPGA resource usage in the \ac{TABF}s.

\begin{table}
  \caption{The FPGA resources used versus their availability, over the ARTS \ac{TABF}.}
  \centering
    \begin{tabularx}{\columnwidth}{lrrr}
    \toprule
            &\hspace{-10mm}\bf{FPGA}&\bf{\# BRAM}&\bf{\# multipliers}\\
    &           & {$\times$ 1000} & {$\times$ 1000}      \\
    \toprule
    16 \acs{UNB} &  128 & 150 / 158 M9K (94\%)  &  122 / 165 (74\%)\\ 
    ~~4 \acs{UNB2} &  16 &  41 / 43 M20K (95\%) &   22 / 24 (90\%)\\ 
     \bottomrule
    \end{tabularx}
    \tablefoot{Listed are the \acf{UNB} and \acf{UNB2} systems, 16 and 4 boards respectively.
      For the commensal search on \acl{UNB2}, resources for the channeliser that also serves the correlator
      are not included. 
      The BRAM and multiplier usage is noted as {\tt Used/Available [Type] (Percentage)}.
      \label{tab:app:fpgaresources}
    }
\end{table}

\section{Detailed Design II: Hardware}
\label{sec:app:design:hw}

\subsection{Relevant general subsystem: Uniboards for \ac{FEBF}}
\label{sec:app:design:hw:febf}

\begin{figure}[b]
  \centering
  \includegraphics[width=0.8\columnwidth]{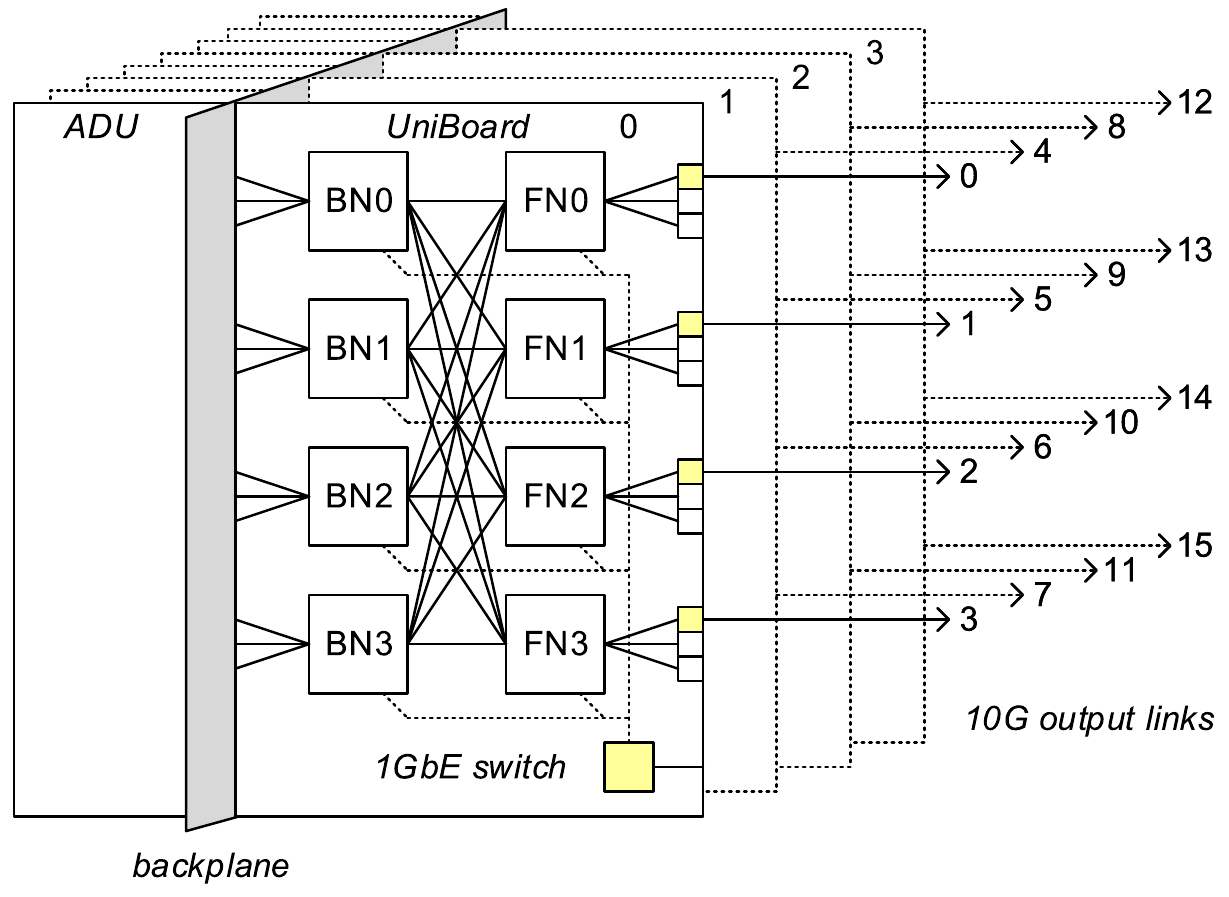}
  \caption{One \ac{FEBF} subrack with 4 \acp{UNB} per telescope path. The FPGAs are marked \ac{BN} and \ac{FN}.
    \label{td-febf-subrack}
    }
\end{figure}

Each \ac{FEBF} subsystem of Fig.~\ref{td-febf-map-v01} 
processes one polarization from one dish for all \acp{CB} on one %
subrack (Fig.~\ref{td-febf-subrack}).
There are thus 24 of these, each containing  4 \acp{UNB}.
Through a backplane the \acp{UNB} connect to each other, and to 8 \ac{ADU} boards that together host 64 \acp{ADC}, to
digitize the data from the 61 antenna elements in the \ac{PAF} \citep{cho+20}.
On each \ac{UNB}, the data is channelized on the 4 \acf{BN} FPGAs and beamformed on the 4 \acf{FN} FPGAs.
The transpose T$_\mathrm{ant}$ is implemented through a combination of the backplane and the mesh interconnects.
The 384 \ac{FEBF} 10G links output the CB480 data to the \ac{TABF}.

\subsection{UniBoards and \acp{OEB} for \ac{TABF}}
\label{sec:app:design:hw:u}

\begin{figure}[t]
  \centering
  \includegraphics[width=0.70\columnwidth]{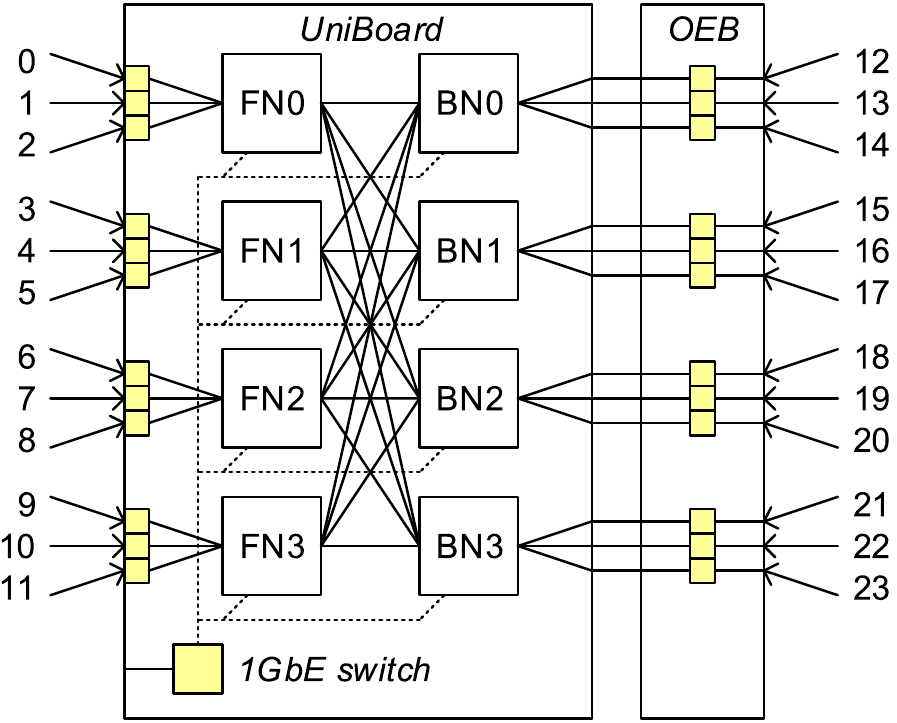}
  \caption{One UniBoard + \acl{OEB}. This unit processes 18.75\,MHz for 24 telescope paths.%
    \label{td-bf-board+oeb}%
    }%
\vspace{-1ex}%
\end{figure}%

\begin{figure}[b]
  \centering
  \includegraphics[width=0.6\columnwidth]{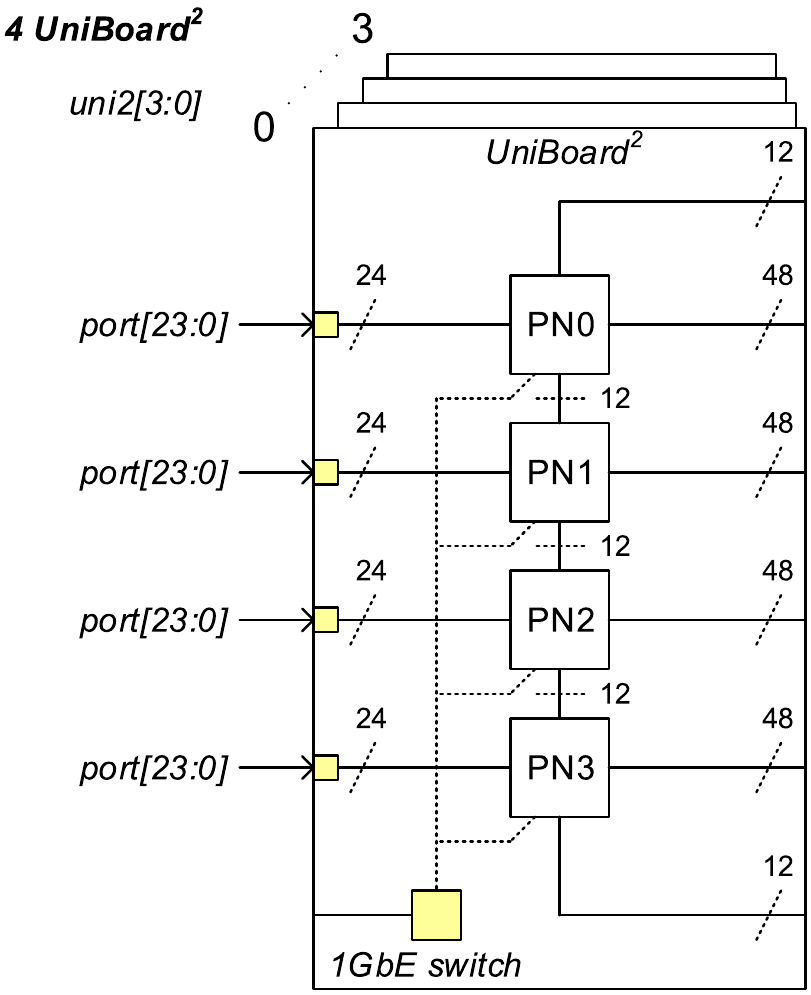}
  \caption{The independent \ac{PN} on the 4 \ac{UNB2}s used for the commensal  \ac{TABF}.
    \label{td-bf-4unb2s-v01}
  }
\end{figure}  

Fig.~\ref{td-bf-board+oeb} shows the \ac{UNB}  with the \ac{OEB}.
The data processing in both the Apertif \ac{FEBF} and the ARTS \ac{TABF} Uniboards is clocked at 200 MHz. 
All are locked to the same 10\,MHz reference as the \acp{ADC} sample clock (Sect.~\ref{sec:app:design:hw:febf}),
ensuring the processing rate in all FPGAs is held fixed
relative to each other.
Through this 10\,MHz reference the long term stability of the ADC sample clocks is guaranteed by the WSRT maser \citep{cho+20}.
The 8 FPGAs on \ac{UNB} are StratixIV type EP4SGX230KF40C2.
In contrast to the \ac{FEBF},
there is no distinction between \ac{FN} and \ac{BN} in the \ac{TABF}.
All FPGAs have the same function.
That is possible because the \acp{OEB} extends the \acp{BN} with fiber optics IO.
Each FPGA supports three 10G links.
On these, each FPGA receives one polarization %
 from three dishes, for the same  18.75\,MHz  band.
On the \ac{UNB},
the mesh, consisting of  5\,Gbps transceiver links,
next interconnects the FPGAs as shown in Fig.~\ref{td-bf-board+oeb}. 
For the dedicated FRB survey, the data are reordered on this mesh such that each \ac{FPGA}
processes the data from five \acp{CB}.
One \ac{UNB} plus \ac{OEB} provides full duplex \ac{IO} for 240\,Gbps.
In total there are 128 \acp{FPGA} on the 16 independently-operating \ac{UNB} in the \ac{TABF}.

\subsection{\aclp{UNB2} for \ac{TABF}}
\label{sec:app:design:hw:u2}

On \ac{UNB2} all \acp{FPGA} are connected identically. There are no \acp{BN} or \acp{FN}, only \acp{PN}.
There are 16 such \acp{FPGA}, type Arria10 10AX115U2F45E1SG,  in the 4 \acp{UNB2} comprising the \ac{TABF}, as shown in
Fig.~\ref{td-bf-4unb2s-v01}. In most respects, such as connectivity, transposing, and signal processing, one \ac{PN} on
\ac{UNB2} replaces an entire \ac{UNB}.

\subsection{Connecting UniBoard to \ac{UNB2}}
\label{sec:app:design:hw:u2u2}

The 128 \acp{FPGA} on the 16 \acp{UNB} are each connected to the Apertif \ac{FEBF} via 3 incoming Rx-only 10G links.
Outgoing data is carried over the corresponding Tx-only link.
The first 10G link transmits the output for the dedicated search.
The other 256 ports connect directly from the \ac{UNB} to the \ac{UNB2}, as shown in Fig.~\ref{td-bf-map-v01}.

\subsection{Transpose using \ac{COTS} 10/40GbE switches}
\label{sec:app:design:transpose}
\begin{figure*}
  \centering
  \includegraphics[width=\textwidth]{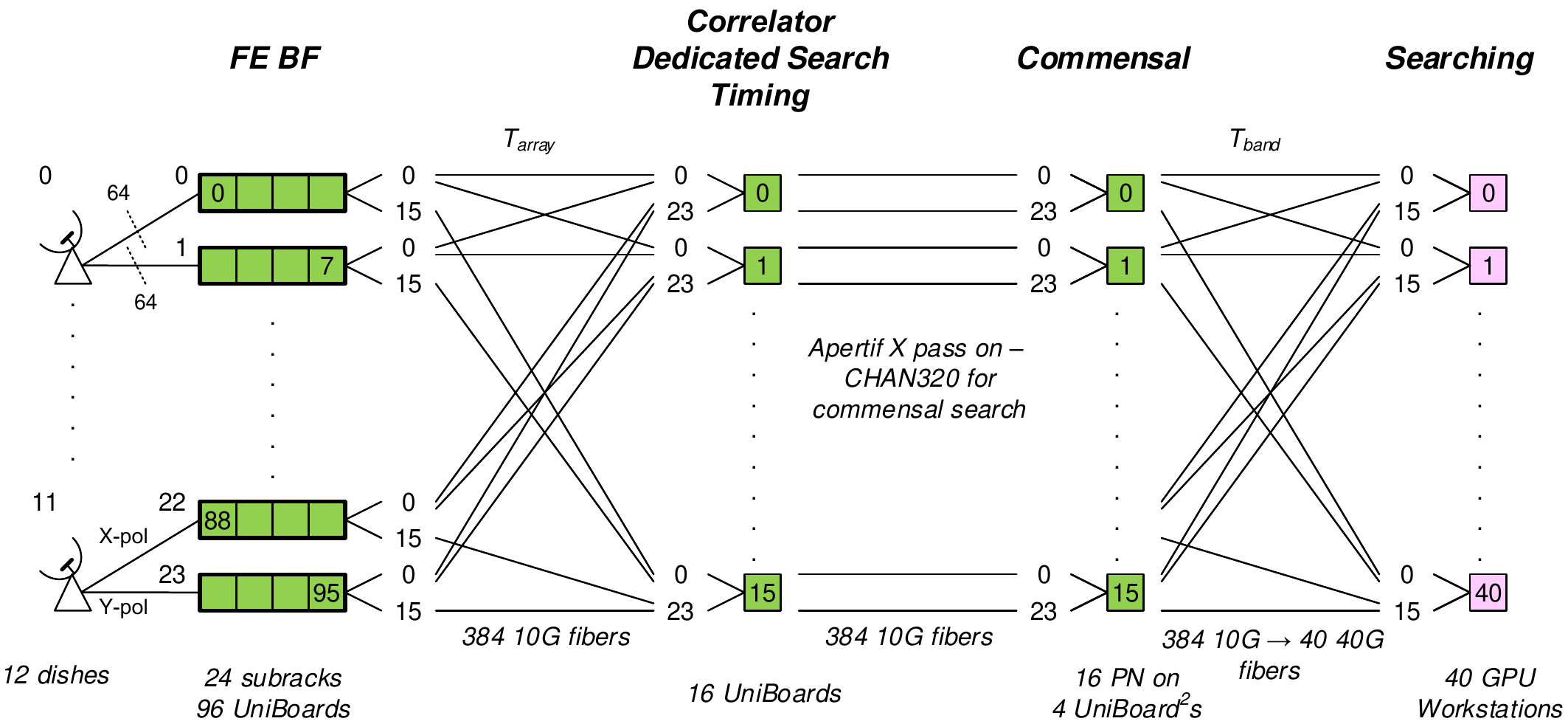}
  \caption{The interconnectivity in the Apertif data streams. Data moves from left to right,
  through the \acl{FEBF}, \acl{X}, \aclp{TABF}, and \acl{PL} on the GPU nodes.
  \label{td-interconnect-map}
  }
\end{figure*}  
The transpose T$_\mathrm{band}$ 
needs to bring together all 16 bands formed in the \ac{TABF},
to have the full bandwidth per \ac{CB} available in a single \ac{PL} workstation (Fig.~\ref{td-interconnect-map}).
Due to the redistribution on the \ac{UNB} mesh, the 40 CBs are already grouped in sets of 5. 
The transposer can thus operate in 8 independent  parts.
Fig.~\ref{td-bf-map-v01}
shows a more detailed diagram for the transposer function. 
Each of four \ac{COTS} 10/40GbE switches handles two of the 8 parallel parts.
Initial designs required a custom transposer, %
but the switches were able to handle the uni-directional links produced by \ac{UNB} and \ac{UNB2}.

In commensal-search mode, the \acp{UNB} operate as correlator, and only pass on the 
CHAN320 data to the \acp{UNB2}.
One third necessarily  passes through the same
 128 10G connections mentioned in Sect.~\ref{sec:4d}, and goes through the switches,
 as indicated by the dashed links Fig.~\ref{td-bf-map-v01}.
The other two thirds is 
 directly passed on from UniBoard
 via 256 10G links.
By thus routing the \ac{UNB} data through \ac{UNB2} the number of required ports in the transposer switches is minimized.
 The \ac{UNB2} output of 270 Gbps is transposed and sent to the \ac{PL}.
This is a factor 9/12 less than for dedicated search, because commensal mode has 9 TABs.

The Dell S6000 switches can handle the necessarily unidirectional traffic. At the output side, 
each switches connects to 2x5 workstations using one 40GbE port per
workstation, and to the other switches.

\section{Detailed Design III: Firmware and Software}
\label{sec:app:design:fwsw}

\subsection{Firmware}
\label{sec:app:design:fw}

\subsubsection{Firmware Timing}
\label{sec:app:design:fw:timing}

The sync interval in the CB480 data is 1.024\,s, which  corresponds to 800000 time samples per beamlet.
An external \ac{PPS} defines the start of the first block of \ac{ADC} data at all \acp{FEBF}.
The timing information is
kept with the streaming data by means of a \acf{BSN} that counts subband periods.
As the PPS is aligned to the top of second of \ac{UTC}, the \ac{BSN} can be 
related to the wall clock time.

\begin{figure}[b]
  \centering
  \includegraphics[width=1\columnwidth]{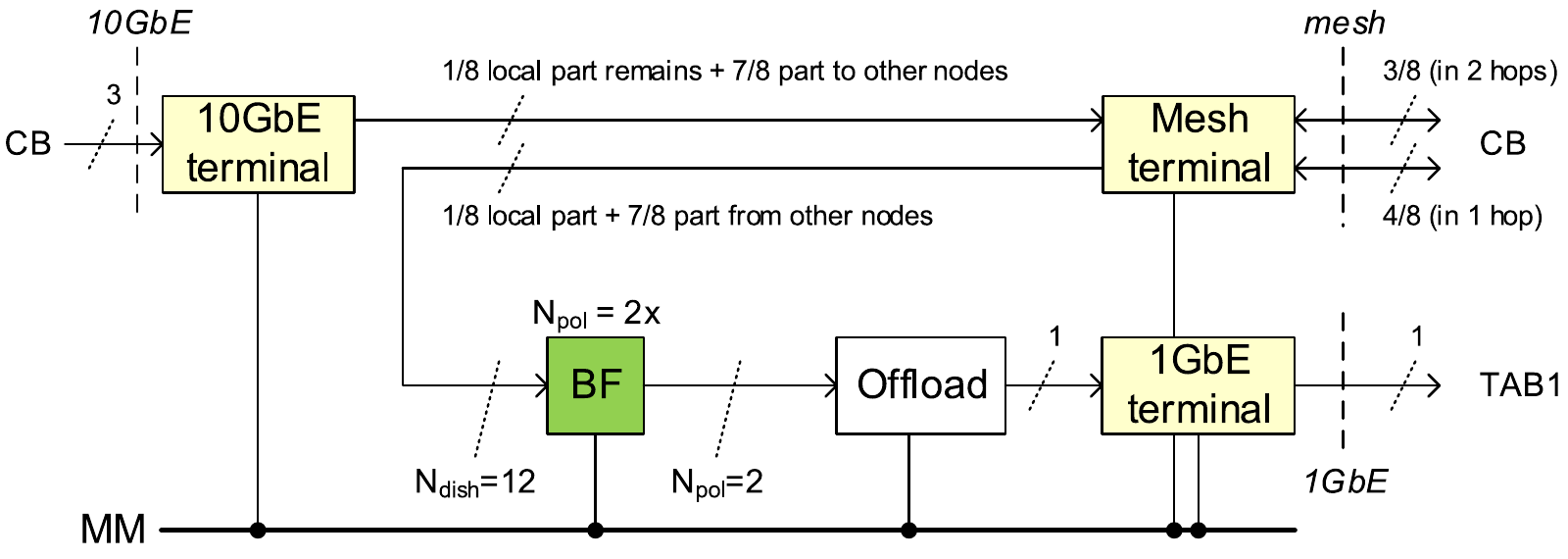}
  \caption{
    Top level block diagram for the single-TAB timing FPGA design on Uniboard.
    \label{td-fw-sc1-v01}
  }
\end{figure}  

\subsubsection{Requantization in the \ac{TABF}}
The CB input is 6 bits wide.
Internally the voltage TAB, power TAB and power IAB calculations are done using up to 18 bit.
Only the channel filterbank \ac{PFB} calculations are done using 9 bit \ac{FIR} filter coefficients, to save multipliers.
The \ac{FFT} part uses 18 bit. 
Generally multiplications are done using the 18x18 bit multipliers, and 18-bit  RAM.
At the output the \acp{TAB} are resized to 8 bit. If overflow occurs, the output value is clipped, preserving sign.

\subsubsection{TAB1 processing on \ac{UNB}}
\label{sec:app:design:fw:TAB1}

The TAB1 beam former
(Fig.~\ref{td-fw-sc1-v01}) 
is based on the \ac{FEBF}.
For its input, the transpose T$_\mathrm{int\_x}$ in the \ac{FEBF} (Fig.~\ref{td-febf-map-v01}) is bypassed.
The UniBoard mesh re-distributes the CB480 input over the \acp{FPGA} as in Sect.~\ref{sec:5b.1}.
Through this programmable routing, the FN0 of all 16 Uniboards together receive the full
300\,MHz for the central CB.
Each of the 16 FN0 output nodes %
offloaded 600\,Mbps of data via the 1GbE control interface of the UniBoards, as shown in Fig.~\ref{td-bf-map-v01}.

\subsubsection{IAB40 processing on \ac{UNB}}
\label{sec:app:design:fw:IAB}

The optional \ac{IAB}  beamformer   sums the input powers from the 12 input dishes.
It operates on subchannels,  integrates these, and  outputs them as channels. 
The IAB function itself does not have weights.

\subsubsection{TAB360 processing  on \ac{UNB2}}
\label{sec:app:design:fw:UNB2}

\begin{figure}[b]
  \centering
  \includegraphics[width=\columnwidth]{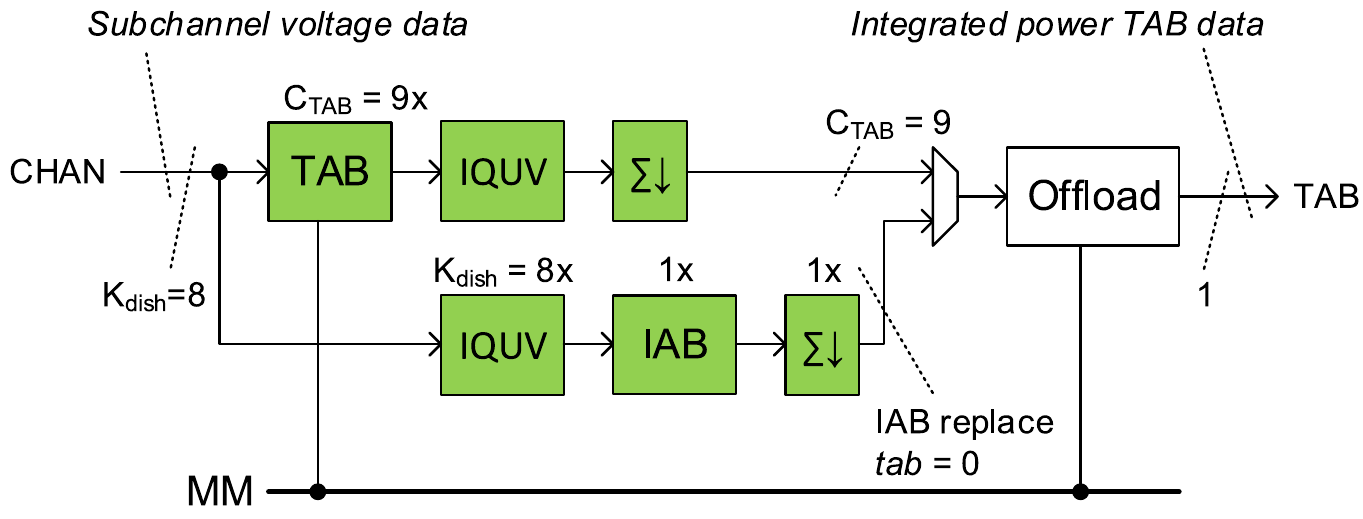}
  \caption{Block diagram for the commensal-search TAB360 processing.
  }
  \label{td-fw-sc3-v01}
\end{figure}

For the commensal search, the same design is used on all \ac{UNB2} FPGAs.
The block diagram for the TAB360 processing (40 CBs $\times$ 9 TABs)
per 10G output is shown in Fig.~\ref{td-fw-sc3-v01},
and is very similar to the \ac{UNB} version  (Fig.~\ref{td-fw-sc4-v01}).

\subsection{Monitoring \& Control software}
\label{sec:app:design:sw}

\subsubsection{Architecture}
\label{sec:app:design:sw;arch}
The Apertif system is built as a set of loosely coupled components that communicate over the network by exchanging
messages over a message bus. %
For this, we employ a messaging middleware layer, which ensures reliable communication.

\subsubsection{Message Bus}
\label{sec:app:design:sw:mb}
The message bus consists of queues, exchanges, and brokers. %
Messages are wrapped in an envelope that contains the address of the recipient, and sent to an exchange.
A broker helps forward the message to a destination queue for pick up by the recipient.
Messages queuing %
provides reliable
communication, guaranteed message delivery, and is easy to use. Routing %
is run-time configurable. %

Apertif uses %
the Advanced Message Queuing
Protocol (AMQP), which is based on IEEE standard (ISO/IEC 19464). Apache Qpid provides an implementation of AMQP, in
both C++ and in Python. The choice for Qpid was driven by experience %
in LOFAR.
We run a broker daemon on every host. %
A single exchange
routes messages to their destination queues.
Each component has queues for difference  message types.

\subsubsection{Layered Design}
\label{sec:app:design:layers}
The Monitoring \& Control software follows a layered design, from
drivers (lowest), to controllers, to orchestration (highest).
The drivers interface with hardware, %
over Ethernet. %
All drivers are written in Python although the UniBoard driver is partly in C++ for performance.
The controllers, next, %
handle Message Bus  commands and serialize access to the drivers. %
Some high-level commands %
are sent to the whole system (e.g.,  "Start Observation"). %
Other, low-level commands are subsystem or driver oriented (e.g., %
"get\_status").
Different applications in the orchestration layer, finally, %
set up components throughout %
the system as a whole.

\subsubsection{Software Components}
\label{sec:app:design:sw:comps}
Each component has %
a functional %
and a management interface
\citep{cho+20}. %
Both  support normal-priority synchronous  function calls,
and high-priority asynchronous calls.
Each component also generates events and notifications.

\subsubsection{Controllers}
\label{sec:app:design:sw:frame}
All basic, shared functions of the controllers 
are implemented through a controller framework, 
that has  three main tasks. %
A RequestHandler places %
new messages from the bus %
into the command queue. An ExecutionHandler %
schedules and executes these queued commands.
A ResponseHandler monitors the resulting %
response queue. %

\subsection{Science-pipeline software}
\label{sec:app:design:sw:science}

 \begin{figure}
   \centering
   \includegraphics[width=1\columnwidth]{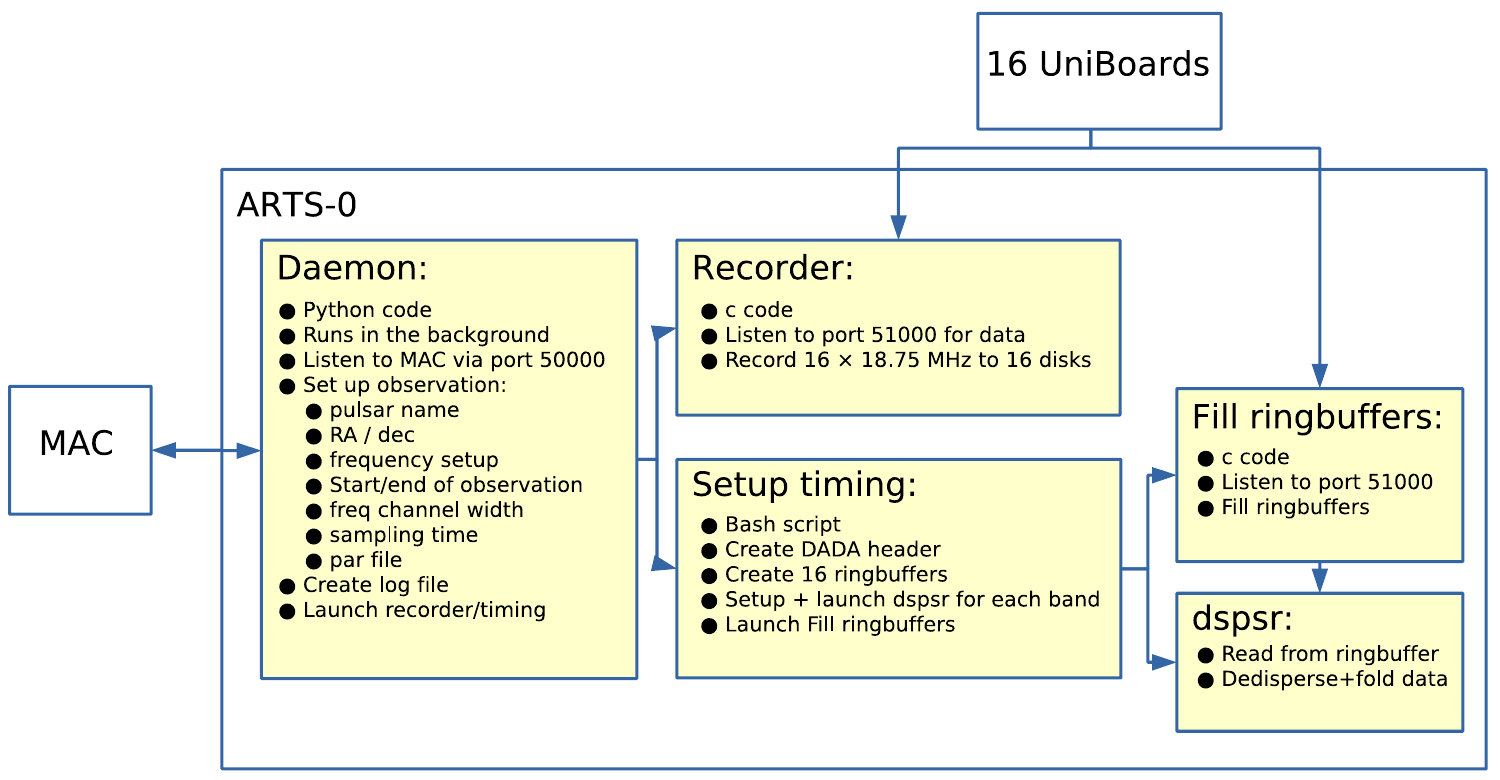}
   \caption{
     Schematic overview of the ARTS timing software pipeline
     \label{app-sw-sci-sc1}
   }
 \end{figure}

\subsubsection{Pulsar timing}
\label{sec:app:design:sw:timing}
The components of the pulsar timing pipeline are detailed in Fig.~\ref{app-sw-sci-sc1}.
The arts-0 machine contains two sets of NIC, CPU, and GPU. Each set is connected by a dedicated interconnect, and
the data handling and reduction processes are pinned to the appropriate network ports, compute cores, and GPUs.
This streamlining was essential for reaching real-time performance without data loss.

\subsubsection{Transient searching}
\label{sec:app:design:sw:arts}

\begin{figure*}
  \centering
  \includegraphics[width=\textwidth]{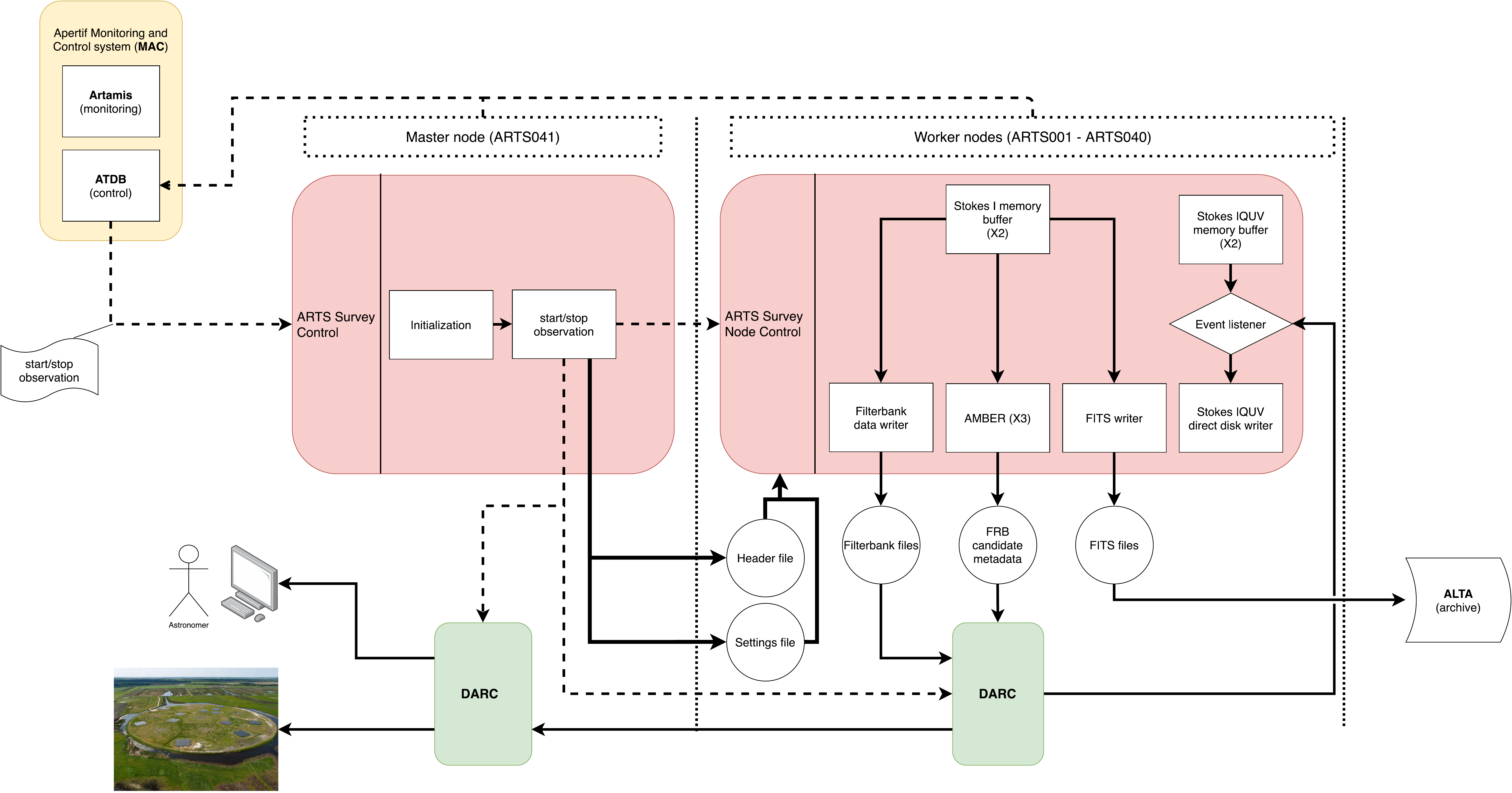}
  \caption{The direct interactions of the \ac{ARTS} specific survey controllers. The controllers are represented by the
    red boxes. Dashed lines represent system control messages. Communication between the monitoring system
    (\acs{Artamis}) and the nodes, as well as any relevant communication between MAC and other components of the system
    (e.g. \ac{ALTA}) are omitted for clarity. Since the stokes IQUV data is post processed before being used, the direct
    disk writer does not show any output. %
  }
  \label{fig:controller}
\end{figure*}

The detailed overview of the \ac{ARTS} cluster survey controllers
described in Sect.~\ref{sec:5c.artsmac} and \ref{sec:6b.1}
 are presented in Fig.~\ref{fig:controller}.

\section{Tied-array beam simulations}
\label{sec:app:beams}

The hierarchical beam forming scheme used for ARTS was introduced at a high level in Sect.~\ref{sec:3d.beam}.
In this appendix we provide a more quantitative discussion;
first on the formation of \acfp{TAB};
next, on combining TABs to \acfp{SB} and \acfp{TB}.

\subsection{Tied Array Beams}
\label{sec:app:beams:TAB}

To determine many TABs are required to cover the \acf{CB},
we will assume that the separation between TABs,
$\theta_\mathrm{sep}$ is equal to their half power beam width (HPBW), which depends on the observing wavelength $\lambda$ and the projected longest baseline $B_\mathrm{max}$ as
\begin{equation}
\theta_\mathrm{sep} = \alpha \frac{\lambda}{B_\mathrm{max} \cos ( \theta_\mathrm{proj})},
\label{eq:theta_sep}
\end{equation}
where $\theta_\mathrm{proj}$ is the angle between the plane perpendicular to the linear WSRT configuration and the line
of sight. We determine $\alpha$ empirically, for $\theta_\mathrm{proj} = 0$. We find that for the 144-m spaced arrays of 8 and 10 dishes
(Apertif-8, dishes RT2$-$RT9, \mbox{$B_\mathrm{max} = 1008$\,m}; and
Apertif-10, comprising RT2$-$RTB, \mbox{$B_\mathrm{max} = 1296$\,m)}, $\alpha = 0.78$ and $\alpha = 0.80$ respectively.

\begin{figure}[b]
    \centering
    \includegraphics[width=\columnwidth]{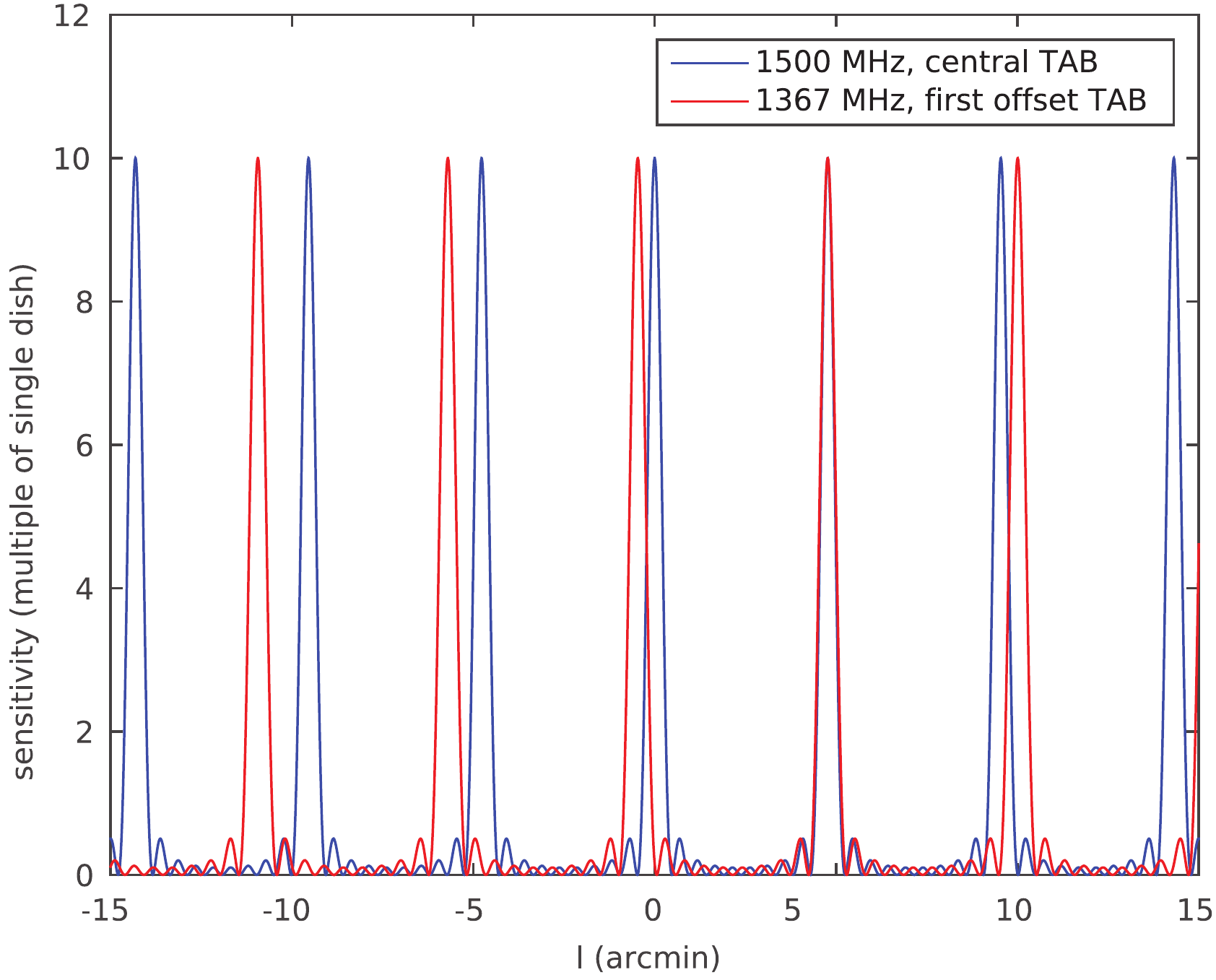}
    \caption{Grating response of the central TAB at 1500\,MHz and the first TAB left from the center at 1367\,MHz for Apertif-10.}
    \label{fig:chromatic_response}
\end{figure}

If the TABs together cover the angular distance equal to that
between a TAB main beam and its first grating response, they provide full coverage of the CB, by a combination of their main beams and first grating response. The grating distance $\theta_\mathrm{grat}$ depends on the observing wavelength and the projected common quotient baseline $B_\mathrm{cq}$ of the regular array as
\begin{equation}
    \theta_\mathrm{grat} = \arcsin \left ( \frac{\lambda} {B_\mathrm{cq} \cos ( \theta_\mathrm{proj} ) } \right ) \approx \frac{\lambda} {B_\mathrm{cq} \cos ( \theta_\mathrm{proj})},
    \label{eq:theta_grat}
\end{equation}
where the approximation holds for small angles.

The number of TABs required per CB, $N_\mathrm{TAB}$, then follows from the ratio of the grating distance and the TAB separation:
\begin{equation}
    N_\mathrm{TAB} = \frac{\theta_\mathrm{grat}} {\theta_\mathrm{sep}} = \frac{B_\mathrm{max}} {\alpha B_\mathrm{cq}}.
    \label{eq:NTAB}
\end{equation}
Note that the wavelength and the projection angle cancel. This implies  the number of TABs required depends on neither the
projection angle nor the frequency.
The number of TABs can be kept unvarying. Using Eq.~\ref{eq:NTAB}, we find that 12 and 9 TABs are needed for Apertif-10 and Apertif-8 respectively.

\subsection{Synthesized Beams}
\label{sec:app:beams:SB}

As shown by Eq.~\ref{eq:theta_grat}, the grating lobe distance is frequency-dependent. This is illustrated in
Fig.~\ref{fig:chromatic_response}, which shows the grating response of the central TAB at 1500\,MHz and the TAB left
from the central beam at 1367\,MHz for Apertif-10. The first grating response right from the center of these two TABs
coincides. This shows that a transient signal received in the first grating response of the central TAB to the right of
the center at
1500\,MHz will not be detected in the corresponding grating response at 1367\,MHz, but by the
corresponding grating response of the first TAB left from the central TAB at 1367\,MHz. For all but the main beam of the
central TAB, we therefore may have to combine chunks of bandwidth from different TABs to form an \ac{SB} at a given distance from the \ac{CB} center.

In principle, there is an optimal combination of chunks of bandwidth from different TABs for each position within the FoV. In practice, the number of Synthesized Beams (SBs) that we can form is limited. A practical minimum number of SBs can be set by considering the TAB separation at the highest operating frequency, i.e., take
\begin{equation}
    \theta_\mathrm{SB} = \theta_\mathrm{sep}(f_\mathrm{max}) = \alpha \frac{\lambda_\mathrm{min}}{B_\mathrm{max}}.
\end{equation}
For Apertif-10, we found $\alpha = 0.80$ and we have $B_\mathrm{max} = 1296$ m. Assuming $f_\mathrm{max} = 1500$\,MHz
and a field-of-view of 30 arcmin, we find that the number of SBs required is 71. 
A grid where SBs are spaced at 0.5~HPBW to improve sensitivity across the field-of-view results in 151 SBs.
While both modes are possible in ARTS, its standard operation uses 71 SBs, based on compute-load trade-offs throughout
the entire search signal chain. 

At a given frequency, the optimal contribution to the SB denoted by index $n_\mathrm{SB}$ is coming from a specific grating response denoted with index $n_\mathrm{gr}$ from TAB with index $n_\mathrm{TAB}$. In the remainder of this section, we determine which grating response from which TAB gives the optimal contribution to a given SB. Since the position shift of the grating responses over frequency increases with distance from the main beam, it is convenient to choose the reference position to be in the center of the CB. This intuitively leads to an indexing scheme in which 0 denotes the central position, a negative index implies a position left or westward of the center and a positive index denotes a position right or eastward of the center. For example, $n_\mathrm{gr} = -1$ denotes the first grating response left from the main beam, $n_\mathrm{TAB} = 0$ denotes the central TAB and $n_\mathrm{SB} = 5$ denotes the synthesised beam located five grid
points to the right of the center of the CB.

We  assign a grating index and a TAB index at the highest frequency to each SB index. We first
determine the closest grating response of the central TAB by
        $n_\mathrm{gr} = [n_\mathrm{SB} / N_\mathrm{TAB}]$,
    where $[\cdot]$ denotes rounding %
    upward towards the next higher integer value.
    Next, we determine the TAB index such that $n_\mathrm{TAB} = \{-N_\mathrm{TAB}/2, -N_\mathrm{TAB}/2+1, \cdots, N_\mathrm{TAB}/2-2, N_\mathrm{TAB}/2-1 \}$ by calculating
    \begin{equation}
        n_\mathrm{TAB} = n_\mathrm{SB} \mod N_\mathrm{TAB}
    \end{equation}
    and subtracting $N_\mathrm{TAB}$ if $n_\mathrm{TAB} > N_\mathrm{TAB}/2-1$.

\begin{figure}
    \centering
    \includegraphics[width=\columnwidth]{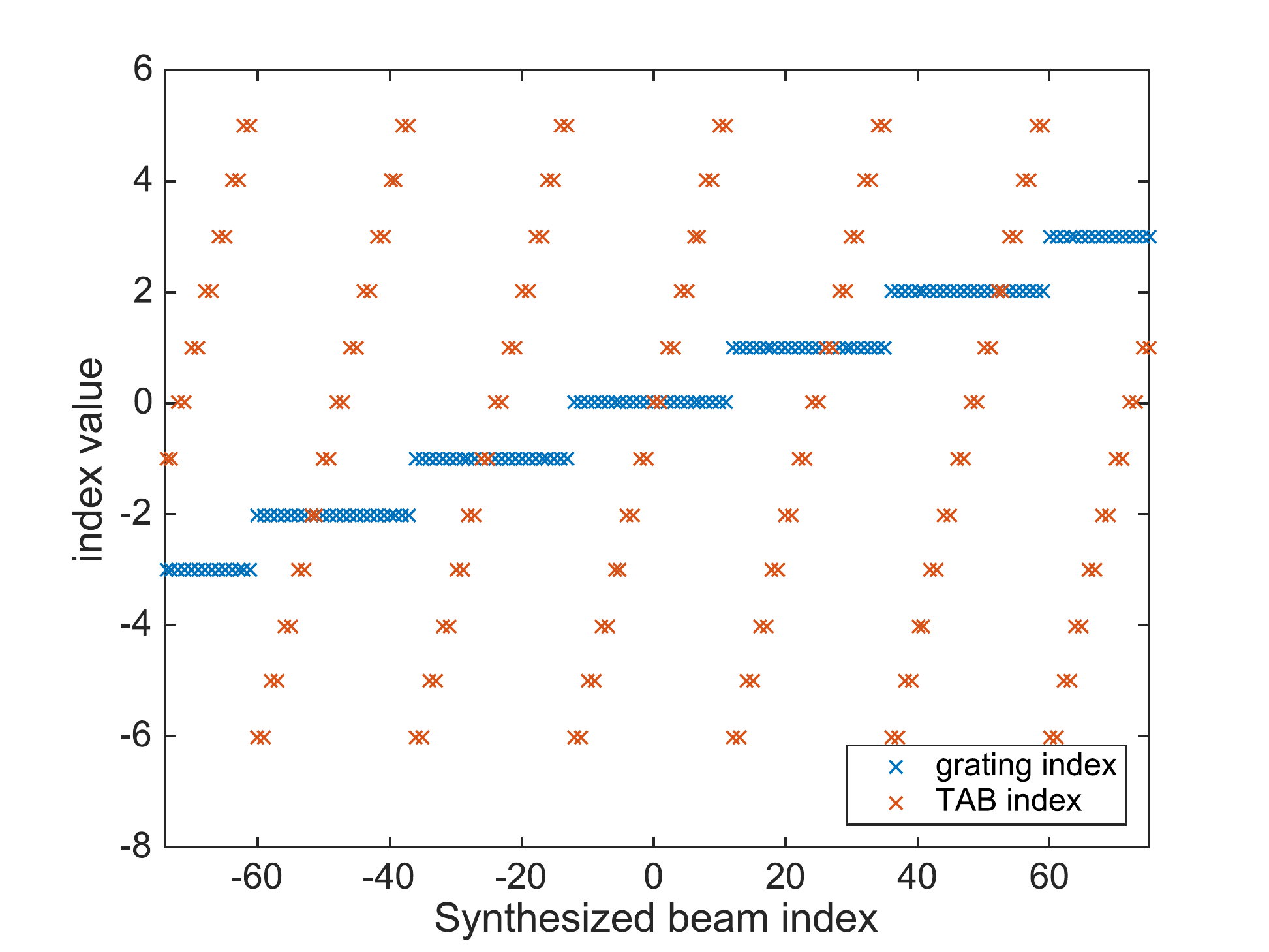}
    \caption{Grating index and TAB index for each SB index at the highest frequency.}
    \label{fig:indices_per_SB}
\end{figure}

Figure \ref{fig:indices_per_SB} shows the resulting grating indices and TAB indices for all SB indices when $N_\mathrm{TAB}=12$. Note that the calculation above holds for even $N_\mathrm{TAB}$. A similar convention can be defined for odd $N_\mathrm{TAB}$. This is not spelled out here to keep this section concise.

The position of the SB with index $n_\mathrm{SB}$ is described by
\begin{equation}
    \theta_{n_\mathrm{SB}} = n_\mathrm{SB} \theta_\mathrm{SB} = n_\mathrm{gr} \theta_\mathrm{grat} (f_\mathrm{max}) + n_\mathrm{TAB} \theta_\mathrm{sep} ( f_\mathrm{max} ).
\end{equation}
Since $\theta_\mathrm{grat}$ and $\theta_\mathrm{sep}$ are frequency-dependent, at a certain (lower) frequency, the
grating response of the next TAB will be at the same position and is therefore the optimal choice at that
frequency.
To find the frequency at which the position of the grating response of the next TAB at frequency $f_0$ coincides with the grating response of the original TAB at $f_\mathrm{max}$, we solve $f_0$ from
\begin{eqnarray}
    \lefteqn{n_\mathrm{gr} \theta_\mathrm{grat} (f_\mathrm{max)} + n_\mathrm{TAB} \theta_\mathrm{sep} ( f_\mathrm{max} ) =} \nonumber\\
    && n_\mathrm{gr} \theta_\mathrm{grat} ( f_0 ) + (n_\mathrm{TAB} - \mathrm{sgn} ( n_\mathrm{SB} ) ) \theta_\mathrm{grat} (f_0),
\end{eqnarray}
where $\mathrm{sgn}$ denotes the signum function. Substitution of Eq.~\ref{eq:theta_sep} and Eq.~\ref{eq:theta_grat} while replacing $\lambda$ by $c / f$ and taking $\theta_\mathrm{proj}=0$, we obtain
\begin{eqnarray}
\lefteqn{n_\mathrm{gr} \frac{c}{f_\mathrm{max} B_\mathrm{cq}} + n_\mathrm{TAB} \frac{\alpha c}{f_\mathrm{max} B_\mathrm{max}} =}\nonumber\\
& & n_\mathrm{gr} \frac{c}{f_0 B_\mathrm{cq}} + (n_\mathrm{TAB} - \mathrm{sgn} ( n_\mathrm{SB} ) ) \frac{\alpha c}{f_0 B_\mathrm{max}},
\end{eqnarray}
which gives
\begin{equation}
    f_0 = f_\mathrm{max} \frac{n_\mathrm{gr} B_\mathrm{max} + (n_\mathrm{TAB} - \mathrm{sgn} ( n_\mathrm{SB} ) ) \alpha B_\mathrm{cq}} {n_\mathrm{gr} B_\mathrm{max} + n_\mathrm{TAB} \alpha B_\mathrm{cq}}.
\end{equation}
Note that a situation in which $n_\mathrm{TAB} - sgn (n_\mathrm{SB}) \notin \{ -N_\mathrm{TAB}/2, -N_\mathrm{TAB}/2+1, \cdots, N_\mathrm{TAB}/2-2, N_\mathrm{TAB}/2-1 \}$ may occur. Since the position of $n_\mathrm{TAB} + N_\mathrm{TAB}/2$ for $n_\mathrm{gr}$ coincides with $n_\mathrm{TAB} - N_\mathrm{TAB}/2$ for $n_\mathrm{gr} + 1$, this can be solved by modifying the indices accordingly. A similar procedure can be followed for $n_\mathrm{TAB} - N_\mathrm{TAB}/2 - 1$.

\begin{figure}
    \centering
    \includegraphics[width=\columnwidth]{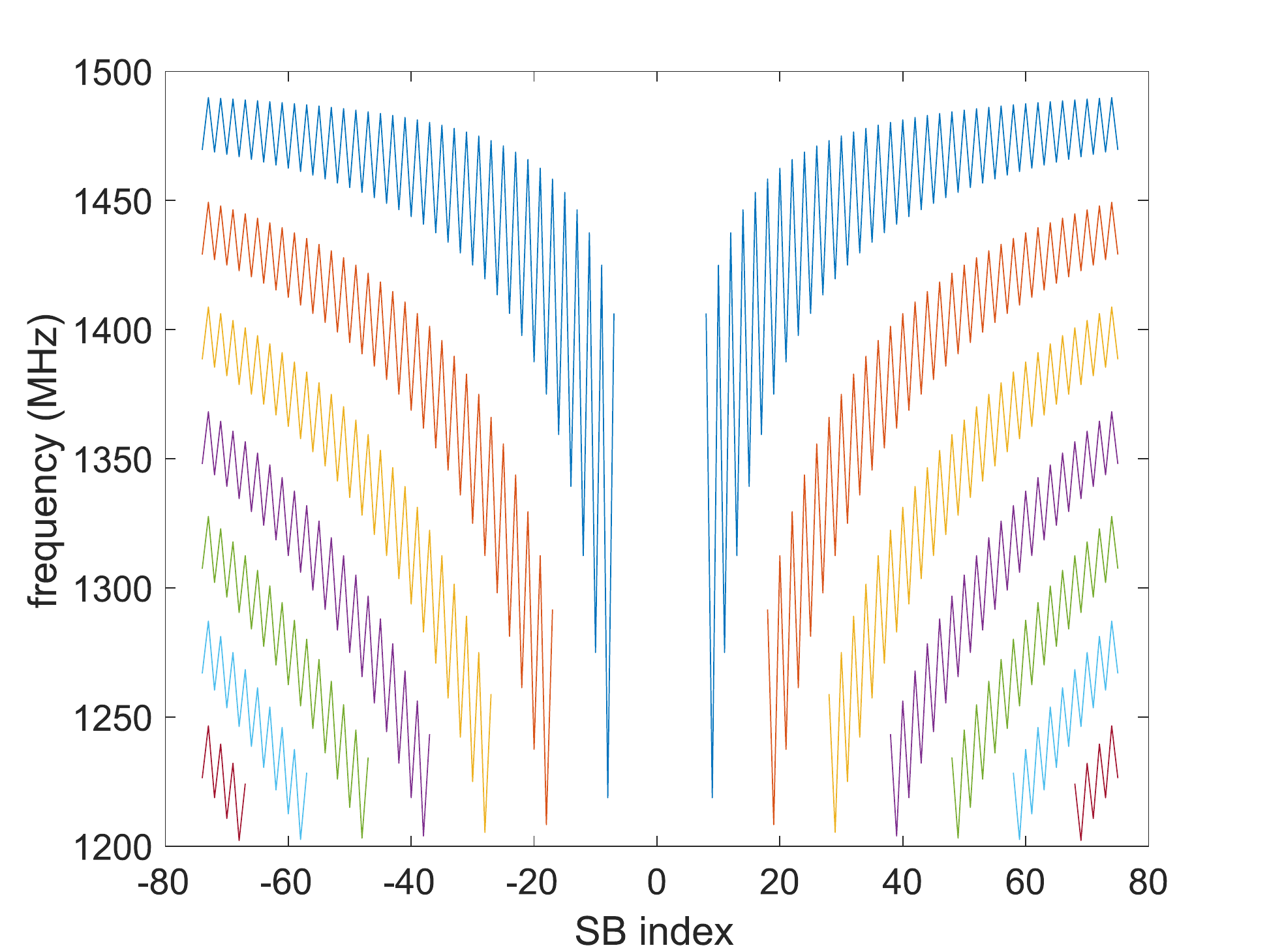}
    \caption{Switching frequencies for all SBs of Apertif-10 covering a FoV of 30 arcmin over the frequency range from 1200 to 1500 MHz.}
    \label{fig:switching_freqs}
\end{figure}

From the discussion above, it is clear that for the $n_\mathrm{SB}$th SB, for which the chunk of bandwidth close to
$f_\mathrm{max}$ is coming from the $n_\mathrm{gr}$th grating response of the $n_\mathrm{TAB}$th TAB, the optimal chunk
of bandwidth around $f_0$ is provided by the $n_\mathrm{gr}$th grating of the $(n_\mathrm{TAB} - \mathrm{sgn}
(n_\mathrm{SB}))$th TAB. Since the HPBW changes only slowly with frequency and the shift of the grating responses scales
linearly with frequency, the optimal switching frequency will be approximate halfway $f_\mathrm{max}$ and $f_0$ if
consecutive SBs are separate by the half power beam width while the optimal switching frequency will be at approximately
a quarter of this frequency interval if consecutive SBs are separated by half the HPBW. The latter applies to ARTS when
using 151 SBs, so we determine  the switch frequency as
    $f_\mathrm{sw} = 0.75 f_\mathrm{max} + 0.25 f_0$
 for even SBs, and as
    $f_\mathrm{sw} = 0.25 f_\mathrm{max} + 0.75 f_0$
for odd SBs. The resulting switching frequencies are shown in Fig.~\ref{fig:switching_freqs}.

\begin{figure}
    \centering
    \includegraphics[width=\columnwidth]{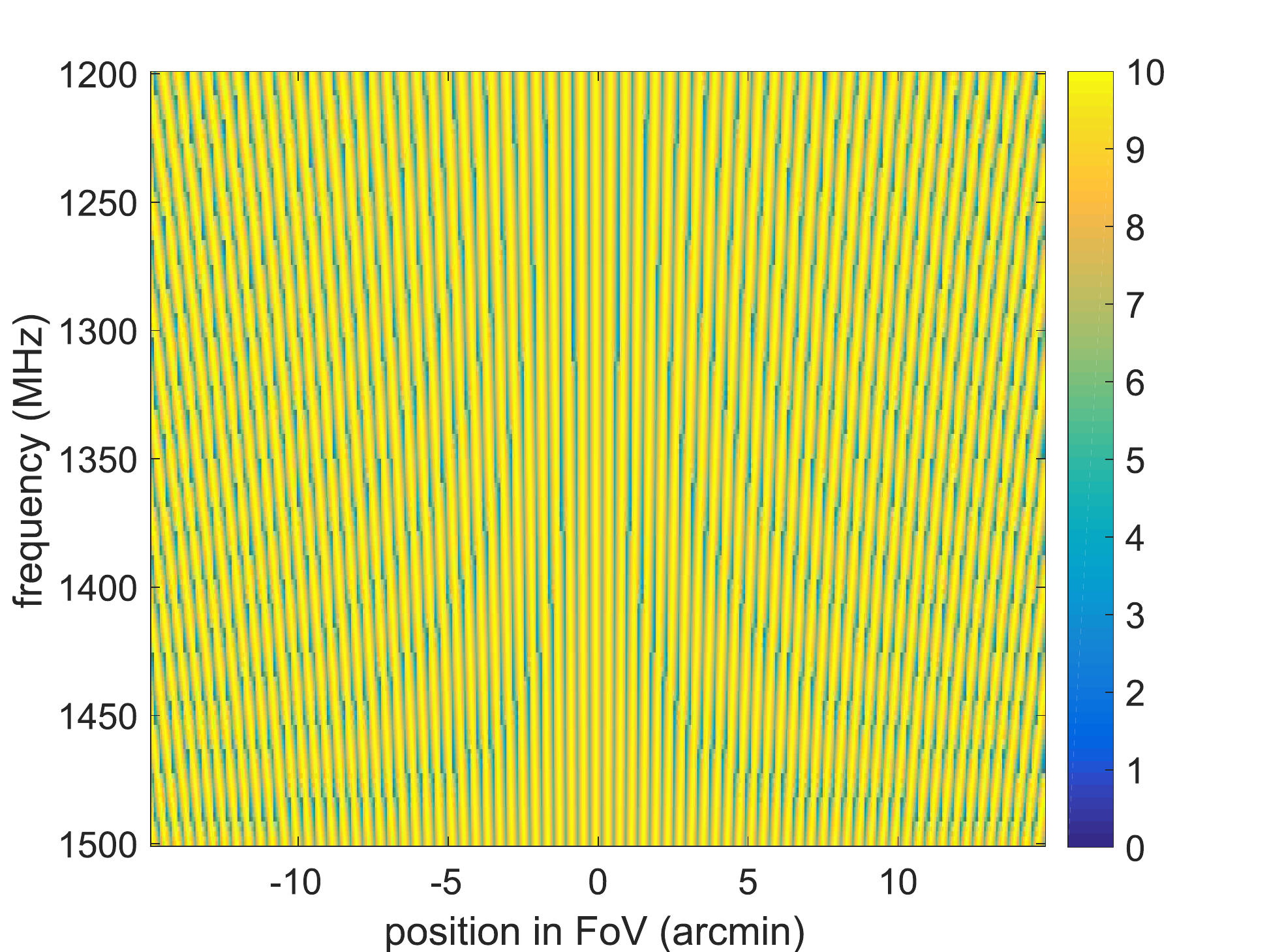}
    \caption{Sensitivity within the CB field-of-view as function of frequency when using the switching frequencies shown in Fig.~\ref{fig:switching_freqs}. The sensitivity is expressed in terms of the sensitivity of a single WSRT dish.}
    \label{fig:sensitivity_over_FoV}
\end{figure}

Fig.~\ref{fig:sensitivity_over_FoV} shows the sensitivity, expressed relative to the sensitivity of a single WSRT dish, within the CB field-of-view as function of frequency when using the switching frequencies calculated above. These sensitivity data indicate that the 151 SBs described above provide an average sensitivity across the full FoV of 81\% of the maximum achievable sensitivity of the WSRT array. The latter would require to form a TAB phase centered at each individual point within the CB field-of-view, i.e., it would, strictly speaking, require the formation of an infinite number of TABs, which is practically infeasible.

\subsection{Tracking Beams}
\label{sec:app:beams:TB}

The grating response of each TAB rotates around the center of the FoV of the CB during an observation. This is illustrated in Fig.~\ref{fig:highlevelbeams}e. A given source may therefore traverse multiple SBs during an observation. To track a specific source (or position within the CB) during an observation, we may thus have to concatenate time domain data from multiple SBs. This section describes a procedure to determine which time intervals from which SBs need to be combined to track a desired position within the CB, i.e., to form a \acf{TB}.

\begin{figure}[b]
    \centering
    \includegraphics[width=\columnwidth]{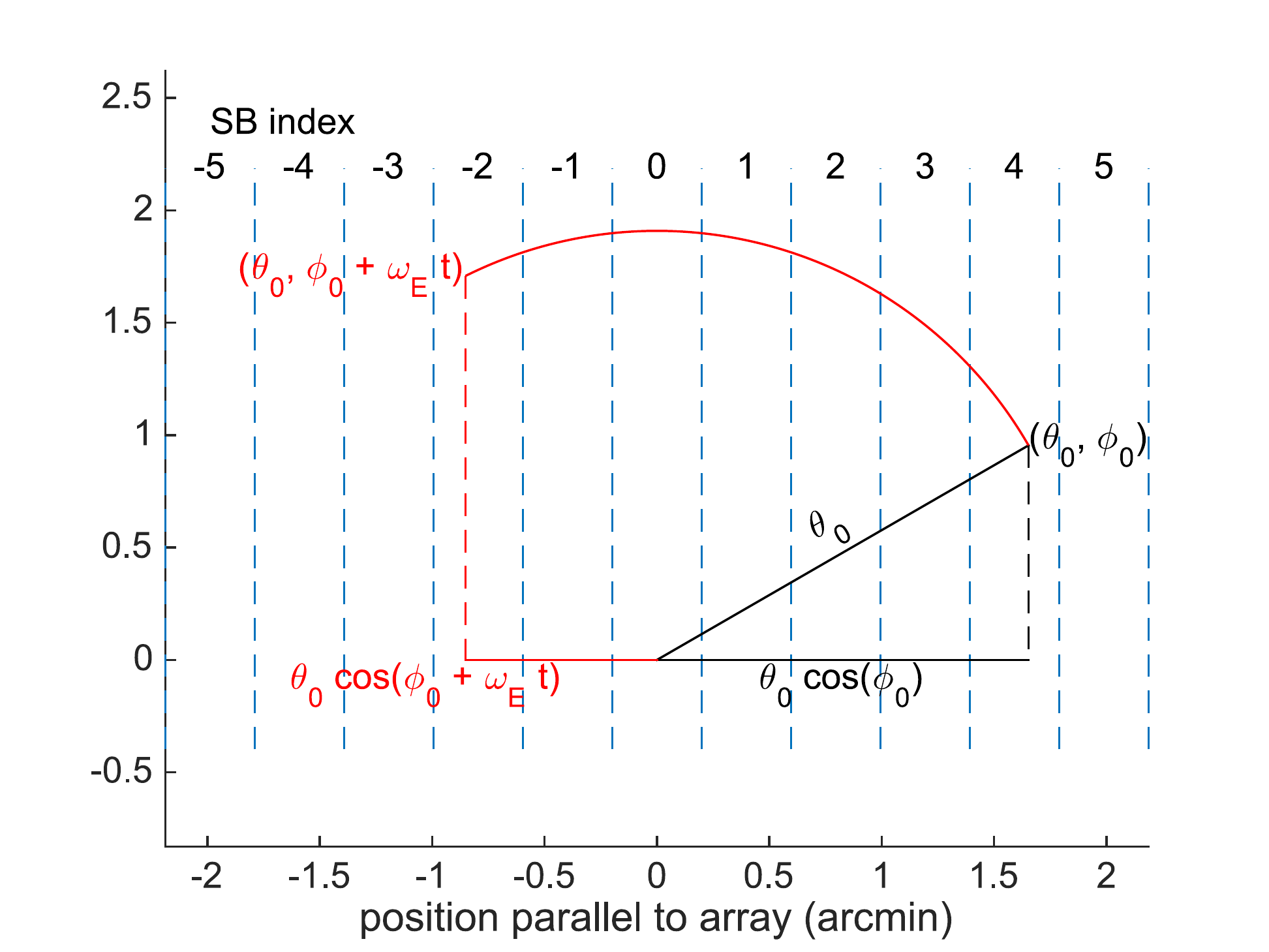}
    \caption{Movement of the locus at cylindrical coordinates $(\theta_0; \phi_0)$ at the start of the observation through the synthesised beams during an observation.}
    \label{fig:locus_TB}
\vspace{3mm}
\end{figure}

Fig.~\ref{fig:highlevelbeams}e illustrates how the TAB gratings rotate through the FoV during an observation.
A specific point within the CB thus moves through the TAB gratings.
This perspective, where the coordinate system is
fixed to the TAB grating response, is shown in Fig.~\ref{fig:locus_TB}. At a specific reference time $t = 0$, a specific locus can be specified by cylindrical coordinates $(\theta_0; \phi_0)$, where $\theta_0$ measures the distance from the field center and $\phi_0$ measures the angle between the line from the field center to the locus and the line parallel to the array, i.e., the line orthogonal to the grating responses. During an observation, this locus will follow a circular path through the CB with an angular velocity given by $\omega_\mathrm{E}$ as indicated by the red track.

The cross-over points between SBs are indicated in Fig.~\ref{fig:locus_TB} by vertical dashed blue lines. The area
between two such lines is associated with a specific SB that can be identified by its SB index as described in
Sec.~\ref{sec:app:beams:SB}.
If we define the central SB as index 0, we can find the SB associated with a specific locus at a specific instant $t$
during an observation by finding out in which SB the point $\theta_0 \cos (\phi_0 + \omega_\mathrm{E} t)$ lies
(Fig.~\ref{fig:locus_TB}):
\begin{equation}
    n_\mathrm{SB} = \left [ \frac{ \theta_0 \cos ( \phi_0 + \omega_\mathrm{E} t )} {\theta_\mathrm{SB}} \right ],
    \label{eq:SB_for_locus}
\end{equation}
where $[\cdot ]$ denotes rounding.

Each point in the compound beam is covered by the grating response of one of the SBs at $t =
0$. Eq.~\ref{eq:SB_for_locus} provides the SB index at each instant $t$ during the observation. From this,
we  construct a \ac{TB} for the full observation for a given point in the CB.
The time series associated with these TBs cover the full length of the observation and therefore allow detection of weaker sources than will be feasible to detect during the drift time for an individual SB.

To cover the full field-of-view of the CB, we can define a hexagonally close-packed grid of TBs. As the CBs themselves are arranged in a similar fashion, the ratio between the hexagonal area covered by a TB in a full synthesis observation and the hexagonal area of a CB is
then equal to  $(\theta_\mathrm{SB}/\theta_\mathrm{CB})^2$. The number of TBs needed to cover a CB will be approximately the inverse, i.e.,  $(\theta_\mathrm{CB}/\theta_\mathrm{SB})^2$. 
If the maximum number of TABs is used to fill the space between the grating responses of Apertif-10, $\theta_\mathrm{SB} = 0.398$ arcmin. If we assume that $\theta_\mathrm{CB} = 30$ arcmin, we will need to synthesise about 5700 TBs. For shorter observations, this number may be reduced as the TBs will have an elongated shape in the direction perpendicular to the array at the mid-point of the observation, which would allow for a larger separation between TBs along that direction.
\end{appendix}

\end{document}